\documentstyle[prd,aps,floats]{revtex}
\input{psfig.tex}

\newcommand{\beq}{\begin{equation}}
\newcommand{\eeq}{\end{equation}}
\newcommand{\bea}{\begin{eqnarray}}
\newcommand{\eea}{\end{eqnarray}}

\def\rp{\rho_\phi}
\def\pp{p_\phi}
\def\wp{w_\phi}
\def\Mp{M_{\rm Pl}}
\def\dl{d_{\rm L}}
\def\Dl{D_{\rm L}}
\def\ol{\Omega_\Lambda}
\def\om{\Omega_{\rm m}}

\def\op{\Omega_\phi}
\def\Dm{\Delta m}
\def\sm{\sigma_{\rm mag}}
\def\ss{\sigma_{\rm sys}}

\def\rom{\rho_{\rm m}}
\def\ddl{\delta\dl}
\def\wt{{\tilde w}}

\draft
\tighten

\begin{document}

\title{Future supernovae observations as a probe of dark energy}
\author{Jochen Weller$^1$ and Andreas Albrecht$^2$\\
$^1$ DAMTP, Centre for Mathematical Sciences, Wilberforce Road,
Cambridge, CB3 0WA, U.K. \\
$^2$ Department of Physics, University of California at Davis, CA
95616, U.S.A.}
\maketitle

\begin{abstract}
We study the potential impact of improved future supernovae data on
our understanding of the dark energy problem. We carefully examine the 
relative utility of different fitting functions that can be used to
parameterize the dark energy models, and provide concrete reasons why
a particular choice (based on a parameterization of the equation of
state) is better in almost all cases. We discuss the details of a
representative sample of dark energy models and show how future
supernova observations could distinguish among 
these. As a specific example, we consider the proposed ``SNAP''
satellite which is planned to observe around 2000 supernovae.  We show 
how a SNAP-class data set taken alone would be a powerful
discriminator among a family of models that would be approximated by a
constant equation of state for the most recent epoch of cosmic
expansion.  We show how this family includes most of the dark energy
models proposed so far. We then show how an independent measurement of 
$\Omega_{\rm m}$ can allow SNAP to probe the evolution of the equation 
of state as well, allowing further discrimination among a larger class 
of proposed dark energy models.  We study the impact of the satellite
design parameters on this method to distinguish the models and compare
SNAP to alternative measurements. We establish that if we exploit the
full precision of SNAP it provides a very powerful probe.

\end{abstract}


\section{Introduction}

One of the most challenging problems in modern cosmology is to
provide an explanation for the recently observed accelerated expansion of
the universe\cite{Perlmutter:97,Riess:98,Perlmutter:99a}. These
observations have reopened the quest for the {\em cosmological
constant} which was introduced by Einstein\cite{Einstein:17}, but later
abandoned\cite{Einstein:31} and infamously cited as his greatest
blunder\cite{Gamow:70}. The cosmological constant can be considered as
new kind of ``world matter''\cite{deSitter:17} and be identified
with the energy density of the vacuum\cite{Zeldovich:68}, explaining
and computing it in terms of particle physics has
been largely unsuccessful\cite{Weinberg:89,Carroll:00} because it is
very difficult to explain the small vacuum energy density of
$10^{-120}\Mp^4$ within fundamental physics; typically it is either much
larger or exactly zero. 

In recent years, the type Ia supernovae (SNe) as standard candles have
been used to
measure the distance-redshift relation in the universe providing evidence
for an energy component in the universe which behaves like a cosmological
constant\cite{Perlmutter:97,Riess:98,Perlmutter:99a}. This means the
pressure of this component is negative and it appears to be dark in a
sense that it is not recognizable by direct
observation\cite{Perlmutter:99b}.  The 
Supernovae Cosmology Project (SCP)\cite{Perlmutter:97,Perlmutter:99a}
found evidence for a positive cosmological constant on the 99\% level. These findings seem to be confirmed if one combines the
most recent cosmic microwave background (CMB) radiation data from the
BOOMERanG (Balloon Observations of Millimetric Extragalactic Radiation
and Geomagnetics)\cite{Boomerang:00,Lange:00,Boomerang:01,Boomerang:01b}, MAXIMA (Millimeter
Anisotropy eXperiment IMaging
Array)\cite{Maxima:00,Balbi:00,Maxima:01} and DASI (Degree Angular
Scale Interferometer)\cite{Dasi:01a,Dasi:01b} experiments 
with observations of rich clusters \cite{Mohr:99,Dodelson:00a}. 

With these observations we need a deeper understanding of the
cosmological constant and attempts have been made to explain the
missing energy as the energy density in a scalar field, which only
interacts with the other fields via gravity. This field is rolling
slowly down
a potential or gets trapped in a local
minimum\cite{Reuter:87,Peccei:87,Wetterich:88,Ratra:88,Peebles:88,Wetterich:95,Frieman:95,Coble:97,Ferreira:97,Ferreira:98,Copeland:98,Caldwell:98,Zlatev:99,Steinhardt:99,Binetruy:99,Brax:99,Choi:00,AS:00}. Therefore the vacuum energy of the
universe becomes important for its evolution and the expansion begins
to accelerate, generalizing the concept of the cosmological
constant. Attempts have been made to connect
this field to fundamental physics\cite{Frieman:95,Brax:99} and resolve
the problem of fine tuning of initial conditions\cite{Steinhardt:99}. The
problem is that there is a plethora of models which can describe
the observed expansion, but with the current available data it is not
possible to distinguish between most of them.

To improve the observational
situation a satellite mission -- the ``SuperNovae Acceleration Probe''
(SNAP) --\cite{SNAP} and other dedicated SNe surveys have been
proposed \cite{Wang:00}. This satellite may
observe about $2000$ SNe within two years and therefore
increase the number of SNe by a factor
of $25$. In this paper we present the details of a representative
sample of dark energy models, discuss how the use of SNe as
standard candles can distinguish the
different models and what can be established about the
equation of state of the dark energy component. Current upper bounds from
SNe observations on the equation of state are $\wp \le -0.6$
\cite{Perlmutter:99b,Garnavich:98,Efstathiou:99}. In order to reconstruct not only
the constant contribution to the equation of state it is convenient to
fit the SNe magnitude - redshift relation with a continuous function
\cite{Starobinsky:98,Efstathiou:99,Huterer:99,Saini:99,Maor:00,Astier:00,Weller:00b}. 

The main
purpose of this paper is twofold: First we compare the quality of two
different fits to the luminosity distance -- redshift relation, where
we emphasize the importance of a ``good'' fit in order to draw
conclusion about the quality of the experiment. Secondly we analyze
how SNe observations can constrain the equation of state factor and
content of dark energy in the universe. A whole section presents the
different dark energy models which we use for this analysis.

The outline of the paper is as follows. In section \ref{curobs} we describe the
current situation of the SNe observations, in section
\ref{snap} we describe briefly the specifics of the proposed SNAP
satellite mission, in section \ref{models} we introduce the commonly
studied dark energy models, the parameters we choose for them and their cosmological evolution. In section \ref{rec} we discuss how to
reconstruct the equation of state 
by expanding the equation of state factor as a power series in
redshift and fit for the expansion coefficients. In section \ref{err}
we discuss the impact of the experimental design and prior constraints
on the matter content. Finally, section \ref{alt} presents alternative
measurements, before we draw our conclusions in \ref{conc}. 

\section{The current observational situation}\label{curobs}

In this paper we concentrate on the results of the distant type Ia
SNe observations \cite{Perlmutter:97,Riess:98,Perlmutter:99a}
and mention other indications just briefly. The SCP and the High-Z
Search Team used bright type Ia SNe as standard candles. These
objects are thought to be thermonuclear explosions of carbon-oxide
white dwarfs\cite{Trimble:82,Trimble:83,Woosley:86}. The correlation
between the peak luminosity and the decline
rate of the luminosity of the
SNe\cite{Barbon:73,Pskovskii:77,Pskovskii:84,Phillips:93} makes it
possible to estimate its magnitude and with spectral information about the host one
can determine its redshift. The correlation can be quantified by
the drop in magnitude $15$ days after the peak luminosity is reached. 
The SNe observations by the SCP are
calibrated using the ``low'' redshift Cal\'an/Tololo
survey\cite{Hamuy:93} which revealed that
they have an excellent distance precision of $\sm = 0.15\; {\rm
mag}$ for the magnitude and therefore can, in fact, be used as {\em standard candles}.

\begin{figure}[!h]
\setlength{\unitlength}{1cm}
\centerline{\hbox{\psfig{file=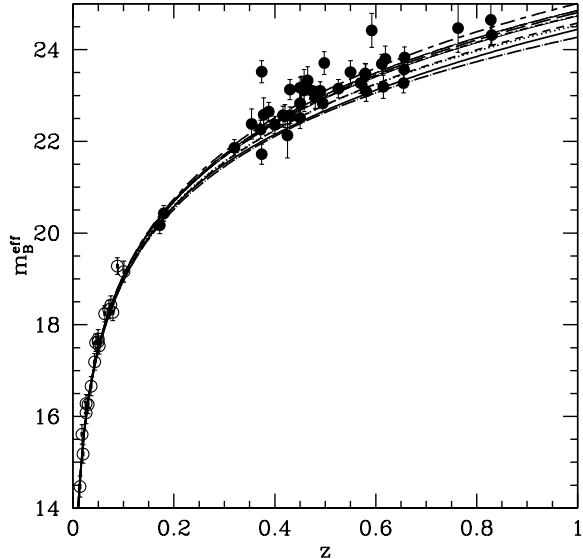,height=8cm,width=8cm}}}
\caption{The Cal\'an Tololo (open circles) and SCP data points (solid
circles). The curves correspond to the theoretical models discussed in
section \ref{models}.}
\label{fig:scp}
\end{figure}

The apparent bolometric magnitude is given by
\beq
m(z) = M + 5 \log \dl(z) +25\, ,
\label{mag}
\eeq
with $M$ the absolute bolometric magnitude and $\dl$ the luminosity
distance, which is usually
defined with distances in units of $10 {\rm pc}$. However, 
cosmological distances are measured in ${\rm Mpc}$ and therefore there is
an additional term $5\log{10^5} = 25$ in
eqn. (\ref{mag}). Furthermore, the luminosity distance depends on the
cosmological evolution and hence on  
the cosmological parameters and is defined by
$\dl^2 = {\cal L}/4\pi{\cal F}$, where ${\cal F}$ is the measured flux
and ${\cal L}$ the absolute luminosity of the object. The luminosity distance $\dl$ can be
expressed in terms of the coordinate distance $\dl(z) = (1+z)r(z)$. As
mentioned before the CMB data of BOOMERanG, MAXIMA and
DASI\cite{Boomerang:00,Lange:00,Boomerang:01,Boomerang:01b,Maxima:00,Balbi:00,Maxima:01,Dasi:01a,Dasi:01b}
in combination with 2dF observations \cite{Efstathiou:01} indicates
strongly that the universe has a flat topology and therefore we
concentrate here on flat cosmologies where the coordinate distance is given by
\beq
	r(z) = \int^z_0 \frac{c}{H\left(z^\prime\right)}dz^\prime\, .
\label{dist}
\eeq
In expression (\ref{mag}) the quantities
$m$, $M$ and $\dl$ depend on the Hubble parameter $H_0$. From
eqn.~(\ref{dist}) we see that $\dl \sim H_0^{-1}$ so we can rewrite $m(z)$
in the following way
\beq
	m(z) \equiv M+5\log \Dl - 5 \log H_0 + 25 \, ,
\eeq 
where we have defined $\Dl \equiv H_0 \dl$. For {\em low redshift}
SNe we can use the linear Hubble relation
\bea
	m(z) & = & M + 5 \log{cz} - 5 \log{H_0} + 25 \nonumber \\	
	     & = & {\cal M} +5\log{cz} \, ,
\label{magh0free}
\eea
where we have defined the magnitude ``zero point'' ${\cal M} \equiv M -
5\log{H_0}+25$. Theoretically, this quantity can be determined by the
survey, but in practise this is
just a statistical nuisance parameter which is marginalized to
estimate the cosmological parameters, so that $\Dl(z)$ can be estimated
without explicit knowledge of $H_0$\cite{Perlmutter:99a}. In fig.\ref{fig:scp} we plot the
effective bolometric magnitude $m_{\rm B}^{\rm eff}$ data points of
the SCP\cite{Perlmutter:97,Perlmutter:99a} and Cal\'an/Tololo 
survey\cite{Hamuy:93} as well as the curves $m(z)$ from the
theoretical models we study in section \ref{models}. The {\em
effective} magnitude refers to the apparent bolometric magnitude which
has been corrected by the lightcurve width-luminosity correction,
galactic extinction and the K-correction from the differences of the
R- and B-band filter\cite{Perlmutter:99a}. If we just allow
a cosmological constant, dark and baryonic matter content in the
universe and assume a flat cosmology, which seems to be confirmed by
recent CMB and large scale structure observations\cite{Dodelson:00a,Lange:00,Balbi:00,Efstathiou:01}, the best fit values are roughly
$\om = 0.28^{+0.09}_{-0.08}$ and therefore $\ol =
0.72$\cite{Perlmutter:99a}. This is in agreement with the
analysis of the High-Z Search Team of $\om = 0.24$ and $\ol =
0.76$\cite{Riess:98}. The low matter density is confirmed
by several observations; by the evolution of the number density
of rich clusters\cite{Bahcall:98}, mass estimates of
galaxy clusters, either by the Sunyaev-Zel'dovich
effect\cite{Carlstrom:99} or through measurements of the X-ray
flux\cite{Mohr:99} and also the shape of the matter power spectrum\cite{Efstathiou:92}.
Visually, one can hardly distinguish the different models in
fig.\ref{fig:scp} in a sense that they all seem to fit well. In
order to be able to tell more we plot the magnitude difference $\Dm(z) =
m(z) - m_\Lambda(z)$, where $m_\Lambda$ is the apparent magnitude of a fiducial
model with $\ol = 0.7$ and $\om  = 0.3$. We show the relative
magnitude in fig.\ref{fig:snap}. We can clearly distinguish most of
the models of section \ref{models}, however we already realize 
the problem with the current observational situation that the available
data can not really differentiate between the particular models.
If we want to distinguish the models in the future we have to be able
to achieve much smaller errorbars than the SCP data has. In the next
section we describe the SNAP satellite project which is able to
achieve this goal.
\section{The SuperNovae Acceleration Probe --  SNAP}\label{snap}
\begin{figure}[!h]
\setlength{\unitlength}{1cm}
\centerline{\hbox{\psfig{file=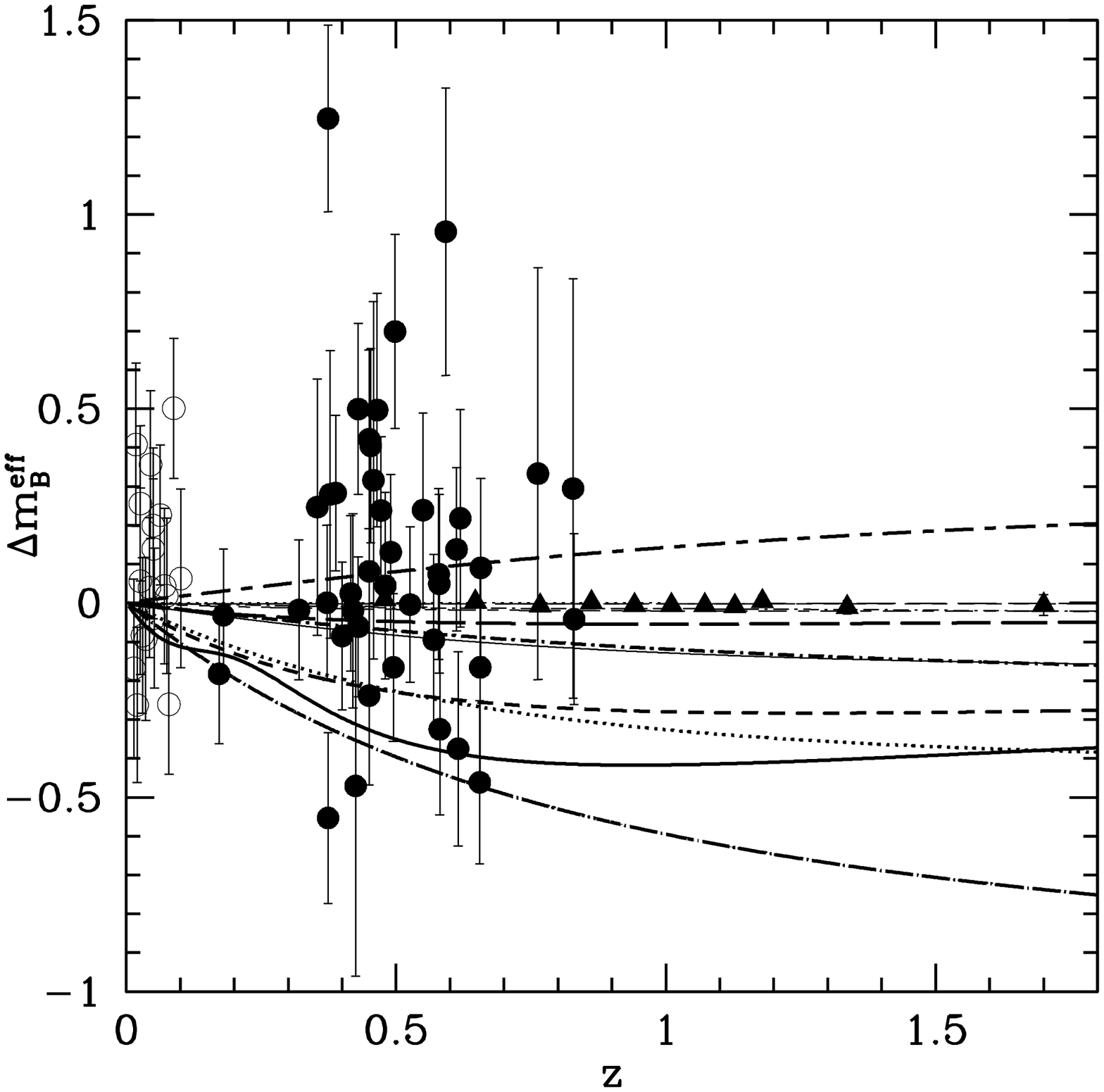,height=8cm,width=8cm}\psfig{file=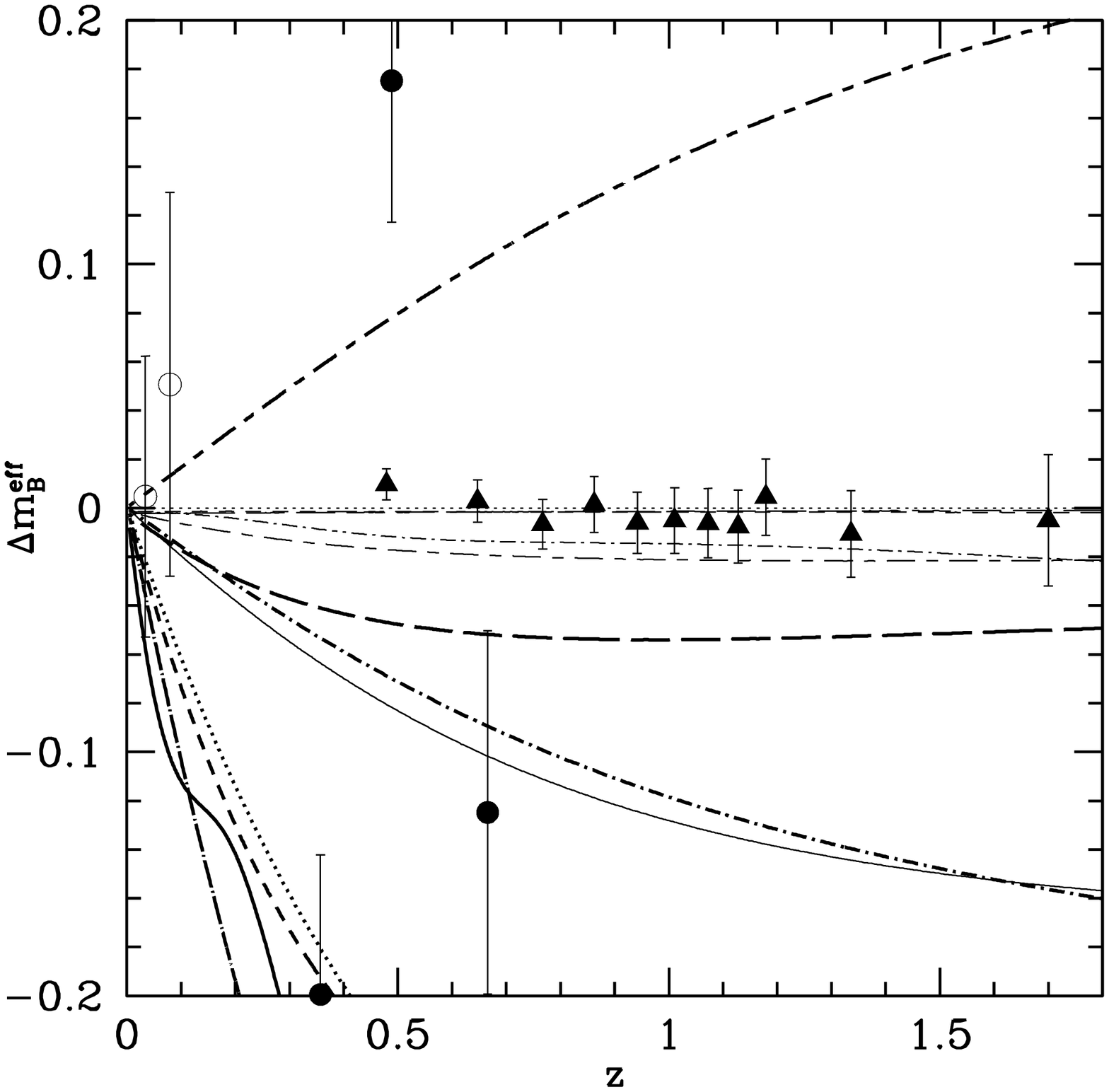,height=8cm,width=8cm}}} 
\caption{The relative magnitude with respect to a cosmology with
$\om=0.3$ and $\ol=0.7$. The SNAP data points are
simulated with this cosmology. The solid triangles
are the {\em binned} data points with 
errorbars from the SNAP type specifications as in table \ref{tab:snap}. We
have not plotted the data in the redshift interval $z=0-0.2$ for the
SNAP experiment. On the left the Cal\'an/Tololo (open circles) and SCP
data points (solid circles) are not binned and in the right figure
they are. The curves 
correspond to the theoretical models discussed in section
\ref{models} and the key to this curves is the same as in fig.\ref{fig:wall}. The thick dot-short dash line is a cosmological constant
model with $\ol = 0.6$ and the thick short dash - long dash line a
model with $\ol = 0.8$. The thick long dash line is the ``Standard
Cold Dark Matter'' model with $\om = 1.0$, which is clearly ruled out
by the current data.}
\label{fig:snap}
\end{figure}

In order to improve the current observational situation significantly
a new satellite, the SuperNovae Acceleration 
Probe -- SNAP  has been proposed\cite{SNAP}, which will be dedicated to the observation of SNe.
The SNAP satellite is equipped with a 2 meter telescope with a $1
\Box^\circ$ optical imager, a $1\Box^\prime$ near-IR imager, and a
three-channel near-UV-to-near-IR spectrograph. Every SNe at
$z<1.2$ will be followed as its brightens and fades. The wide-field
imager makes it possible to find and follow approximately 2000 SNe Ia
in two years, and the $1.8-$ to $2.0-$ aperture of the mirror allows this
data set to extend to redshift $z=1.7$. Furthermore, systematic uncertainties
will improve considerably compared to the current situation. The
uncertainty due to the 
Malmquist bias, the fact that the most distant SNe are only the ones
with large intrinsic brightness and therefore represent a very {\em
biased} sample in brightness, will also improve since each SNe
will be observed 3.8 magnitudes below peak brightness. For a large
subsample spectral time series and cross-wavelength flux calibration
will reduce the uncertainties from the K-correction and cross-filter
calibration. The Sloan Digital Sky Survey (SDSS)\cite{SDSS}, the Space
InfraRed Telescope Facility (SIRTF)\cite{SIRTF}
observation and SNAP spectra of host galaxy subdwarfs will improve the
systematic uncertainty due to the Milky Way Galaxy extinction. The
uncertainty due to gravitational lensing by clumped masses will be
averaged out due to the large statistics. The error due to extinction
due to ordinary dust outside the Milky Way will be reduced due to the
cross-wavelength calibrated spectra. Also the uncertainties due to
Non-SNe contaminations will decrease. Gray dust uncertainties can be
addressed due to large observed redshifts, with $z>1.4$, and with broad
wavelength measurements into the near-IR. Due to the large sample
size and detailed lightcurve and spectral information, SNAP will
provide sufficient data to measure second order effects like 
uncorrected evolution of the SNe. These systematic errors lead to
an absolute uncertainty of $\ss = 0.02 \, {\rm mag}$ at redshift $z=1.5$, while the statistical 
calibrated uncertainty is $\sm = 0.15\, {\rm mag}$ which corresponds
to approximately $7\%$ uncertainty in the luminosity distance. The redshift coverage of
the SNAP satellite within two years is shown in table \ref{tab:snap}.
\begin{table}[!h]
\centering
\begin{tabular}{lcccc}
redshift interval & $z=0-0.2$ & $z=0.2-1.2$ & $z=1.2-1.4$ &
$z=1.4-1.7$ \\
\hline
number of SNe & 50 & 1800 & 50 & 15 
\end{tabular}
\caption{SNAP specifications for a two year period of observations, with a 
statistical uncertainty of $\sm = 
0.15\, {\rm mag}$ and an uncertainty limit of ${\hat \ss} = 0.02\,
{\rm mag}$ at redshift $z=1.5$.\label{tab:snap}}
\end{table}
The numbers in table \ref{tab:snap} are based on the {\em observed} rates of SNe out
to redshift $z=1.7$. In the redshift interval $z=0.2$ to $z=1.2$ 
we assume that SNAP will observe a 20 $\Box^\circ$ field within two
years. At high redshifts there are many more
SNe, but SNAP will not have the time for a spectroscopic follow
up on all of them. Likewise, at the lowest redshifts there will be more type Ia
SNe, but the limiting factor here is the sky coverage of SNAP.
We use these specifications and simulate the SNAP experiment assuming
a background cosmology with $\om=0.3$ and $\ol = 0.7$. In the
simulation we assume a Gaussian distribution of the uncertainties and
an equidistant sampling of the redshifts in the four ranges. We
further neglected the errors in redshift, since they are expected to
be of the order $\delta z = 0.002$ and therefore relatively small. In
fig.\ref{fig:snap} we show the results of this simulation. For
plotting purposes we bin
the data points so the resulting uncertainty is $\sigma=0.02$. The number
of data points in {\em one} bin, $N_{\rm bin}$, is given by $N_{\rm bin}
\le \sm^2/\ss^2$. However, the realistic situation is a bit more tricky
since the systematic error is drifting from $\ss=0$ at $z=0$ to
$\ss=0.02$ at $z=1.5$. For the discussion of systematic errors we
assume a linear drift and use 
$\ss=z{\hat \ss}/1.5$. In the right plot in 	 
fig.\ref{fig:snap} we also bin the SCP and Cal\'an/Tololo data.
One clearly recognizes in fig.\ref{fig:snap}, that one can
distinguish some of the models with a SNAP type observation, while the
current data does not allow any differentiation. In the next section we
will present the dark energy models we studied and then in section
\ref{rec} we will quantify how SNe observations can distinguish these models.

\section{Dark Energy Models}\label{models}
As mentioned in the introduction a possibility to generalize the concept of a
cosmological constant is by introducing a scalar field which only
gravitationally interacts with the other fields. The dark energy field
is supposed to slowly roll down the potential or is trapped in a local
minimum. This leads to a vacuum dominated state of the universe which
hence leads to an accelerated expansion. The energy
density of the field is given by its kinetic and potential
component,
\beq
	\rp = \frac{1}{2}\dot{\phi}^2 + V(\phi)\, ,
\eeq
while the pressure is given by the difference,
\beq
	\pp = \frac{1}{2}\dot{\phi}^2 - V(\phi)\, .
\eeq
Note that we assume that the field is homogeneous on large scales.
The proportionality factor
\beq
	\wp \equiv \frac{\pp}{\rp}\, ,
\eeq
in the equation of state, $\pp = \wp\rp$, is $\wp=-1$ if the kinetic term
$\dot{\phi}^2/2$ is negligible. This is exactly the equation of state
for a cosmological constant term. In this paper we
study the behaviour of the magnitude -- redshift relation and therefore,
we have to solve the Friedmann equation
\beq
	H^2(z) \equiv \left(\frac{\dot{a}}{a}\right)^2 =
\frac{1}{3}\left[\rho_{\rm other}+\frac{1}{2}\dot{\phi}^2+V(\phi)\right]\,
\eeq
where we have used the Planck mass $\Mp = 2.44 \times
10^{18} {\rm GeV}$ as a unit, $a$ is the scale
factor of the Robertson-Walker metric and $\rho_{\rm other}$ is
the total energy density of the other contributing fields or energy
components, like dark and baryonic matter and radiation. The evolution of
the dark energy field is given by the field equation
\beq
\ddot{\phi}+3H\dot{\phi}+V^\prime(\phi) = 0\, ,
\eeq
with $V^\prime(\phi)=d{\rm V}/d{\rm \phi}$. If $V(\phi)$ is
approximately constant and the other energy components are negligible,
the solution for the scale factor is
$a\sim\exp[\sqrt{V}t]$ and hence, the expansion of the universe 
is accelerating. This is the same concept as
inflation\cite{Guth:81,Albrecht:82,Linde:82}, which also exploits the rapid
expansion rate. However, in the context of the cosmological constant
and dark energy we are interested in solution where the universe is
vacuum dominated only in recent times and not in the early universe as
in inflationary models.

\begin{figure}[!h]
\setlength{\unitlength}{1cm}
\centerline{\hbox{\psfig{file=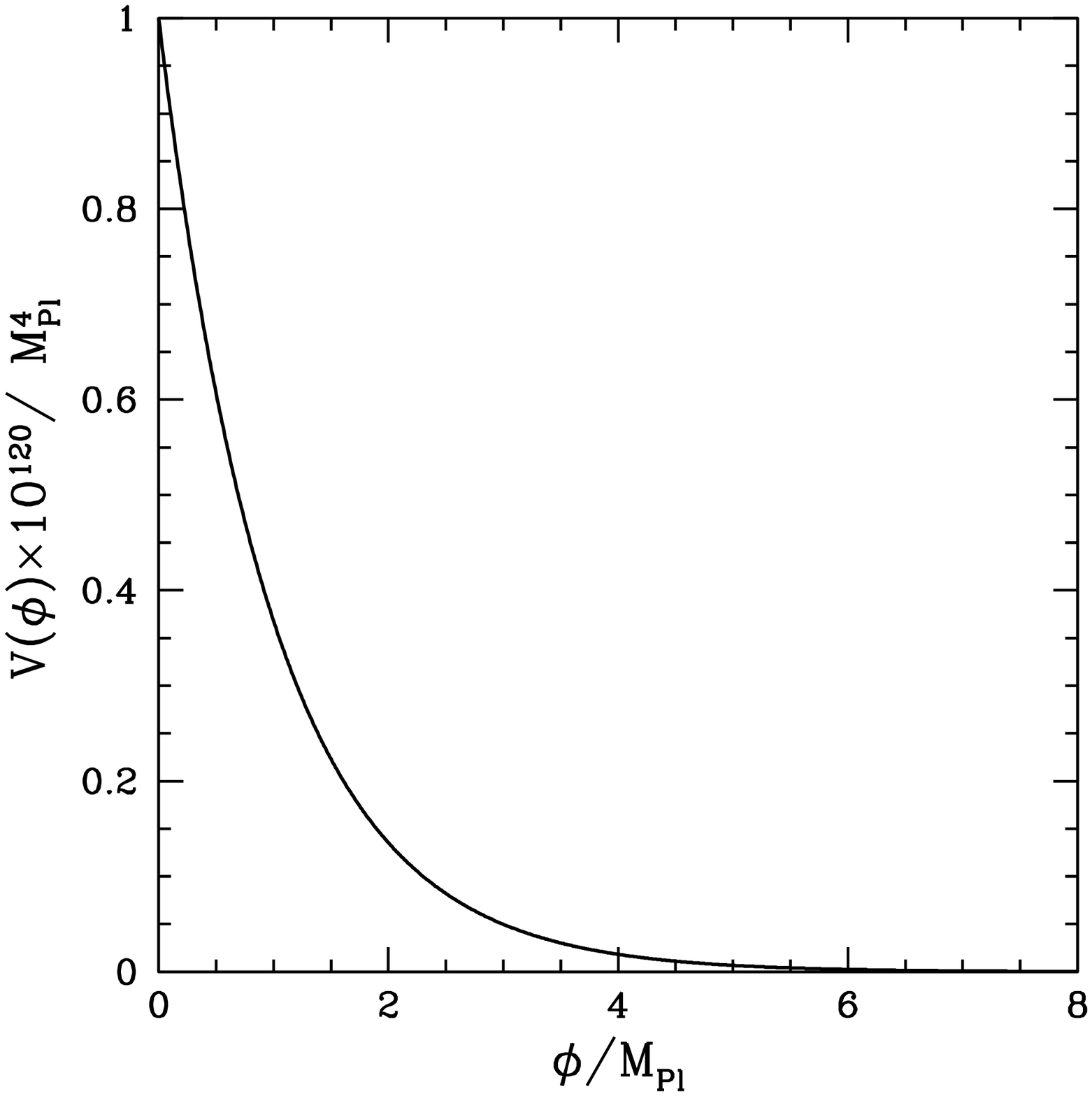,height=8cm,width=8cm}\psfig{file=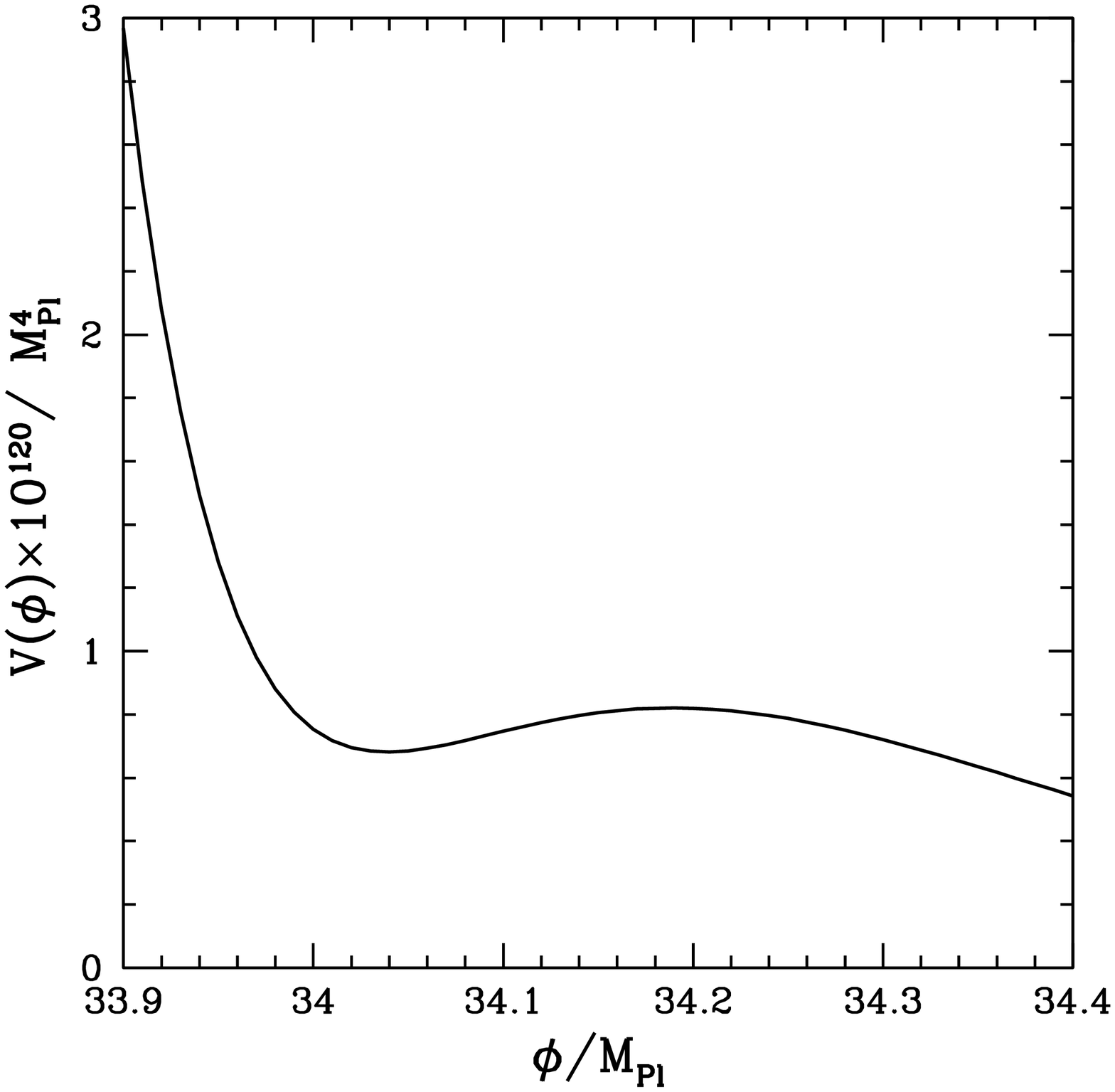,height=8cm,width=8cm}}}
\caption{On the left the pure exponential potential 
\protect\cite{Wetterich:88,Ratra:88,Peebles:88,Wetterich:95,Ferreira:97,Ferreira:98,Copeland:98} which is an 
example for a slow roll dark energy model and on the right the
exponential with a polynomial prefactor as proposed in\protect\cite{AS:00}
which gives rise to a local minimum in which the field is trapped.}
\label{fig:pot}
\end{figure}

There are two possibilities to neglect the kinetic energy
$\dot{\phi}^2/2$: either the field rolls down very slow the potential
``hill'' or it is trapped in a local minimum, which is illustrated in fig.\ref{fig:pot}.
On the left side we plot an exponential potential as discussed
in\cite{Wetterich:88,Ratra:88,Peebles:88,Wetterich:95,Ferreira:97,Ferreira:98,Copeland:98} which gives
rise to a slow roll of the dark energy field. On the right side we
plot the model proposed in\cite{AS:00} which is a potential with a
local minimum. 
\begin{figure}[!h]
\setlength{\unitlength}{1cm}
\centerline{\hbox{\psfig{file=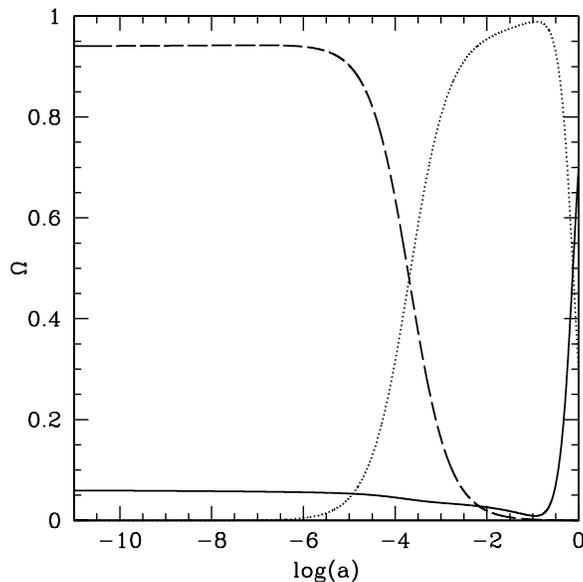,height=8cm,width=8cm}}}
\caption{The evolution of the densities relative
to the critical density for the trapped minimum
model{\protect\cite{AS:00}}. The long dashed line is $\Omega_{\rm r}$,
the energy density in the radiation,
the dotted line $\om$, for the matter fields, and the solid line
$\op$, the dark energy contribution.}
\label{fig:om}
\end{figure}
In fig.\ref{fig:om} we show the evolution of the densities
relative to the critical density  $\rho_c = 1/3H^2(z)$ for the model
described in \cite{AS:00}. This example shows the generic feature that
the universe first is radiation dominated (dashed line), then matter
dominated (dotted line) and finally becomes dominated by
dark energy just before the present day (solid line).

In the following discussion we present the dark energy
models we will use to test the opportunities of future SNe observations. The
specified parameters of the models below all lead to $H_0 = 65\,{\rm
km}/{\rm sec}/{\rm Mpc}$, $\om = 0.3$ and $\op =0.7$, where $\op$ is
the relative energy density of the dark energy component today. 

\begin{figure}[!h]
\setlength{\unitlength}{1cm}
\centerline{\hbox{\psfig{file=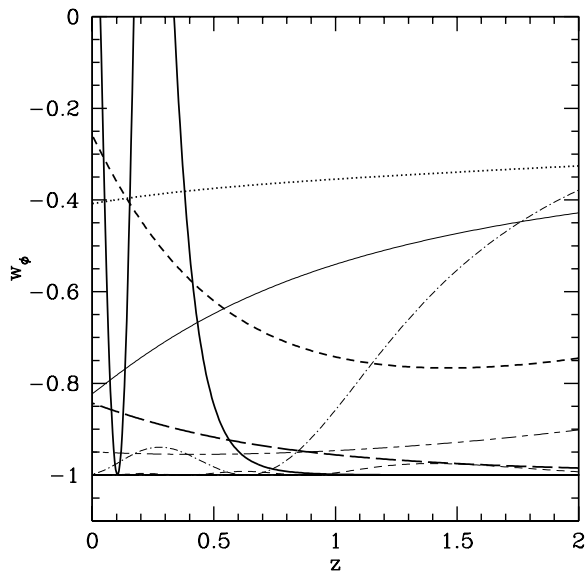,height=8cm,width=8cm}}}
\caption{The redshift evolution of equation of state factor $\wp = \pp/
\rp$ for the discussed models. The thin short dash line is the
trapped minimum model, the thin dot - short dash line is from the
brane inspired potential, the thin short dash - long dash line from
the potential which involves two exponentials, the thick short dashed
line from the periodic potential, the thick long dashed line from the
pure exponential, the thick solid line from the Pseudo Nambu-Gotu
Boson potential, the thin solid line from the Supergravity inspired
potential, the thin long dashed line from the exponential tracker
solution (underneath $w=-1$), and the thick dotted line from the inverse tracker.}
\label{fig:wall}
\end{figure}

\smallskip
{\noindent\it Pure exponential} - This potential appears naturally in
compactified higher dimensional Kaluza-Klein theories as well as in
certain Supergravity models. This model was discussed in the context
of a dark energy field
in\cite{Wetterich:88,Ratra:88,Peebles:88,Wetterich:95,Ferreira:97,Ferreira:98,Copeland:98,Barreiro:00} and is given
by the potential
\beq
	V(\phi) = V_0e^{-\lambda\phi}\, .
\eeq 
For a range of parameters and initial conditions this solution exhibits an
attractive behaviour whereby the field tracks the dominant
component of the background cosmology, i.~e.~matter or
radiation. However, in order to fulfill the observational constraint
of the SNe experiments the model parameters have to be chosen
from the transient, non-attractive, branch and the model needs fine
tuning of initial conditions: in our discussion we used the values
$V_0 = 10^{-120}\Mp^4$, $\lambda = 1 \Mp^{-1}$, $\phi(0) = 0.135 \Mp$
and $\dot{\phi}(0) = 0$. The magnitude - redshift relation for this model
corresponds to the thick long-dashed line in
figs.~\ref{fig:scp},\ref{fig:snap},\ref{fig:wall}. Note that
this model is slightly below the zero line of the $\Lambda$ model in
fig.\ref{fig:snap} and just outside the $1\sigma$ errorbars of the
SNAP satellite.

\smallskip
{\noindent\it Pseudo Nambu-Goldstone Boson (PNGB)} - This potential can arise
as potential energy of very light axions if the $U(1)$ Pecci-Quinn
symmetry is broken \cite{Frieman:95,Coble:97,Caldwell:98,Choi:00}. The potential is given by
\beq
V(\phi) = M^4\left[\cos\left(\phi/f\right)+1\right] \, ,
\eeq
where $M$ is of the order of a very light neutrino mass ($M \sim
0.001 - 0.01 {\rm eV}$) and $f$ is the symmetry breaking scale ($f
\sim 10^{15}-10^{19}{\rm GeV}$). For a wide range of parameters the
model can behave like a pure cosmological constant and does not
require fine tuning. However, we studied a parameter branch where the
equation of state factor oscillates, as seen in
fig.~\ref{fig:wall}. We use this particular setup, because it will be used later to
illustrate a case in which the reconstruction of the equation of state
could be troublesome. We used $M^4 = 1.001 \times 10^{-120} \Mp^4$ and
$f = 0.1 \Mp$. In this parameter branch it is also necessary to
tune the initial conditions to fulfill the observational constraints
to $\phi(0) = 1.184 \times 10^{-4} \Mp$ and $\dot{\phi}(0) = 0$. In
the magnitude -- redshift relation of figs.\ref{fig:scp},\ref{fig:snap},\ref{fig:wall} the thick solid
line represents to this model. We note the oscillatory
nature of the potential is also observed in the apparent magnitude
$m(z)$. For this choice of parameters the model is almost ruled out
already by the current SCP and Cal\'an/Tololo data.

\smallskip
{\noindent\it Cosmological tracker solutions} - These solutions are
a generalization of the attractor behaviour of the pure exponential
potential\cite{Peebles:88}. The potentials have a functional form
$f(M/\phi)$ and the most studied examples are the inverse tracker
potential
\beq
	V(\phi) =  \frac{M^{4+\alpha}}{\phi^{\alpha}}\, ,
\eeq
and the exponential tracker potential
\beq
	V(\phi) = M^4e^{M/\phi} \, .
\eeq
The notion of {\em tracker} solutions refers to the fact that these
solutions evolve on a common evolutionary track independent of the
initial conditions\cite{Steinhardt:99,Zlatev:99}. The inverse tracking
potential is motivated by supersymmetric QCD. The common feature of
these models is that the density in the dark energy field at late times dominates over all
the other energy contributions and therefore the expansion of the
universe begins to accelerate. The cosmic coincidence problem
\cite{Zlatev:99} is the fact that one still has to adjust the
parameters of the model to determine the time when the dark energy
component begins to dominate. However, as mentioned above, the initial
conditions are almost arbitrary. For the inverse tracker potential
we used the parameters $M=2.11 \times 10^{-12}\Mp$ and $\alpha =
6$. In figs.\ref{fig:scp},\ref{fig:snap},\ref{fig:wall} the inverse tracker model
corresponds to the thick dotted line. This model seems also to be
marginally disfavoured by current data as evident from the
right panel of fig.~\ref{fig:snap}. The only parameter to adjust for the
exponential tracker potential is $M=9.09 \times 10^{-31} \Mp$ and this
model is plotted as a thin long dashed line in
figs.\ref{fig:scp},\ref{fig:snap},\ref{fig:wall} and behaves in the shown redshift
range almost entirely like the cosmological constant model.

\smallskip
{\noindent\it Supergravity potential} -
This model is inspired by supersymmetry breaking in type I string
theory and Supergravity \cite{Binetruy:99,Brax:99} with the potential given by
\beq
	V(\phi) =
\frac{M^{4+\alpha}}{\phi^\alpha}\exp\left[\frac{1}{2}\left(\frac{\phi}{\Mp}\right)^2\right]\,
.
\label{eqn:potsugra} 
\eeq
Since supersymmetry breaking
should occur above the electroweak scale and in order to avoid fine
tuning of initial conditions, the parameters of this model have to
fulfill the constraints $\alpha \ge 11$ and $M \gtrsim
10^{-8} \Mp$. These requirements seem to lead to a ``unnatural'' way
of supersymmetry breaking \cite{Kolda:99} but nevertheless this model
is rare in that it is at least related to a fundamental
theory, and recent work shows that SUGRA may prevent this type of
difficulty \cite{Brax:01}. For small values of $\phi$ the exponential in
eqn.(\ref{eqn:potsugra}) is approximately constant so at early times
the evolution behaves like an inverse tracker model and has, therefore, all the advantages of the tracking solutions. The parameters
we chose for our discussion are $M=1.611 \times 10^{-8} \Mp$ and
$\alpha = 11$ and the model is plotted as a thin solid line in
figs.\ref{fig:scp},\ref{fig:snap},\ref{fig:wall}. We recognize that although this
model is clearly different from a cosmological constant, we can not
distinguish this model from a pure cosmological constant with the
current data.

\smallskip
{\noindent\it Exponential with polynomial and rational prefactor} -
The problem with the models discussed so far is that the involved mass
scales seem not to be {\em natural} in terms of Planck scale
physics. The model proposed by Albrecht and Skordis\cite{AS:00}
addresses this issue by multiplying an exponential potential by a
polynomial prefactor
\beq
	V(\phi)=V_p(\phi)e^{-\lambda\phi}\, ,
\eeq
where $V_p(\phi)$ is chosen to be
\beq
	V_p(\phi) = A+\frac{\left(\phi -
B\right)^\alpha}{\Mp^{\alpha-4}}\, .
\eeq
In the right panel of fig.\ref{fig:pot} we show this potential for
$\alpha = 2$. In this example the field gets trapped in the local
minimum of the potential independent of the initial conditions, so
no fine tuning of them is necessary. In
\cite{Weller:00} it is shown that this false vacuum state of the field
is stable to quantum decay, while \cite{Bean:00a} discusses the
possibilities of a classical roll through this potential. 
The parameters used here are $\lambda = 8 \Mp^{-1}$, $B = 33.989 \Mp$ and
$A=0.01\Mp^4$. This model is shown as a thin short dashed line in
figs.\ref{fig:scp},\ref{fig:snap},\ref{fig:wall} and we recognize that it is
completely indistinguishable from a pure cosmological constant. It is
possible to generalize the polynomial prefactor and allow rational
functions and there might be a possibility to connect the dark energy
potential to the interaction of two separated 3-dimensional branes
from string theory\cite{Dvali:99,Weller:00}. A promising candidate for such a
potential is 
\beq
	V_p\left(\phi\right) =
\frac{\Mp^6}{\left(\phi-B\right)^2+\delta} \, ,
\eeq
where $\delta$ regularises the singularity at $\phi=B$. As with the trapped minimum
model this potential also has 
the feature that the field gets trapped in a false vacuum state. We
take the parameters of the model to be
 $\lambda = 8 \Mp^{-1}$, $=B=35.1628 \Mp$ and $\delta =
0.01\Mp^2$. This brane model is depicted as a thin short-dash dotted
line in figs.\ref{fig:scp},\ref{fig:snap},\ref{fig:wall} and is also not
distinguishable from a pure cosmological constant model. A model with different parameters but very similar behaviour is discussed in \cite{Skordis:00}. In both
models the involved parameters are of order ${\cal O}(1)$ in units of
the Planck mass $\Mp$. The parameter $B$ needs to be {\em adjusted}
that the field gets trapped in the local minimum at the right time to
account for the observed density in the dark energy field of $\op =
0.7$ today.

\smallskip
{\noindent\it Two exponentials} - This type of potential could arise
in string theory as a possible result of Kaluza-Klein type
compactification and is given by
\beq
	V(\phi) = V_0\left[e^{\lambda \phi} + e^{\beta \phi}\right]
\eeq
and there is {\em no} fine tuning problem of the initial
conditions\cite{Barreiro:00}. The parameters in this model are chosen to be $V_0 = 8.2 \times
10^{-121} \Mp^4$, $\lambda = 20 \Mp^{-1}$ and $\beta = 0.5 \Mp$. In
\cite{Barreiro:00} other possible parameter choices are
discussed. In figs.\ref{fig:scp},\ref{fig:snap},\ref{fig:wall} the model is drawn as a
thin short-dashed - long-dashed line and we note that this model's
apparent magnitude evolves almost like a cosmological constant in
the observed redshift range.

\smallskip
{\noindent\it Periodic Potential} - The common feature of most of the
models discussed so far is that the parameters have to be adjusted in
a way that the 
dark energy component only becomes dominant {\em today}, which means
we live in a special epoch. The only exception is the {\em pure
exponential}\cite{Peebles:88,Ferreira:98} which has an attractor
behaviour and follows the dominant component of the background
component, that is matter or radiation. However, the pure exponential
models in the attractor branch of the parameter space are ruled out by Big
Bang Nucleosynthesis (BBN) and the SNe observations. A sinusoidal modulation of
the pure exponential can resolve this problem\cite{Dodelson:00b} and
such potential is given by
\beq
	V(\phi) = V_0 \left[1+\delta\sin\left(\beta
\phi\right)\right]e^{-\lambda \phi}\, .
\eeq
There is only an adjustment of the parameters necessary to fulfill the
BBN constraint and the parameters used in our discussion are $V_0 =
2.55 \times 10^5 \Mp^4$, $\lambda = 4.0 \Mp^{-1}$, $\delta=0.98$ and $\beta
= 0.51 \Mp^{-1}$. This model corresponds to the thick short-dashed
line in figs.\ref{fig:scp},\ref{fig:snap},\ref{fig:wall}.

In fig.\ref{fig:wall} we show the evolution of the equation of state
factor $\wp$ of the dark energy component. We recognize that most of
the models have a smooth behavior, apart from the PNGB
model. For this model $\wp$ oscillates between $-1$ and $1$. We have
now a fairly representative sample of dark energy models though it seems
impossible to include the rapidly increasing number of {\em all} suggested
models. Two classes of models which are entirely missing in our
discussion are the one where
the dark energy field is non-minimally coupled to
gravity\cite{Uzan:99,Chiba:99a,Bean:00}, where the field is directly
coupled to matter\cite{Amendola:99,Amendola:00} and where the dark energy field is
kinetically driven\cite{Chiba:99b,Picon:00}. The reconstruction of the
equation of state for these models is discussed in
\cite{Boisseau:00,Barger:00}. In the following we will
show which of the presented models can be distinguished by SNAP type
SNe observations. 

\section{A fit designed to reconstruct the equation of
state}\label{rec}
In order to distinguish between different dark energy models we have to be
able to quantify how well SNe observations with a SNAP type experiment can map out the magnitude -- or 
luminosity distance -- redshift relation. There has been a suggestion to fit
the luminosity distance by a
polynomial\cite{Huterer:99}. This fit is motivated by the need of a
smooth function to reconstruct the equation of state factor of the
dark energy component and by its obvious simplicity. More recent work suggest to fit the luminosity
distance by a rational function with three free
coefficients\cite{Saini:99}. This fit has the advantage that in
extreme cases it behaves like the analytical solutions for a pure
cosmological constant cosmology or a completely matter dominated
universe. There have also been other suggestions with different
fitting functions \cite{Nakamura:01} and it may to be possible to fit
directly for the evolution of the dark energy density \cite{Wang:01}.

The polynomial fit of the luminosity distance is defined by
\beq
	\dl(z)=\sum_{i=0}^{N}c_i z^i\, ,
\label{polfit}
\eeq
where we will truncate the power series at an appropriate $N$.  Since the
luminosity distance $\dl(0)=0$, for all cosmological models, we can
set $c_0 = 0$. In order to study this fit we use the
proposed dark energy models from section \ref{models} and create data
sets with the SNAP type
specifications from table \ref{tab:snap} with a Monte Carlo
simulation. We assume that the errors $\delta m$ in the magnitude are
Gaussian distributed with a zero mean and a variance of
$\sm$. Furthermore we impose an equidistant sampling in redshift,
which seems to be the 
optimal sampling \cite{Huterer:00}, and neglect the
uncertainties in redshift space. The simulation is repeated and the
fitting procedure $N_{\rm stat} = 1000$ times to obtain the 
appropriate statistics and we find that the distribution of the
coefficients $c_i$ is Gaussian.

We will now discuss how to reconstruct the equation of state factor
$\wp$ from the measured magnitude or luminosity distance. The
conservation of energy in the dark energy component yields
\beq
	\frac{{\dot \rho}_\phi}{\rp} = - 3 H (1+\wp)\, ,
\label{rp}
\eeq
with $H = {\dot a}/a = (\rom + \rp)/3$, $a^{-1} = 1+z$ and
the definition of the coordinate distance $r(z)$ in eqn.~(\ref{dist})
we obtain\cite{Huterer:99}
\beq
	1+\wp = \frac{1+z}{3}\frac{3\om H_0^2
\left(1+z\right)^2+2\frac{r^{\prime\prime}}{r^{\prime 3}}}{\om H_0^2
\left(1+z\right)^3 - \frac{1}{r^{\prime 2}}}\, ,
\label{wrec}
\eeq
where $r^\prime$ denotes the derivative of the coordinate distance $r$
with respect to redshift $z$. Since the coordinate distance is
$r(z)=\dl(z)/(1+z)$ we can calculate the derivatives from the fit in
eqn.~(\ref{polfit}), although we will need to quantify the matter content $\om$
in order to do this.  Since the errors in the coefficients $c_i$ are
Gaussian we can calculate the error in the reconstructed $\wp$ by
ordinary Gaussian error propagation
\beq
	\delta\wp^2 = \sum_{ij}\frac{\partial \wp}{\partial c_i}\frac{\partial
\wp}{\partial c_j} \sigma_{ij}\, ,
\eeq
with $\sigma_{ij}$ the covariance matrix of the simulated sample of $c_i$.

\begin{figure}[!h] 
\setlength{\unitlength}{1cm}
\centerline{\hbox{\psfig{file=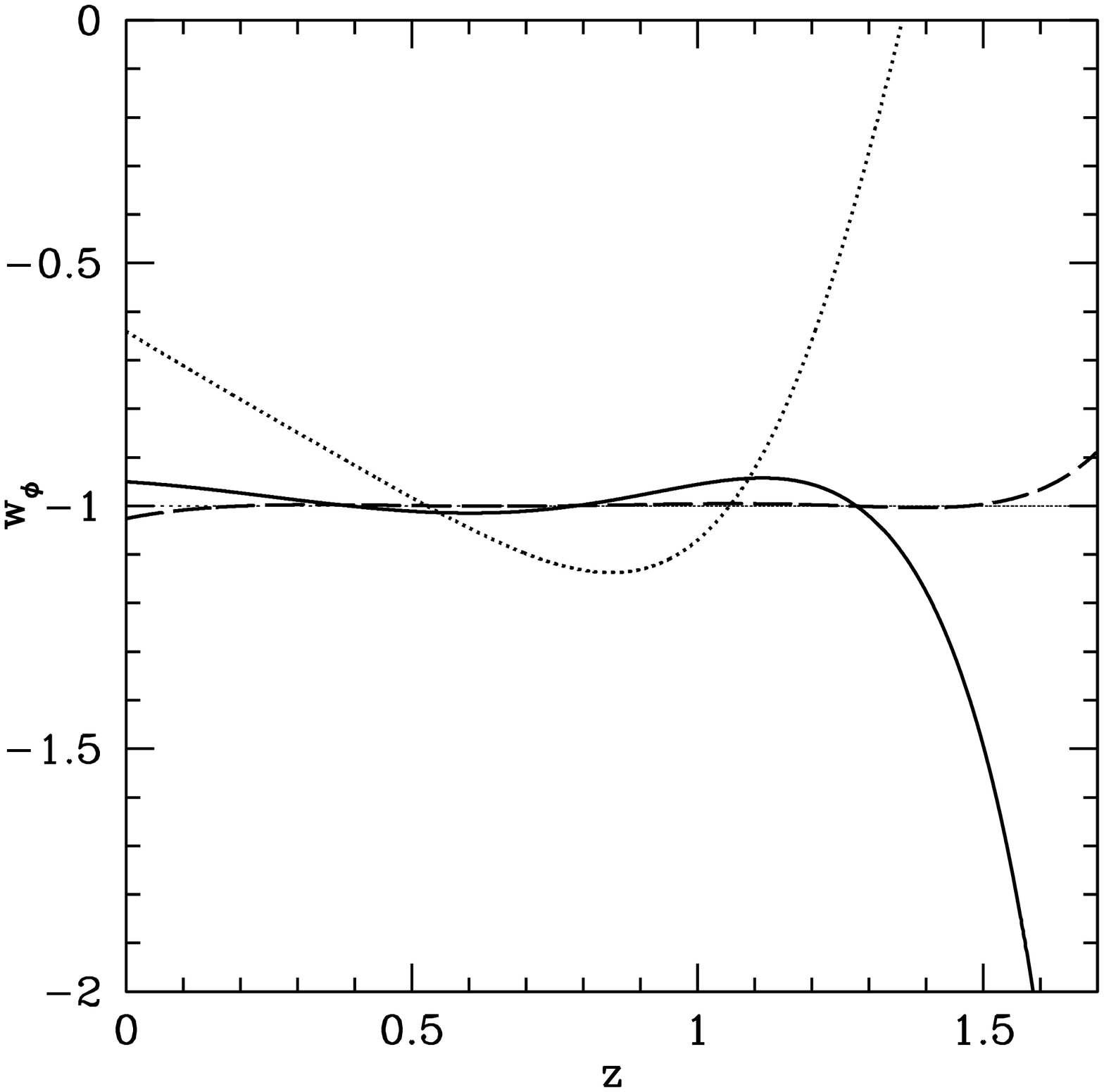,height=8.0cm,width=8cm}\psfig{file=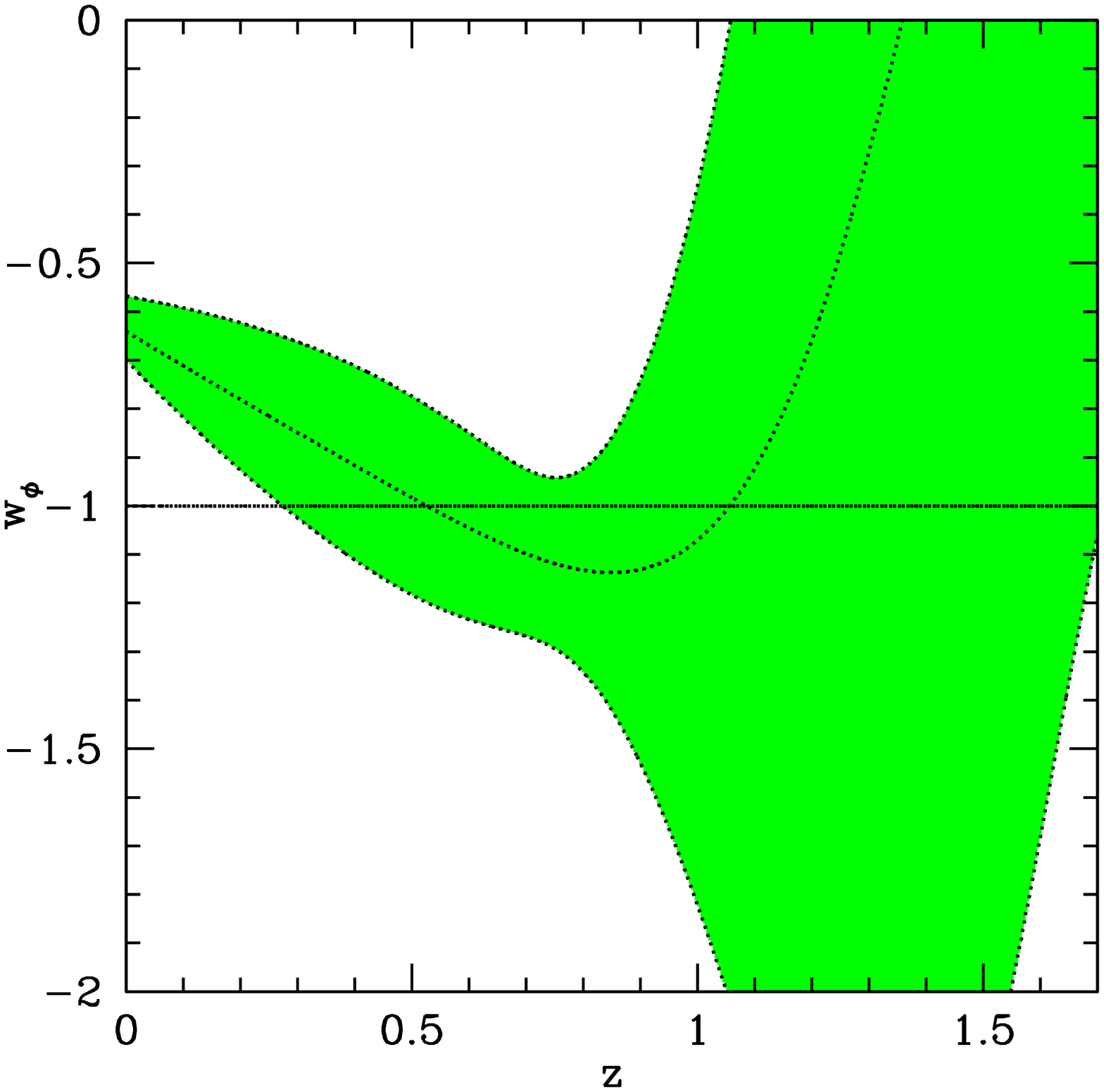,height=8.0cm,width=8cm}}} 
\centerline{\hbox{\psfig{file=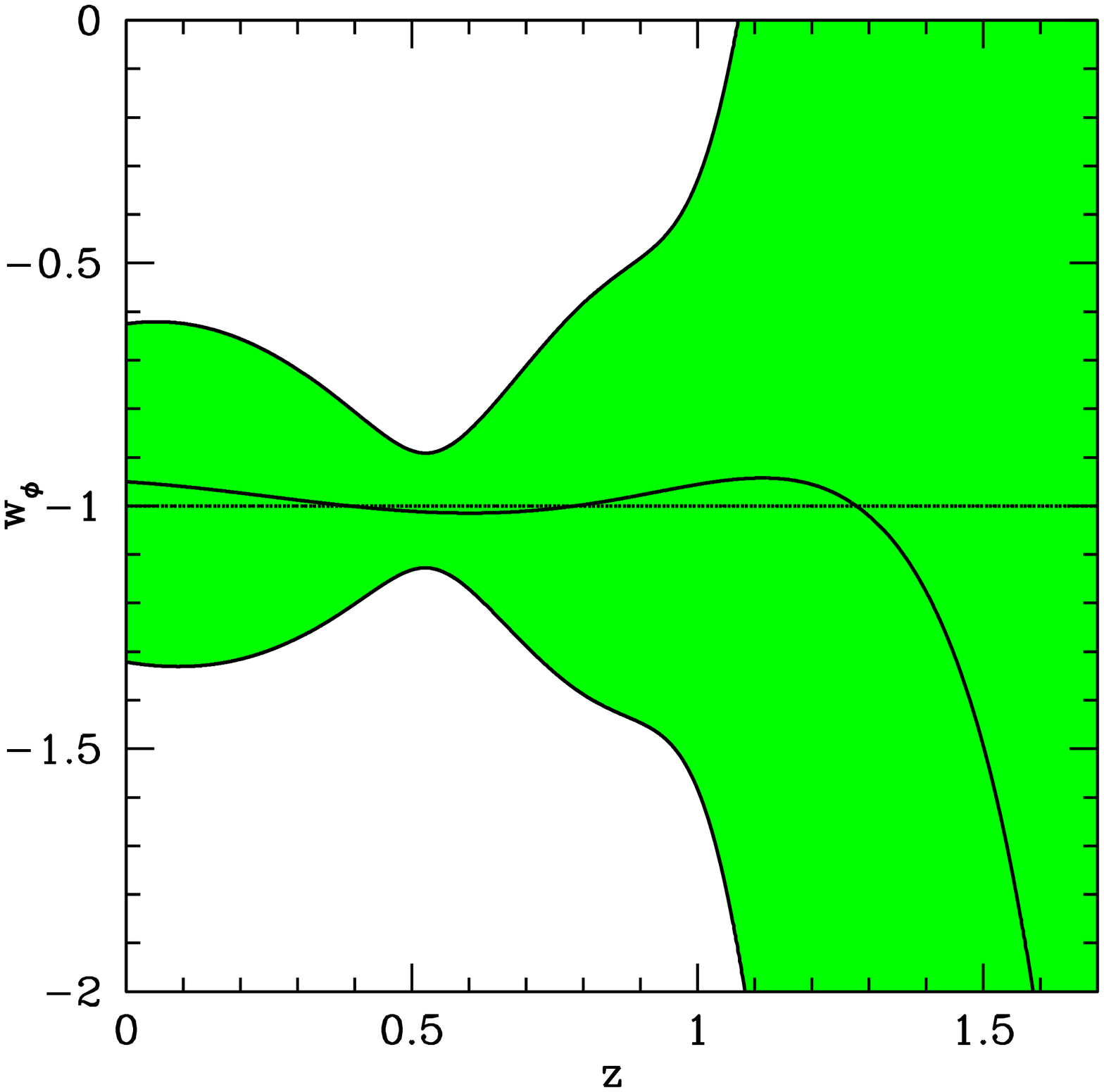,height=8.0cm,width=8cm}\psfig{file=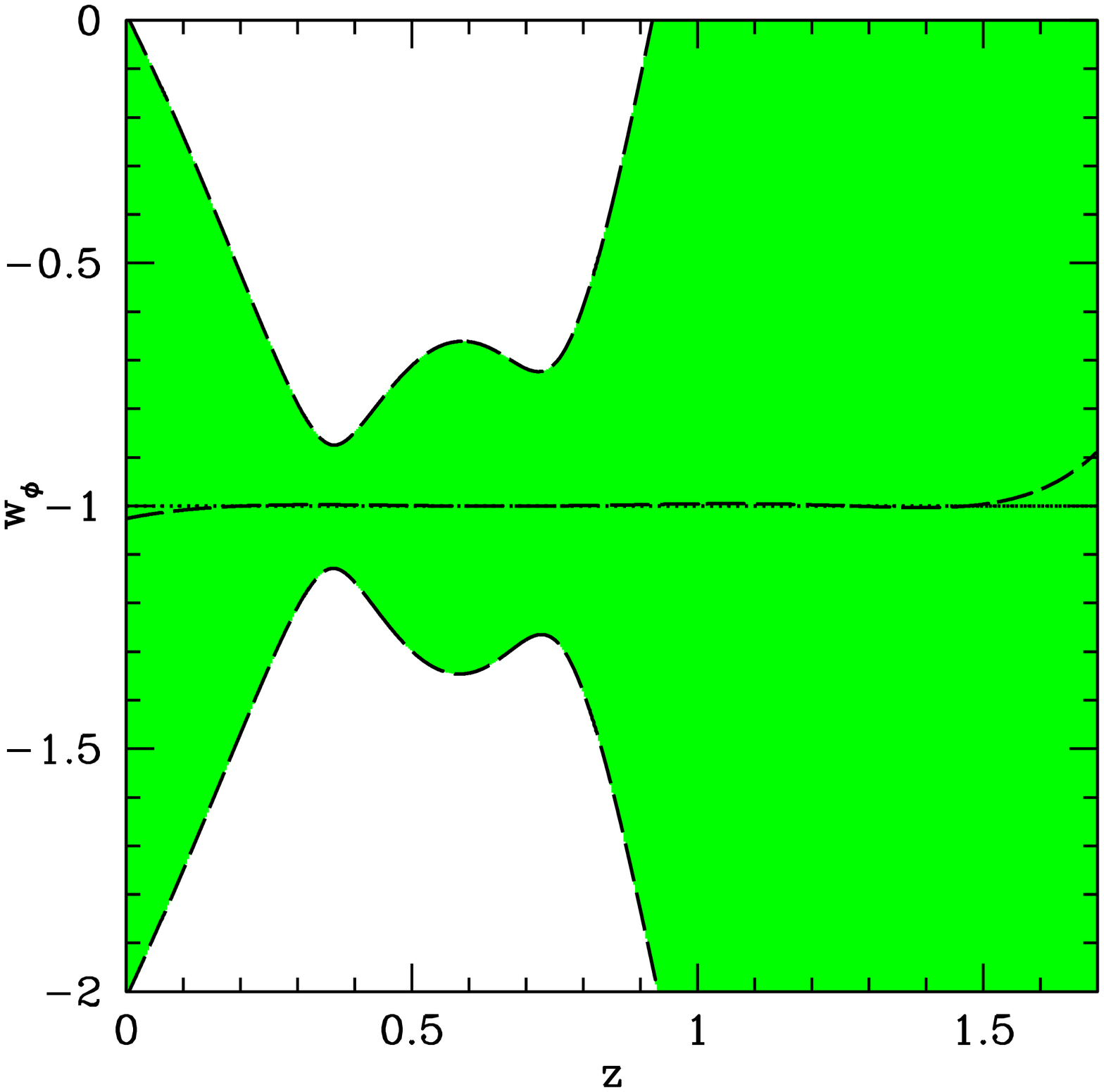,height=8.0cm,width=8cm}}}
\caption{The reconstructed equation of state factor $\wp$ for the
cosmological constant model with the theoretical value of $\wp(z) =
-1$. In the top left panel we show the mean values for the
reconstructed $\wp$ with the dotted line for the $N=3$ polynomial fit,
the solid line the $N=4$ and the dashed line the $N=5$ fit. The top
right panel is the $N=3$ fit with the shaded region representing the
$1\sigma$ uncertainty levels. The lower left panel is the same plot
for $N=4$ and the lower right panel for $N=5$.}
\label{fig:w_pol}
\end{figure}
In fig.~\ref{fig:w_pol} we show the reconstructed equation of state
factor $\wp$ for different values of $N$. The background cosmology is
a cosmological constant model with 
$\om=0.3$ and $\ol=0.7$. In the top right panel we plot the mean
values of the reconstructed $\wp$. We note that for $N=3$ fit
(dotted line and top right panel) the mean value does not represent the
theoretical $\wp$ well. This is because in order to reconstruct $\wp$
we need the second derivative of the coordinate distance $r(z)$
(eqn.~\ref{wrec}). The coordinate distance is already reduced by one
order in $z$ compared to the luminosity distance so the $N=3$ fit
might not represent sufficiently the evolution in $r(z)$ to produce a
second derivative in $r(z)$ which represents the theoretical value at
least roughly. The $N=4$ (solid line) and $N=5$ (dashed line) fit
however reproduce, at least in the relevant redshift range, the
theoretical value of $\wp$ to a satisfactory level. The $N=5$ fit
naturally reproduces the $\wp$ better then the $N=4$ fit, however the
errorbars for the $N=5$ fit (lower right panel) are much larger then
for the $N=4$ fit (lower left panel). A fourth order polynomial,
therefore, yields the best reconstruction for $\wp$ with the SNAP type specifications
from table \ref{tab:snap}. We note, however, from the lower left
panel in fig.~\ref{fig:w_pol} that for the 4th order polynomial fit we
can only reproduce $\wp$ in the required range if $-1.3 < \wp < -0.7$, at the
$1\sigma$ level.\\

As we saw in the previous section it seems nearly impossible to make
any predictions about the equation of state factor $\wp$ with the SNAP
type
specifications. However, the problem could be the polynomial fit and
{\em not} the SNAP experiment. If we assume for example a constant
$\wp$ and try to reconstruct it with a polynomial fit of the
luminosity distance, we can show analytically that {\em no} finite
order polynomial can do this exactly. Even in the extreme case of just
a pure cosmological constant model with $\ol=1$ or the SCDM model with
$\om = 1$ the polynomial can not fit it exactly. This problem was 
recognized in \cite{Huterer:99} and they suggested to use Pad\'e
approximants or even splines. In \cite{Saini:99} a rational
function is used which at least allows that the extreme cases of $\ol=1$ and
$\om=1$ to be fitted with an exact relation. This method has been
improved due to the introduction of a more complicated rational
function with more free parameters \cite{Chiba:00}. A problem with all these
reconstruction methods is that we need, the matter density $\om$ as
another input parameter as seen in eqn.~(\ref{wrec}). A further problem is that
in order to reconstruct $\wp$ we have to calculate {\em second} order
derivatives in $r(z)$ which will in general, increase the errorbars on the
reconstructed equation of state factor. We propose a fit which has
been discussed recently \cite{Maor:00,Astier:00,Weller:00b}   	
allowing one to read off the equation of state factor directly with {\em no}
reconstruction as in eqn.~(\ref{wrec}) and also includes the
possibility to fit for $\om$ in a more natural way.

We expand the equation of state factor into its redshift evolution
\beq
	\wp = \sum_{i=0}^{N}w_i\left(1+z\right)^i \, ,
\label{wexp}
\eeq
where we chose the expansion in $(1+z)$ for computational
convenience. With this expansion and eqns.~(\ref{dist}),(\ref{rp}) we
obtain the luminosity distance in a flat universe
\beq
	\dl^{\rm fit}(z) =
\frac{c(1+z)}{H_0}\int\limits_0^z\frac{\left(1+z^\prime\right)^{-3/2}}{\sqrt{\om+\op\left(1+z^\prime\right)^{3w_0}\exp\left\{3\sum\limits_{i=1}^N\frac{w_i}{i}\left[\left(1+z^\prime\right)^i-1\right]\right\}}}\,
dz^\prime \, ,
\label{dfit}
\eeq
with $\op = 1-\om$. We note that for $w_i = 0$ for $i\ge1$ and
$w_0=-1$ we obtain the standard result for the cosmological
constant. In the following discussion we take $\om=0.3$ fixed and
later we examine the fit with different prior information on $\om$. 

The fit is done by minimizing the $\chi^2$ function
\beq
	\chi^2\left(\left\{w_i\right\}\right) = \sum_{k=0}^{N_z}
\left[\frac{\dl(z_k)-\dl^{\rm fit}(z_k)}{\ddl(z_k)}\right]^2\, ,
\label{wfit}
\eeq
with $\ddl(z_k) = \sm\dl(z_k)\ln (10 /5)$, the uncertainties in
$\dl(z)$ as the weight on the particular data points. The sum runs over the
whole redshift range from table \ref{tab:snap}, with $N_z$ the overall
number of measurement redshifts. Again we assume that
the redshift range is split into equidistant samples in the four ranges from table
\ref{tab:snap}. We minimize this rather complicated expression with
the {\tt MINUIT} routine from the CERN program library which also
delivers the covariance matrix on the parameters.
\begin{figure}[!h] 
\setlength{\unitlength}{1cm}
\centerline{\hbox{\psfig{file=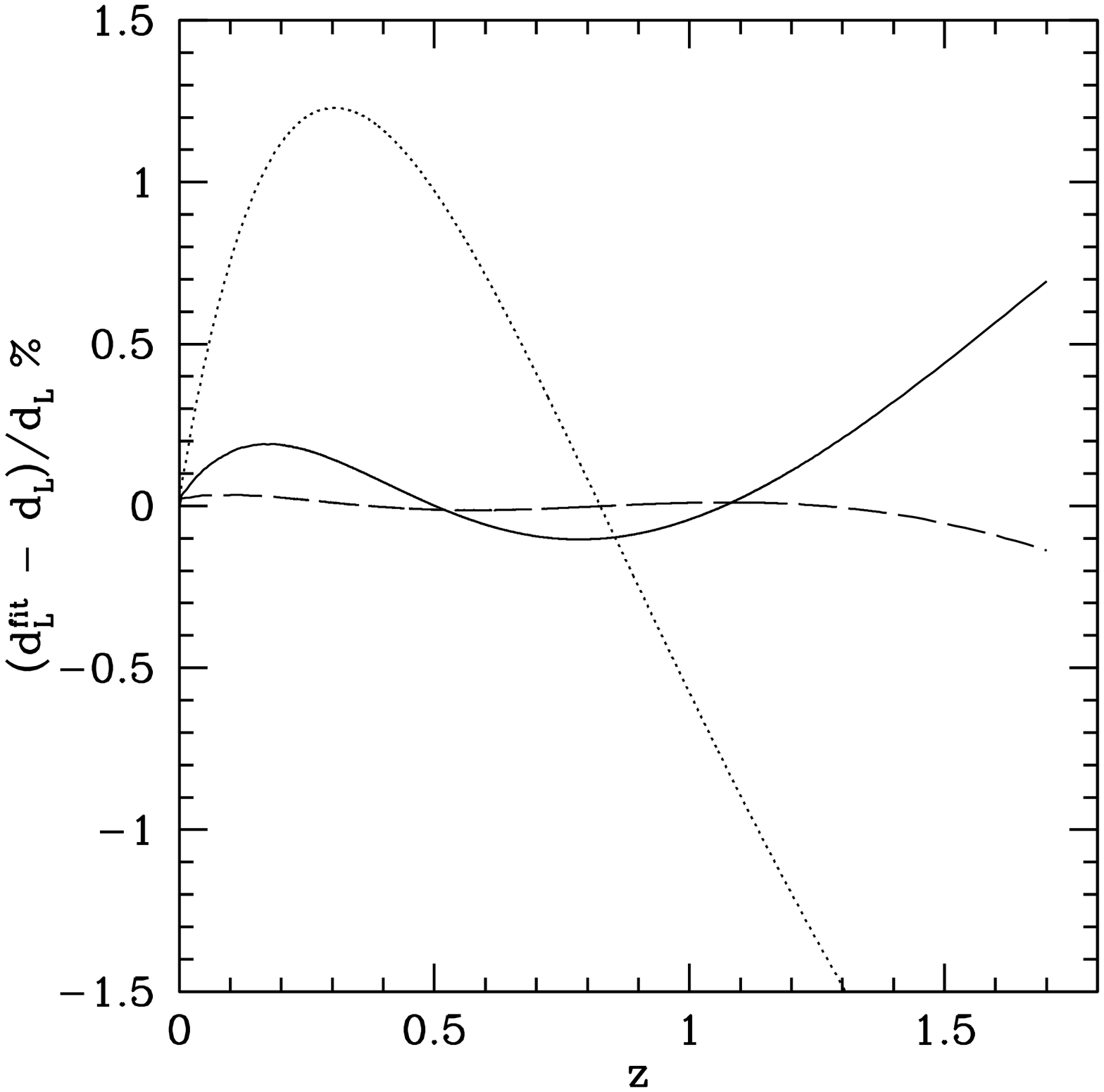,height=8.0cm,width=8cm}\psfig{file=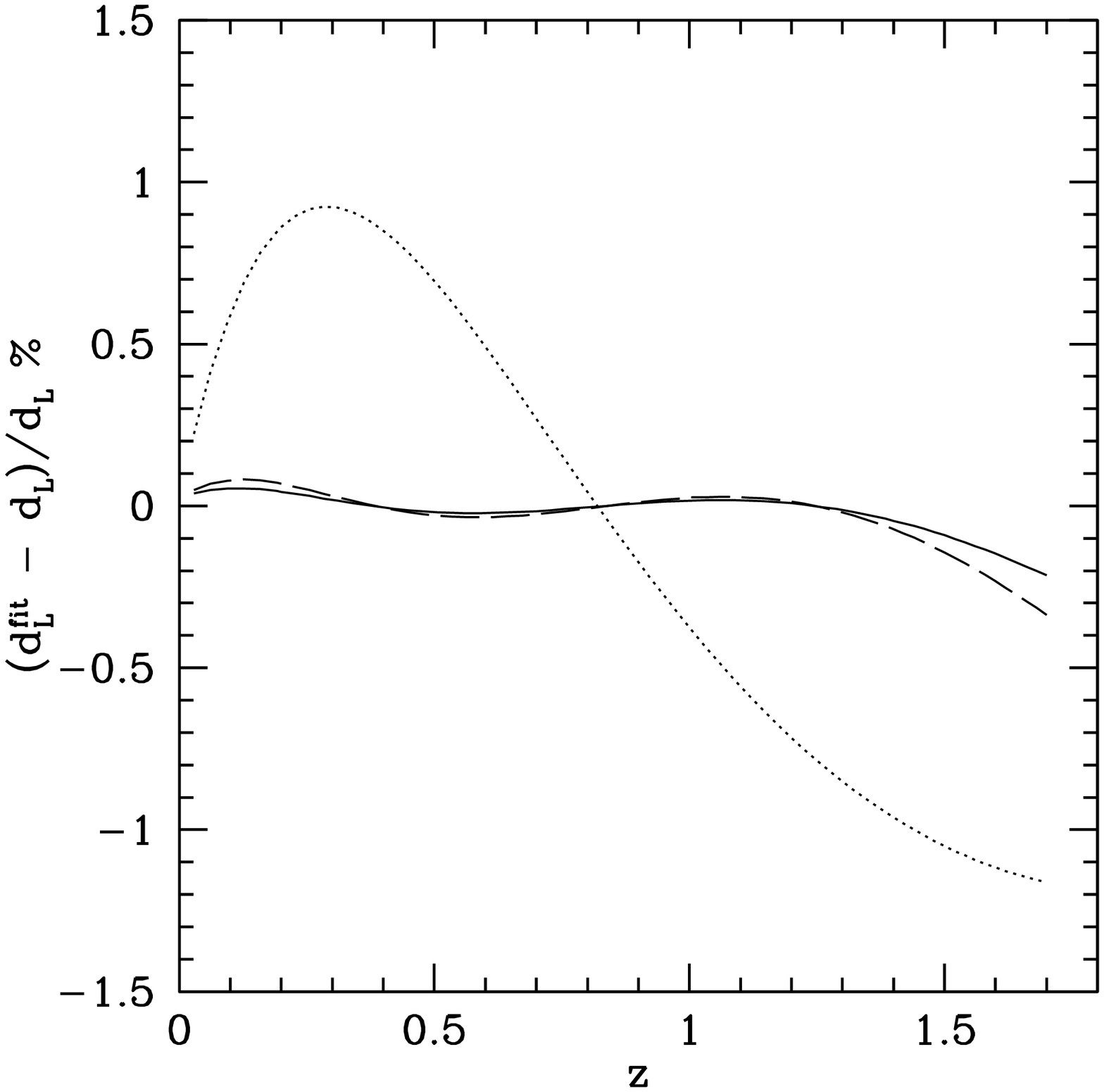,height=8.0cm,width=8cm}}}
\caption{The fit from eqn.~(\ref{dfit}) for the periodic potential as
the cosmological background. We plot the fitted $\dl^{\rm fit}$ with
respect to the theoretical luminosity distance $\dl$ in percent (\%).
The dotted line is the $N=0$ fit, the solid line the $N=1$ fit and the
dashed line the $N=2$ fit. The left panel is the one with $\om=0.3$ as
a prior and the right one has just the constraint that $0\le\om\le1$.}
\label{fig:wfit}
\end{figure}
In fig.~\ref{fig:wfit} we plot the fit for different values of
$N$ for the periodic potential as the dark energy model. We choose
this model because we see in fig.~\ref{fig:wall} that the equation of
state factor for this model is evolving within the relevant range. For a
cosmological constant model a fit with $N=0$ should be the best, because it
is exact and we do not gain any information by going to higher
order. In fig.~\ref{fig:wfit} we note that the $N=0$ fit, with
$\chi^2 \approx 31$,
is a relatively poor fit of the theoretical values if we set a prior of
$\om=0.3$ (left panel). 
The first order results in a satisfactory
fit with $\chi^2 \approx 0.47$ and $N=2$
improves this result only slightly, with $\chi^2 = 7.3 \times
10^{-3}$. So in order to study the 
luminosity distance $\dl$ the first order fit seems to be
sufficient. If we release the prior on $\om$ and just constrain the
matter contents of the universe 
by $0\le\om\le1$ 
the $N=1$ and $N=2$ fit are indistinguishable and fit reasonably
well. It appears  
that the $\wp$ fit (eqn.~\ref{dfit}) leads to more accurate results than the polynomial
fit (eqn.~\ref{polfit}) with less free parameters!
\begin{figure}[!h]
\setlength{\unitlength}{1cm}
\centerline{\hbox{\psfig{file=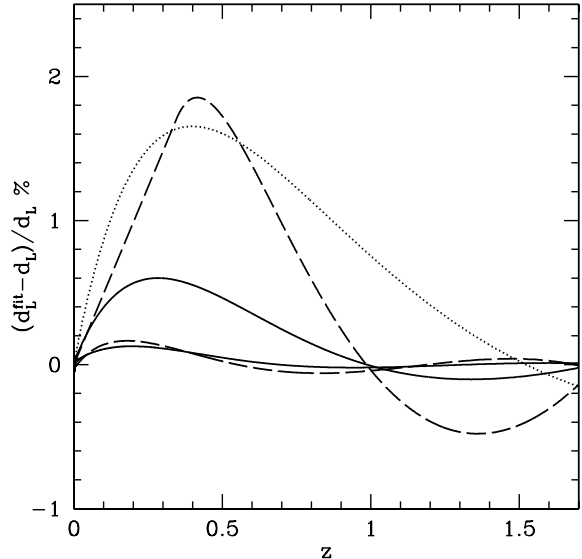,height=8.0cm,width=8.0cm}}}
\caption{Relative accuracy of the rational fit (dotted line), the
quadratic and cubic polynomial fit (dashed line) and the linear and
quadratic $w$-expansion (solid line). The cosmology is taken from the periodic potential.}
\label{fig:comp_fit}
\end{figure}

In fig.~\ref{fig:comp_fit} we plot the relative accuracy of the
different fits, for an unweighted sampling points. We recognize
that the cubic polynomial- and quadratic $w$-expansion lead to the best
fit. This results holds also if we use the weights from the SNAP
specifications. We performed this comparison for the SUGRA and the
periodic potential models as well as for a toy model \cite{Weller:00b}. For the
cosmological constant the $w$-expansion results by construction to the
best results because the fit is ``exact''. We expect that this
behaviour holds as well for the nearly constant models, like the two
exponentials model.

\begin{figure}[!h] 
\setlength{\unitlength}{1cm}
\centerline{\hbox{\psfig{file=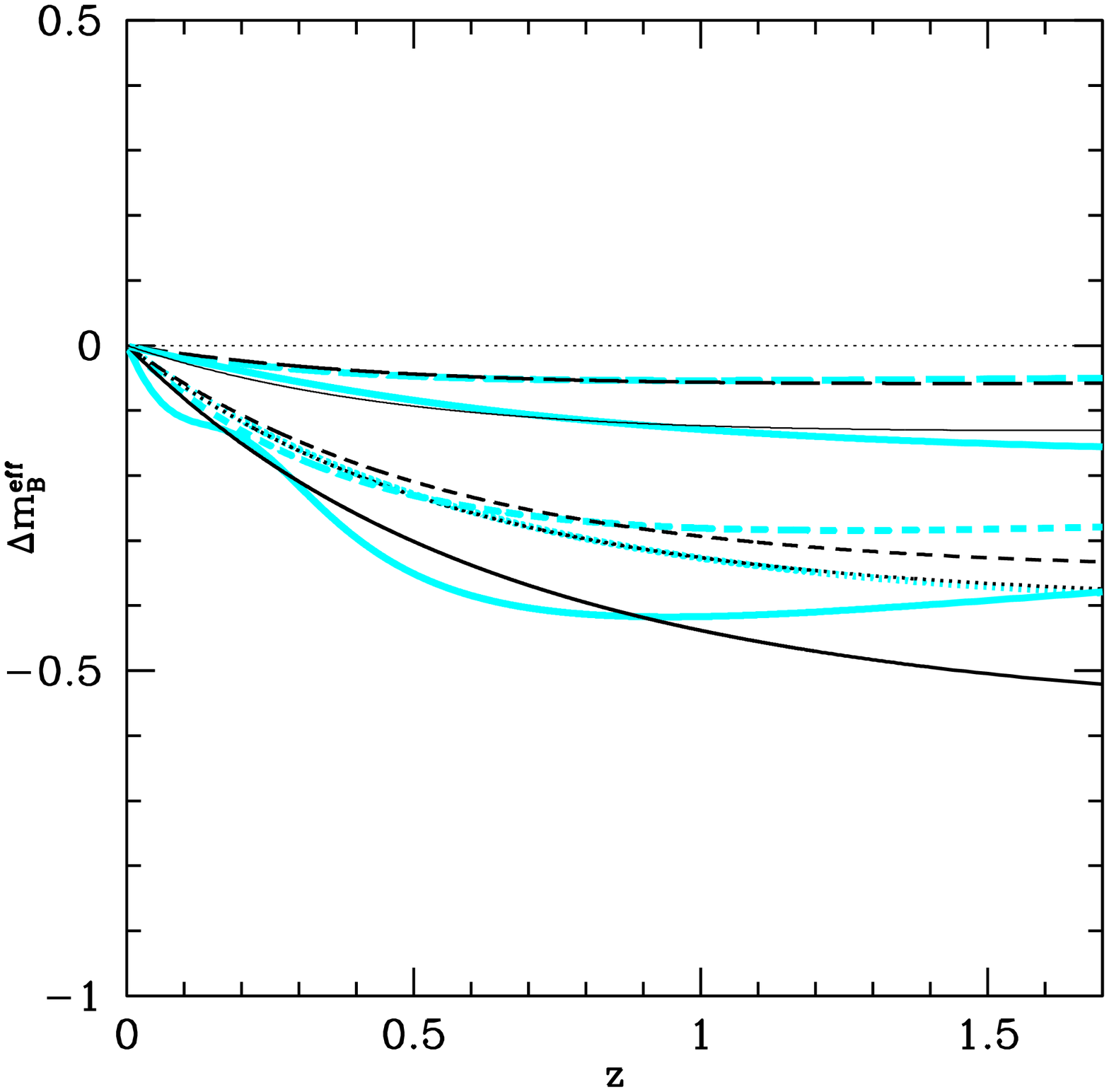,height=8.0cm,width=8cm}\psfig{file=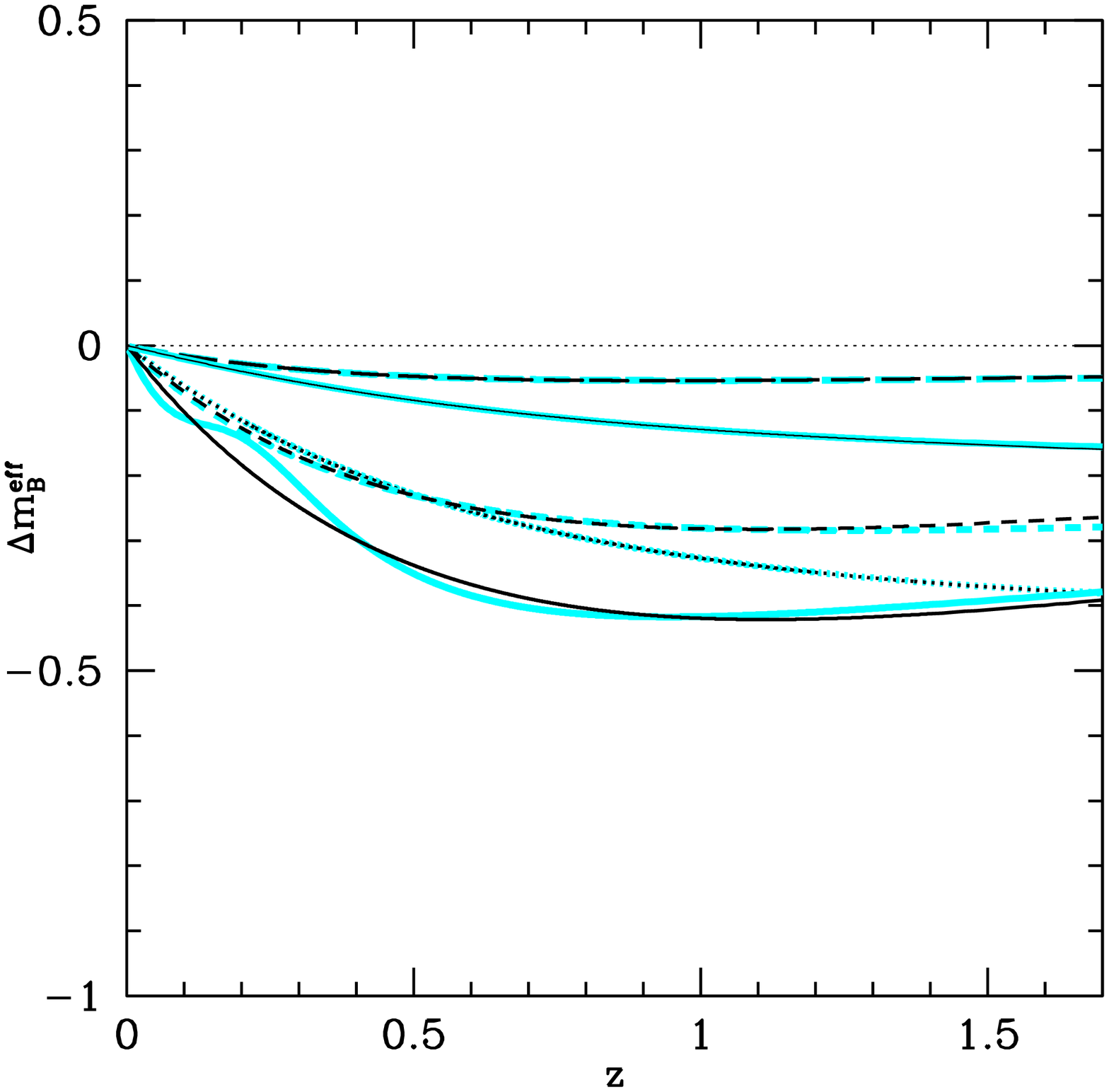,height=8.0cm,width=8cm}}}
\caption{The relative magnitude plots for the $N=0$ fit in the left
panel and the $N=1$ fit in the right panel. The light grey lines are
the theoretical values and the dark lines are the fitted results. The
line styles are the same as in fig.~\ref{fig:wall}.}
\label{fig:relmag_fit}
\end{figure}
In fig.~\ref{fig:relmag_fit} we plot the results of the constant,
$N=0$, fit and the linear, $N=1$, fit for a few of the sample of dark
energy models which we have discussed (black lines) and their
theoretical values (grey 
lines). We note that for the models which do not evolve much such as 
the inverse tracker (dotted line) and the pure exponential (long
dashed line), the constant fit seems to be sufficient. However, for the
SUGRA model (thin solid line) and the periodic potential (short dashed
line) only the first order fit is acceptable. Note that both orders
are not a good enough to fit the PNGB model (thick solid line). 

In the following we will discuss the resulting $\wp(z)$ from the
fit. First, we have to know the error matrix which is calculated as the
inverse of the second derivative of the $\chi^2$-function at its
minimum
\beq
\sigma_{ij}^{-1} =
\left.\frac{\partial^2\chi^2\left(\left\{w_i\right\}\right)}{\partial w_i
\partial w_j}\right|_{\left\{w_{i,j}=w_{i,j}^{\rm min}\right\}}\, ,
\label{wfit_sig}
\eeq
which is a valid approximation if the $\chi^2$ function has an
approximately parabolic shape around its minimum. {\tt MINUIT} also
calculates the marginalized errors on the parameters by calculating
the values of the parameters for $\chi^2_{\rm min} +\Delta\chi^2$ with
$\Delta\chi^2 = 1$. We used
both methods and found that they give consistent results. The
errors on $\wp$ are then given by Gaussian error propagation
\beq
	\delta\wp^2 = \sum_{ij}\frac{\partial\wp}{\partial
w_i}\frac{\partial\wp}{\partial
w_j}\sigma_{ij}=\sum_{ij}\left(1+z\right)^{i+j}\sigma_{ij}\, .
\label{wfit_err}
\eeq 
\begin{figure}[!h] 
\setlength{\unitlength}{1cm}
\centerline{\hbox{\psfig{file=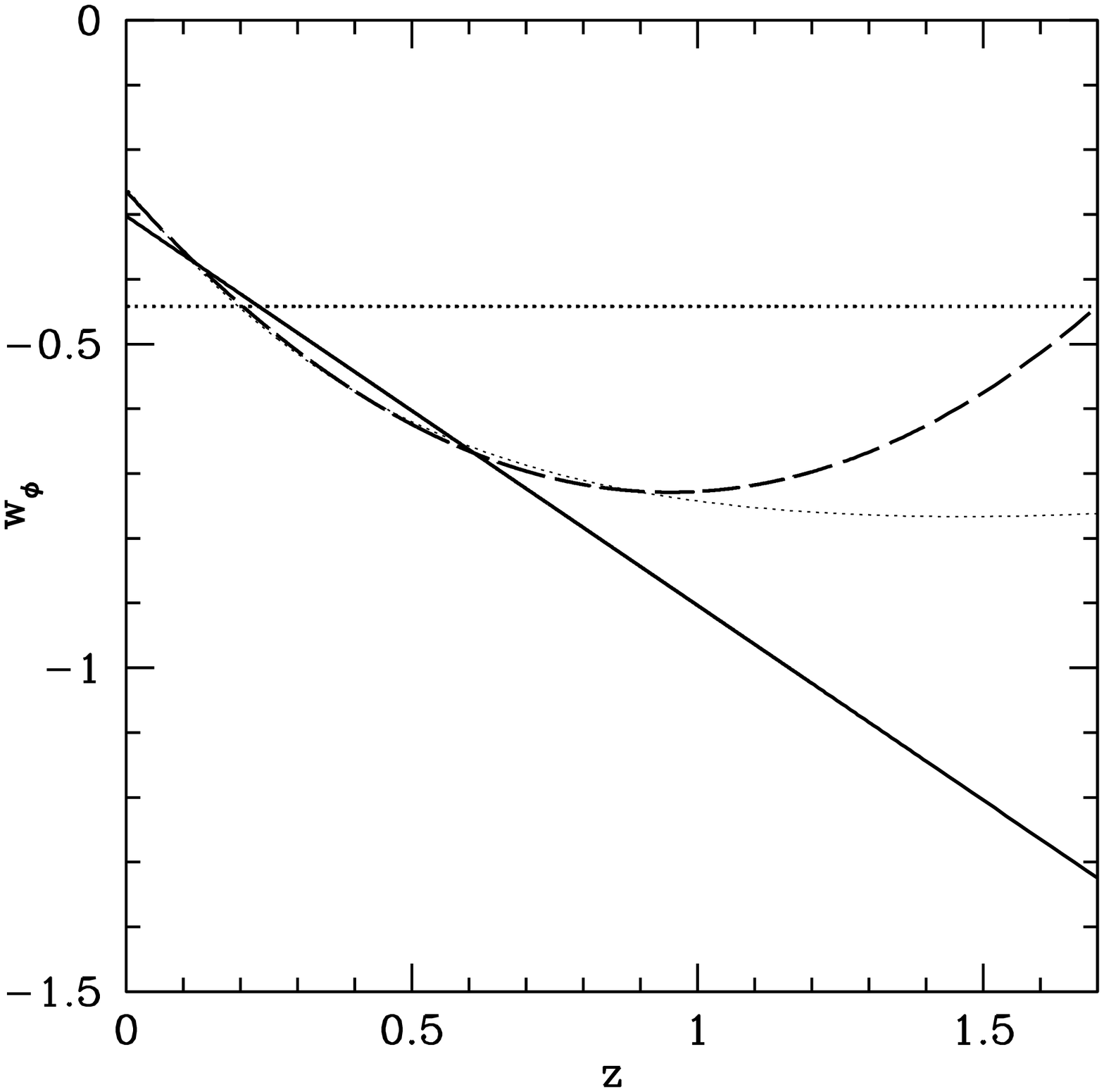,height=8.0cm,width=8cm}\psfig{file=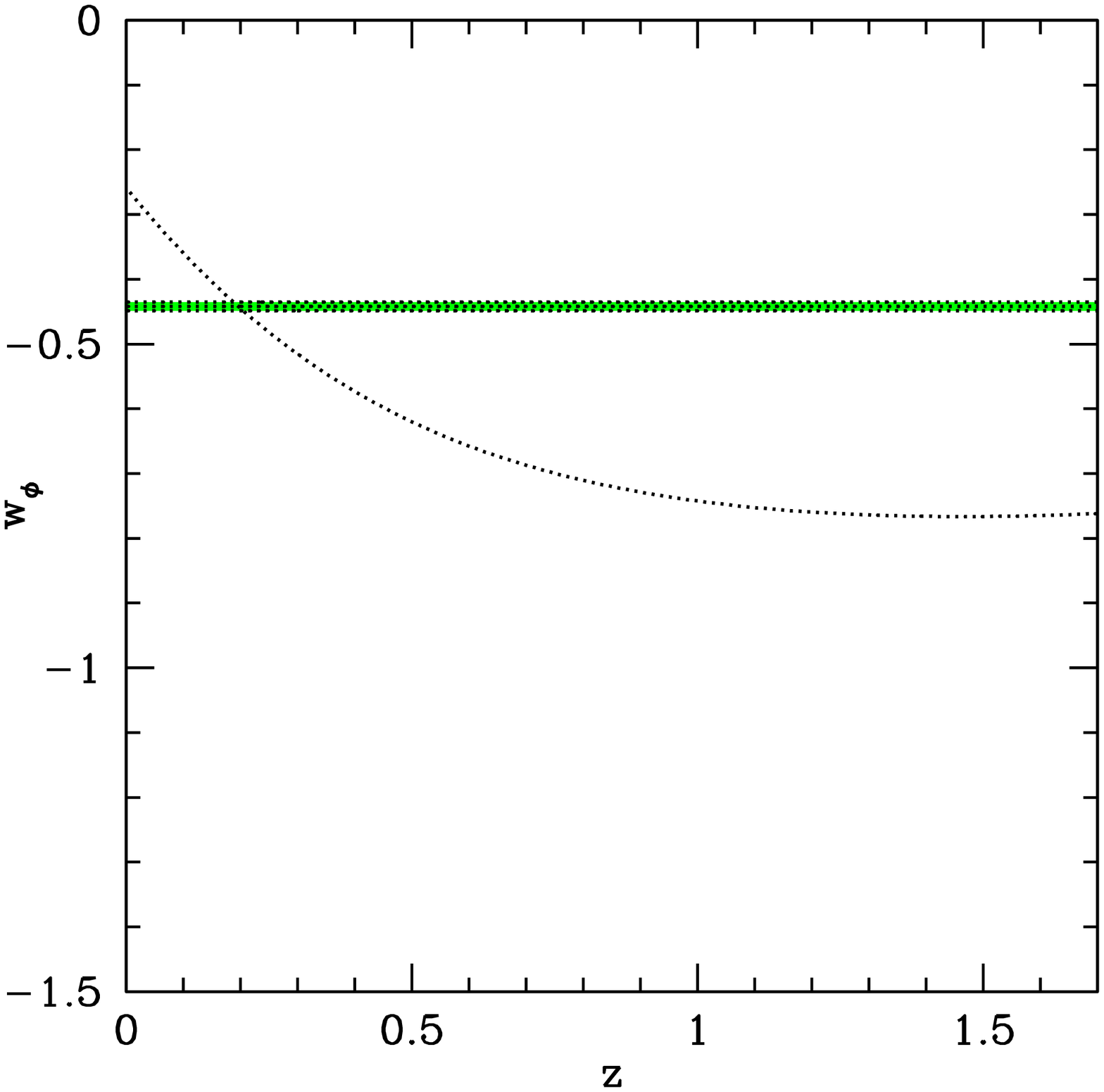,height=8.0cm,width=8cm}}} 
\centerline{\hbox{\psfig{file=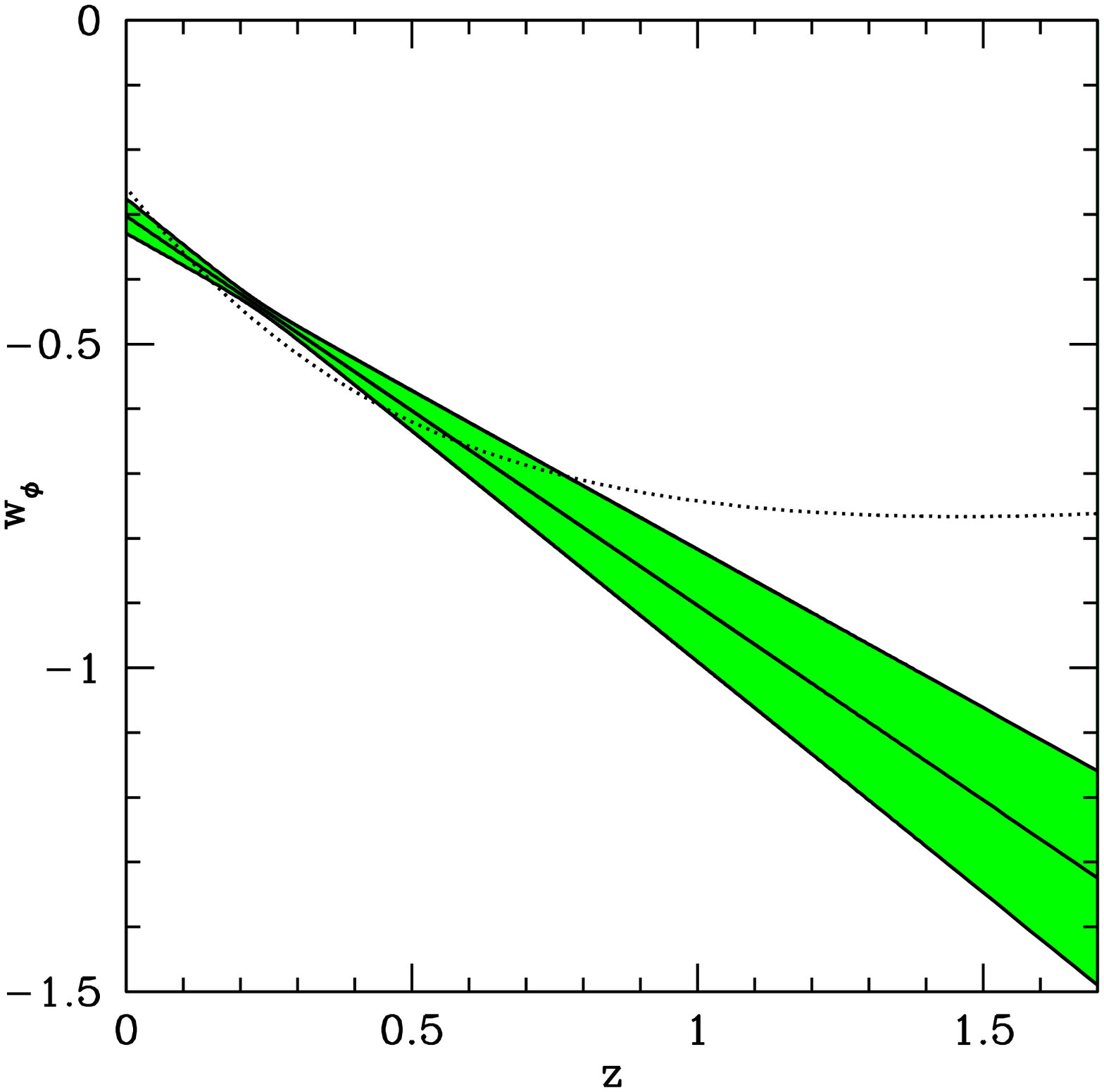,height=8.0cm,width=8cm}\psfig{file=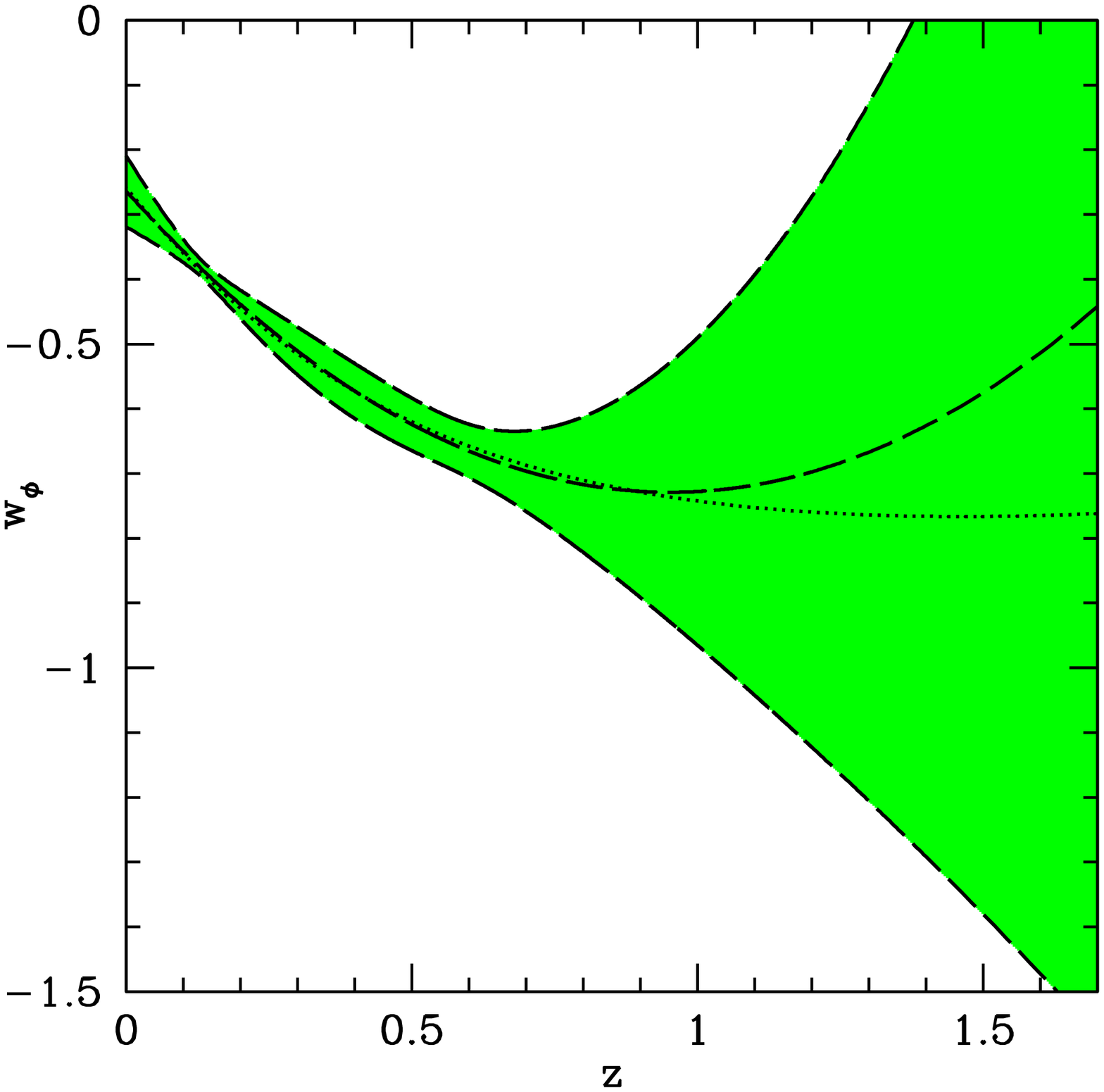,height=8.0cm,width=8cm}}}
\caption{The equation of state factor $\wp$ for the
periodic model. The theoretical value is $\wp(z)$ is given by the thin
dotted line.
In the top left panel we plot the mean values for the
fitted $\wp$ with the dotted line for the $N=0$ fit,
the solid line for the $N=1$ and the dashed line the for $N=2$ fit. The top
right panel is the $N=0$ fit with the shaded region representing the
$1\sigma$ uncertainty levels. The lower left panel is the same plot
for $N=1$ and the lower right panel for $N=2$.}
\label{fig:w_wfit}
\end{figure}
In fig.~\ref{fig:w_wfit} we plot the resulting $\wp$ for different
orders $N$ of the fit in eqn.~(\ref{dfit}) with the periodic potential
as background cosmology. In the top left panel we plot the mean values
and the theoretical curve (thin dotted line). The dashed line is for
the $N=0$ fit and we recognize that a $\wp = w_0 = const.$ fit can not
reproduce the evolving model. The $N=1$ fit (solid line) already
represents some evolution, so for $z>0.6$ the fit becomes fairly poor.
The dashed line is the quadratic, $N=2$, fit, which leads to a
better result than the linear fit. In the top right panel we show the
constant, $N=0$ fit with it's very small errorbars, in the lower left
panel the $N=1$ fit and in the lower right panel the $N=2$ fit with
the errorbars. We recognize that the errorbars for the $N=1$ and
$N=2$ fit are roughly on the same level for $z<0.7$, but then the
errorbars of the $N=2$ fit increase rapidly. In general the errorbars on
the $N=1$ fit are smaller then the ones on the $N=2$ fit, since there
are less degrees of freedom. 

\begin{figure}[!h] 
\setlength{\unitlength}{1cm}
\centerline{\hbox{\psfig{file=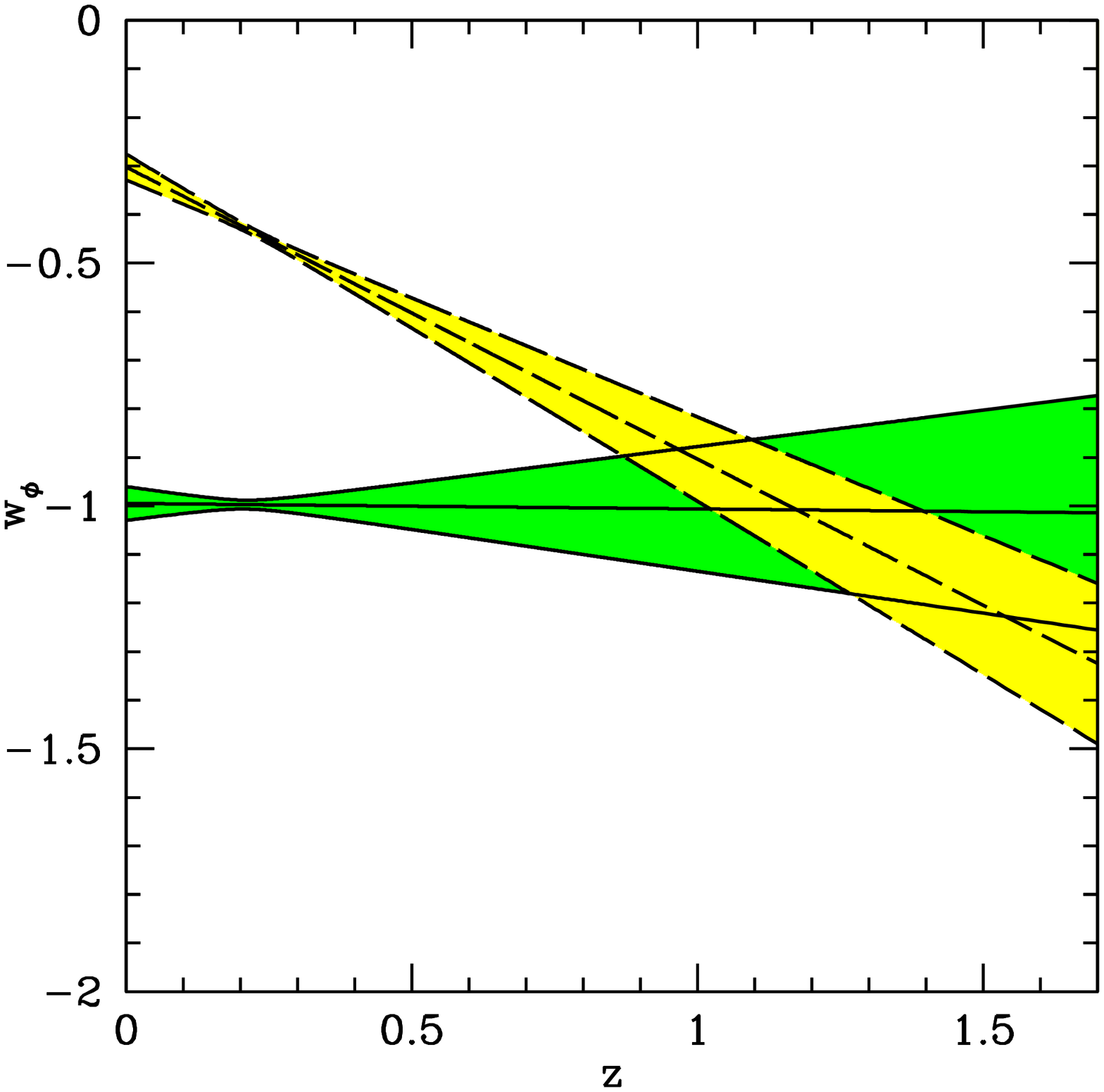,height=8.0cm,width=8cm}\psfig{file=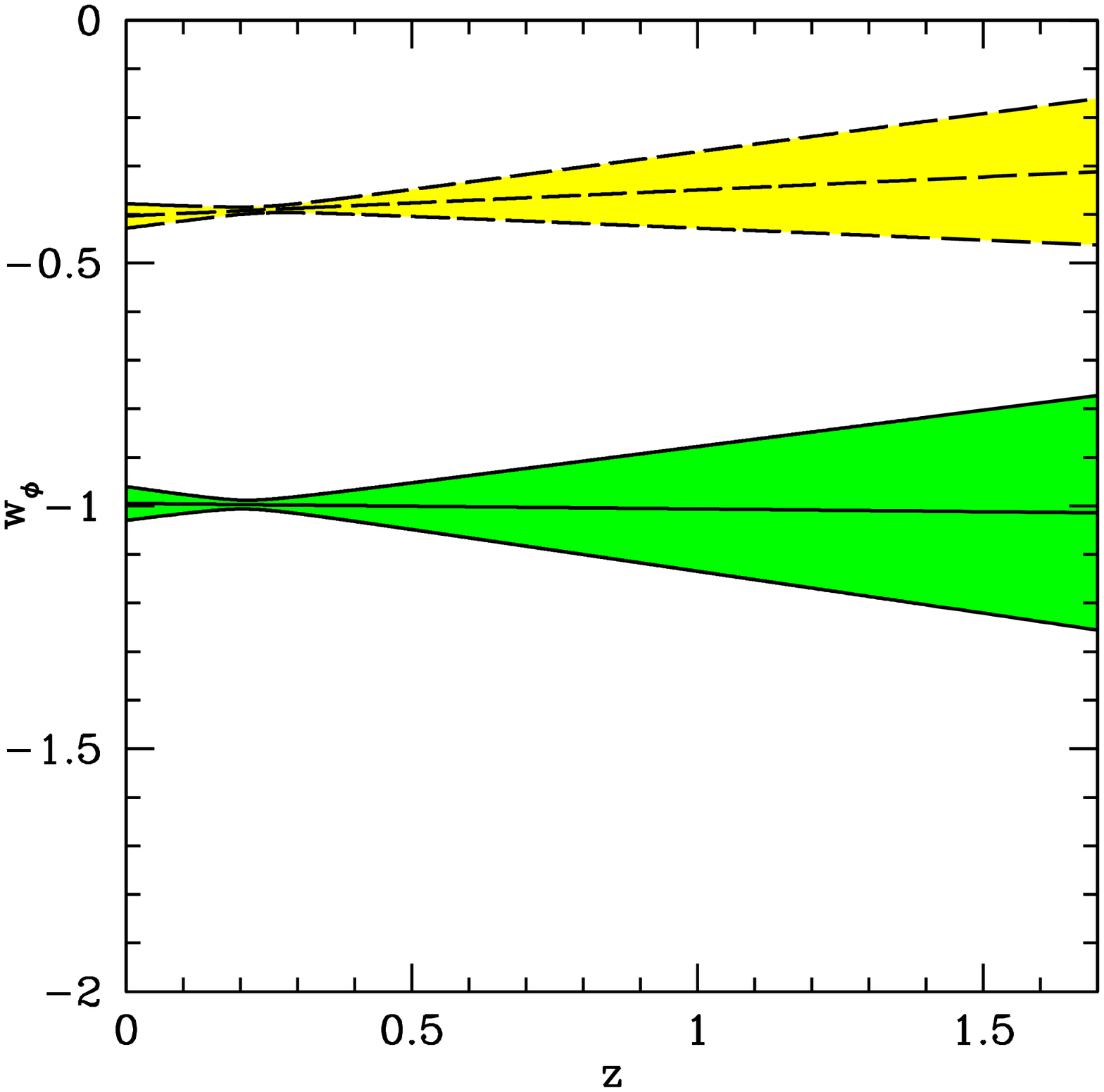,height=8.0cm,width=8cm}}} 
\centerline{\hbox{\psfig{file=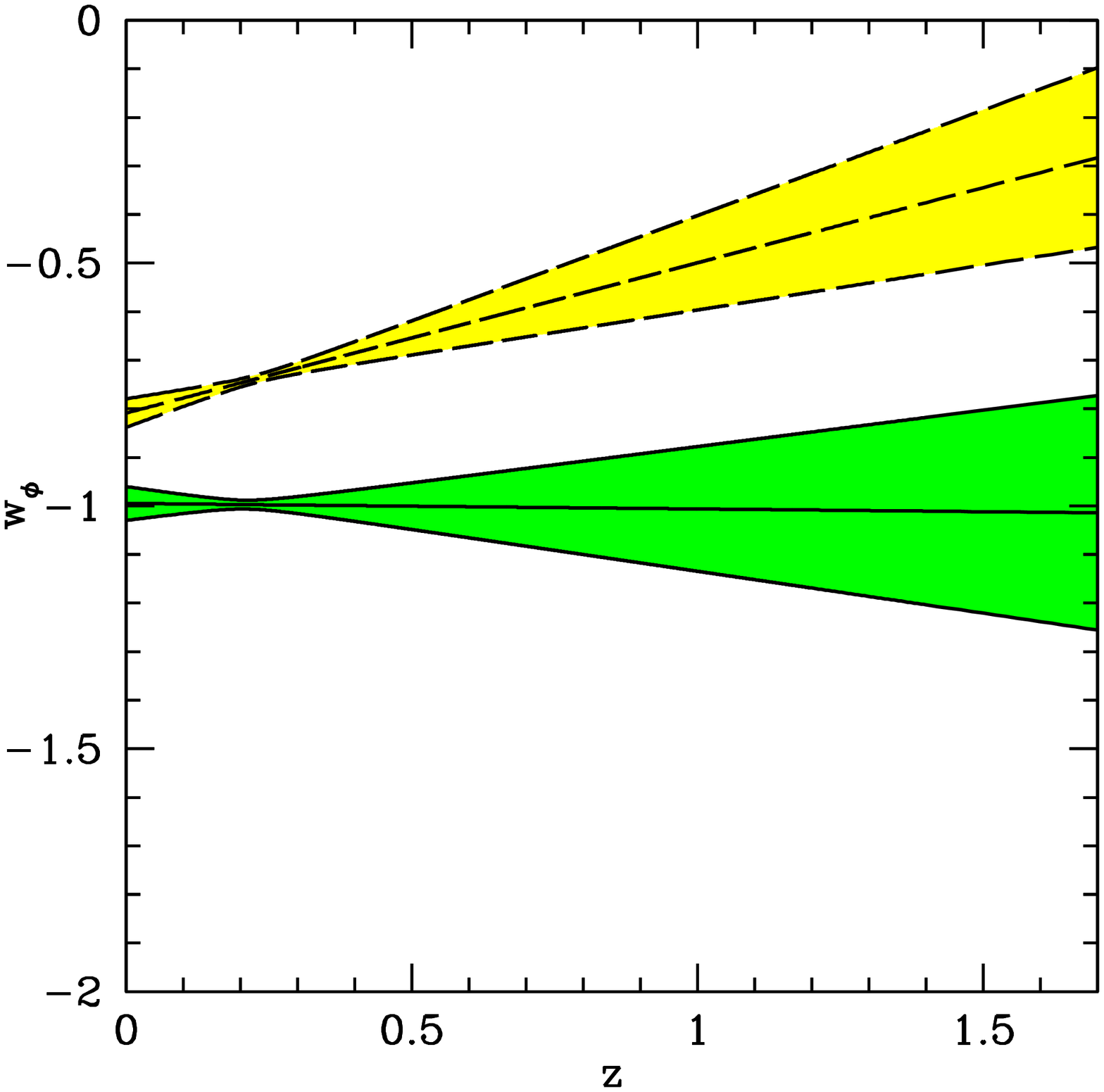,height=8.0cm,width=8cm}\psfig{file=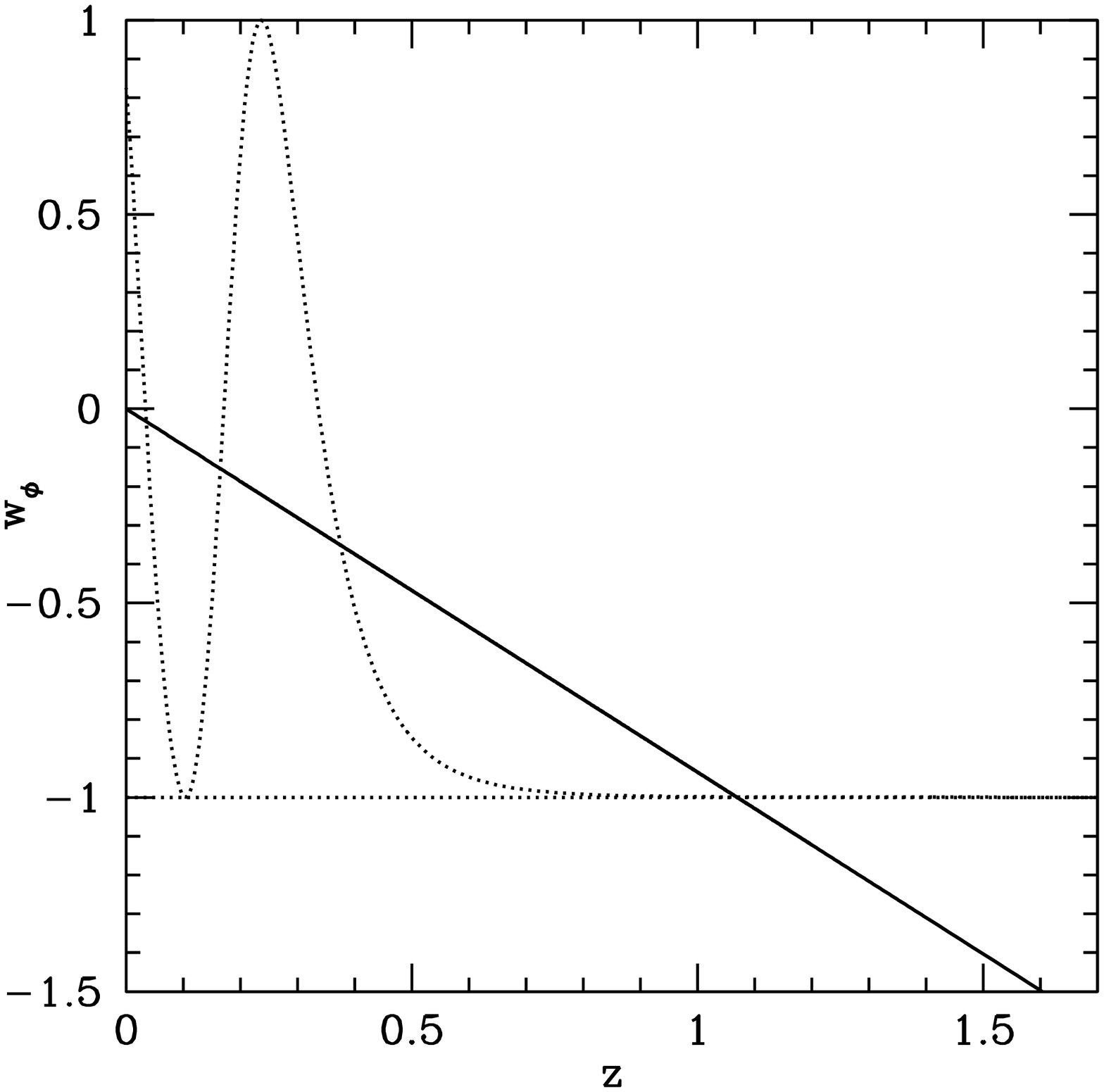,height=8.0cm,width=8cm}}}
\caption{The fitted $\wp$ for different dark energy models. The
solid lines and the dark shaded region corresponds to the mean and
$1\sigma$ error regions of the pure
cosmological constant model. The dashed lines and the light shaded
region to the periodic potential (top left panel), the inverse tracker
potential (top right panel) and the SUGRA potential (lower left
panel). In the lower right panel we show the theoretical (dotted line)
and reconstructed (solid line) $\wp$ of the PNGB model.}
\label{fig:w_wfit_mod}
\end{figure}
In fig.~\ref{fig:w_wfit_mod} we plot the periodic potential model (top
left panel), the inverse tracker (top right panel) and the SUGRA
inspired dark energy model (bottom left panel). The light shaded regions are
the $1\sigma$ confidence levels. In each plot we show the cosmological
constant model for comparison (dark shaded region). We recognize that
for the fit from
eqn.~(\ref{dfit}) we can distinguish the $\wp(z)$ evolution from a
cosmological constant on the $1\sigma$ level. We also performed the fits for
the $N=2$ approximation and got still better results as for the
polynomial fit, despite the increased size of the errorbars. In the
lower right panel we show the mean of the fit for the PNGB model
(solid line). We recognize that the $\wp$ fit is not appropriate for
the oscillating PNGB model (dotted line). This behavior does not
improve much if we apply a second order fit.
The polynomial fit hardly can distinguish the reconstructed
$\wp$ of the inverse tracker, the periodic and the SUGRA potential
from a cosmological constant, although their theoretical $\wp$ is completely
different from $\wp = -1$ as evident from fig.~\ref{fig:wall}.

We will now examine the question of whether we can reconstruct an
evolving $\wp$ with the SNe
observations. Since the $\chi^2$ values for the
$N=1$ fit where sufficient and the errorbars on this fit are
relatively small we will concentrate in the following on this linear fit.
In order to be able to decide if a model is evolving we perform a
change of variable to more convenient expansion. We can rewrite
eqn.~(\ref{wexp}) as
\beq
	\wp(z) = \sum_{i=0}^{N}w_i\left(1+z\right)^i =
\sum_{i=0}^N\wt_iz^i\, ,
\eeq
with
\beq
 \wt_i = \sum_{k=0}^{N} {k \choose i} w_k\, .
\eeq
For the $N=1$ fit this leads to $\wt_0 = w_0+w_1$ and $\wt_1 = w_1$.
The errors in the new expansion coefficients $\wt_i$ can, again, be found by
Gaussian error propagation
\beq
	\delta\wt_i^2=\sum_{kl}{k \choose i}{l \choose i}\sigma_{kl}\,
.
\label{eqn:transerr}
\eeq
We calculated these expansion coefficients and their $1\sigma$ errors
for all the models.
\begin{table}[!h]
\centering
\begin{tabular}{ccccccc}
 & $\wt_0$ & $\delta\wt_0$ & $\wt_1$ & $\delta\wt_1$ & theoretical evolution &
 evolution reconstructed\\
\hline
$\Lambda$ & $-1.00$ & $0.035$ & $-0.011$ & $0.16$ & $-$ & $-$ \\
\hline
trapped minimum & $-0.99$ & $0.035$ & $-0.0057$ & $0.16$ & $-$ &
 $-$ \\
\hline
Brane & $-0.97$ & $0.034$ & $0.028$ & $0.16$ & $+$ & $-$ \\
\hline
two exp. & $-0.95$ & $0.034$ & $-0.016$ & $0.16$ &$-$ & $-$ \\
\hline
periodic & $-0.30$ & $0.027$ & $-0.60$ & $0.11$ & $+$ & $+$ \\
\hline
pure exp. & $-0.84$ & $0.033$ & $-0.14$ & $0.15$ &$0$ & $0$ \\
\hline
PNGB & $-0.00$ & $0.025$ & $-0.94$ & $0.10$ & $+$ & $+$ \\
\hline
SUGRA & $-0.81$ & $0.029$ & $0.31$ & $0.13$ & $+$ & $+$  \\
\hline
exp. tracker & $-1.00$ & $0.035$ & $-0.011$ & $0.16$ & $-$ & $-$ \\
\hline
inv. tracker & $-0.40$ & $0.025$ & $0.054$ & $0.10$ & $-$ & $-$
\end{tabular}
\caption{The evolution coefficients with errorbars for the linear fit
$\wp = \wt_0 +\wt_1z$. ``$+$'' denotes evolution, ``$-$'' no evolution
and ``$0$'' marginal evolution.}
\label{tab:wcoef}
\end{table}
In table \ref{tab:wcoef} we present the expansion coefficients for
the dark energy models we have discussed. We note that we obtain evidence
at the $1\sigma$ level for evolution for the SUGRA and periodic potential. The only model
which is not reproduced correctly in the context of an evolving
$\wp$ is the brane-inspired model. This is because equation of state factor for
the brane model changes only relatively early ($z> 0.8$),
where the data does not have so much weight. Note although 
we get a consistent result for the PNGB model, the fit of
$\wp$ is relatively bad (fig.~\ref{fig:w_wfit_mod}, lower right
panel) and it has with $\chi^2 \approx 24.8$ an unacceptable
luminosity distance fit. We also note in fig.~\ref{fig:w_wfit_mod} that the
inverse tracker 
model is not only evolving, but also does not behave like a cosmological
constant, since $\wp \approx -0.4$, which is consistent with the
result in table \ref{tab:wcoef}. 
\begin{figure}[!h] 
\setlength{\unitlength}{1cm}
\centerline{\hbox{\psfig{file=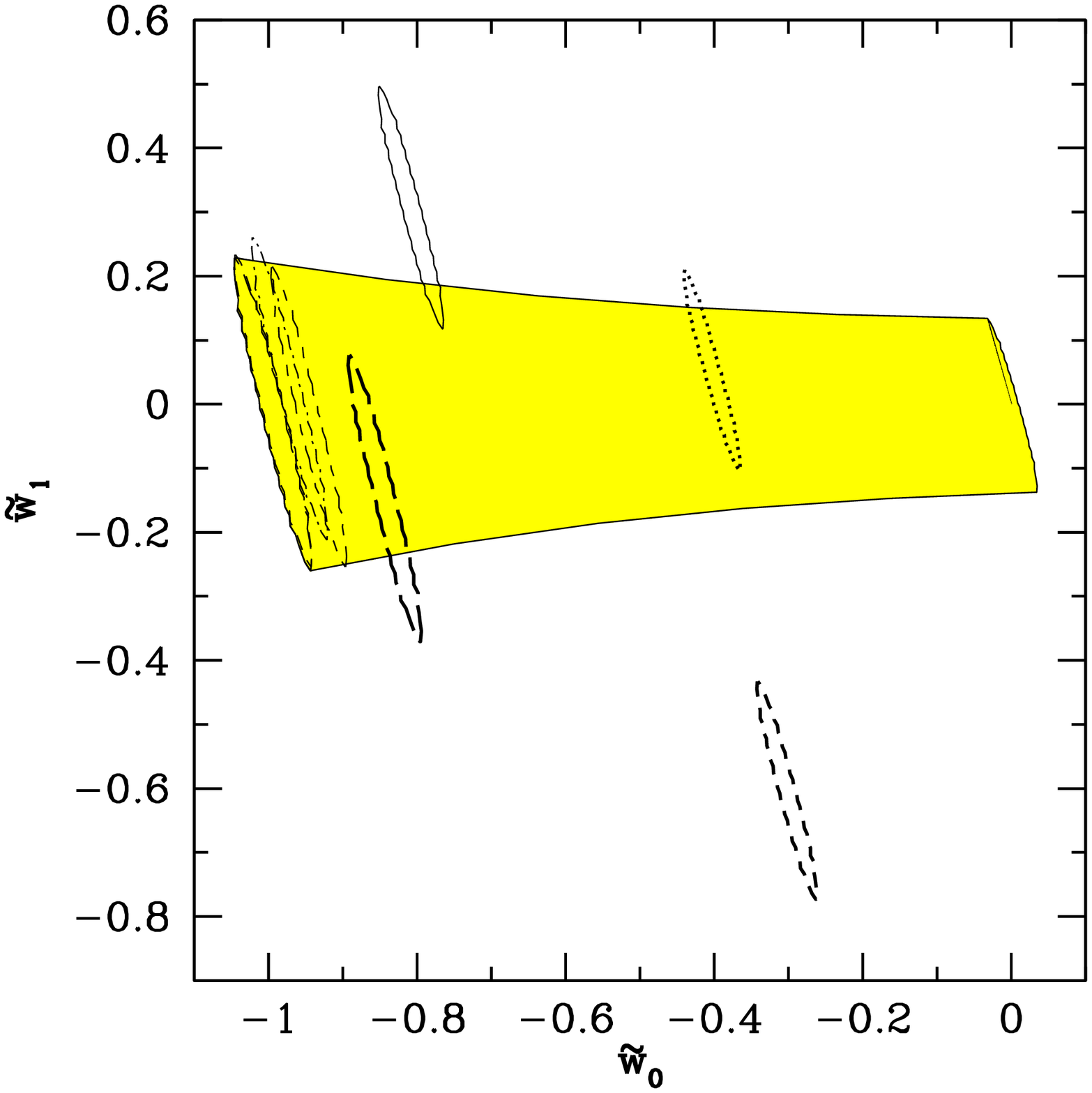,height=8.0cm,width=8cm}\psfig{file=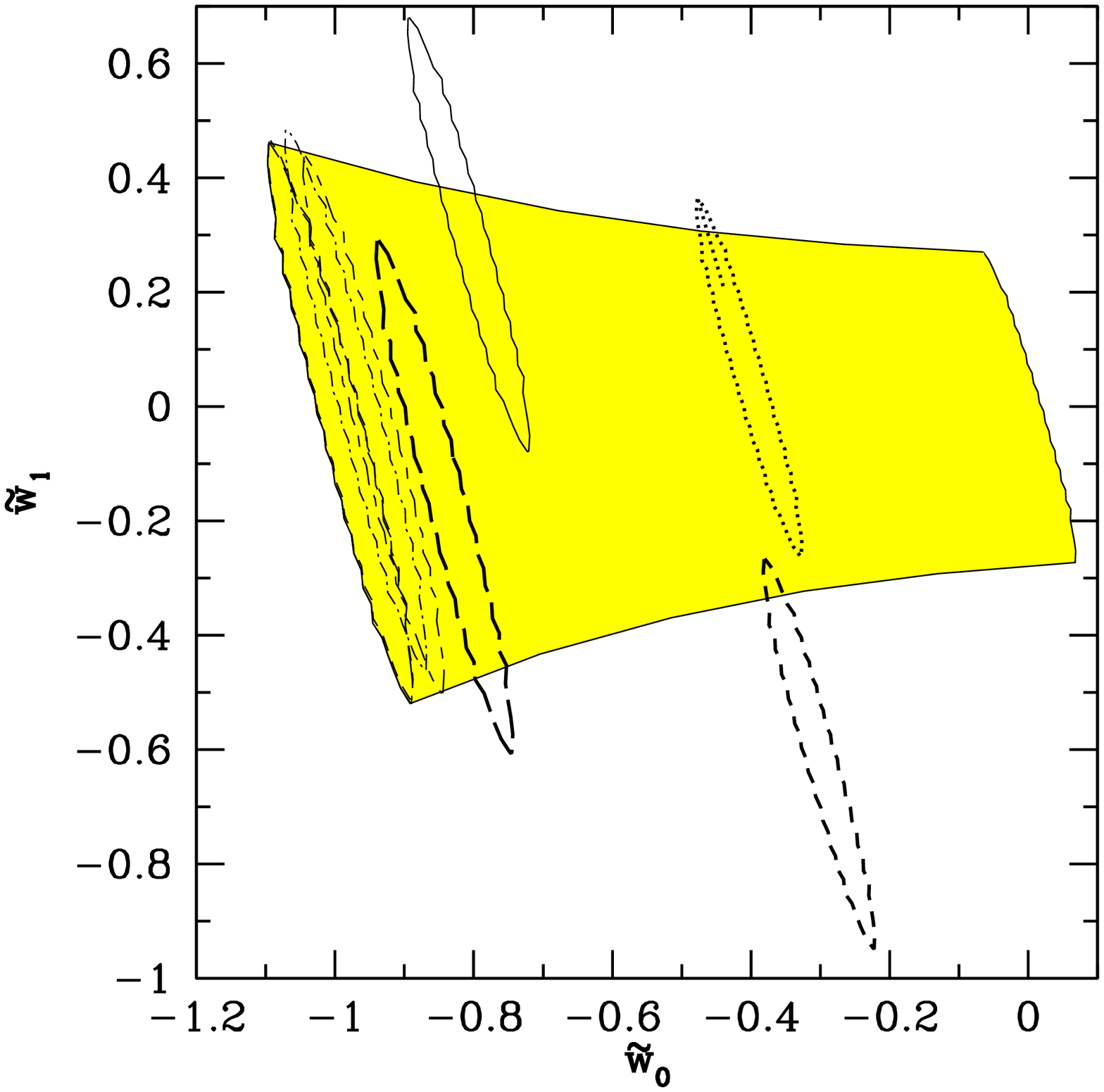,height=8.0cm,width=8cm}}} 
\caption{The joint confidence regions in the $\wt_0 - \wt_1$
plane. In the left panel we show the $68.3\%$ confidence regions and
in the right panel the $99\%$ region. The shaded region is the
uncertainty region for an arbitrary but constant equation of state
factor with $-1<\wp<0$. The thin short dash ellipse is for the
trapped minimum models, the thin dot - short dash ellipse is for the
brane inspired potential, the thin short dash - long dash ellipse for
the potential which involves two exponentials, the thick short dashed
ellipse for the periodic potential, the thick long dashed ellipse for the
pure exponential, the thin solid ellipse for the Supergravity inspired
potential, the thin long dashed ellipse for the exponential tracker
solution, and the thick dotted ellipse for the inverse tracker.}
\label{fig:w0w1}
\end{figure}
In fig.~\ref{fig:w0w1} we show the joint probability contours in the
$\wt_0 - \wt_1$ plane, for the $68.3\%$ and $99\%$ confidence
levels. The confidence levels for the joint probability are calculated
as the regions with $\chi^2=\chi^2_{\rm min} + \Delta\chi^2$ with
$\Delta\chi^2 = 2.3$ and $\Delta\chi^2 = 9.21$ respectively, which is
valid since the errors are symmetric. The
shaded region is the $68.3\%$, resp. $99\%$, 
confidence region for $-1 < \wp = const < 0$. In this plot we see
that only the SUGRA (thin solid ellipse) and periodic potential (thick
short dashed ellipse) can be
distinguished from a constant $\wp$ at the $1\sigma$ level (left
panel). At the $99\%$ level (right panel ) it is even harder to extract evolving
models and only the periodic potential can be distinguished from
$\wp=const$. However, if we are just interested in whether a model is evolving
we have to concentrate on the parameter $\wt_1$ and marginalize over
$\wt_0$, which corresponds to the projection of the confidence region
with $\Delta\chi^2 = 1$ for the $1\sigma$ errorbars\cite{NumRec,Cowan:98}. But
even if we consider the marginalized errors on $\wt_1$ in table
\ref{tab:wcoef} the only model for which we can find evidence for
evolution with $99\%$ confidence is the periodic potential. Note that
we have omitted the PNGB model in this discussion because the
reproduced $\wp$ for this model is not valid.

Up to now we have used a fixed prior on $\Omega_m$ to fit the luminosity
distance and reconstruct the equation of state factor $\wp$. We will
now discuss how our results will change if we have no prior
information on $\om$ and use just the constraint $0\le\om\le1$. 
\renewcommand{\arraystretch}{1.5}
\begin{table}[!h]
\centering
\begin{tabular}{ccccccc}
 & $w_0$ & $\delta w_0$ & $w_1 = \wt_1$ & $\delta w_1 = \delta\wt_1$ & $\om$ & $\delta\om$ \\
\hline
$\Lambda$ & $-0.94$ & ${}_{-0.46}^{+1.08}$ & $-0.063$ & ${}_{-1.10}^{+0.64}$ & $0.31$ & ${}_{-0.31}^{+0.07}$ \\
\hline
inv. tracker & $-0.44$ & ${}_{-0.15}^{+1.53}$ & $-0.01$ & ${}_{-0.99}^{+0.99}$ & $0.38$ & ${}_{-0.38}^{+0.23}$ \\
\hline
periodic & $0.75$ & ${}_{-0.65}^{+0.86}$ & $-1.10$ & ${}_{-0.83}^{+0.69}$ & $0.41$ & ${}_{-0.21}^{+0.08}$ \\
\hline
SUGRA & $-1.10$ & ${}_{-0.20}^{+1.31}$ & $0.21$ & ${}_{-1.42}^{+0.29}$
& $0.34$ & ${}_{-0.34}^{+0.13}$
\end{tabular}
\caption{The expansion parameters from eqn.~(\ref{wexp}) for a $N=1$ 
fit, where $\om$ is only constrained by $0\le\om\le1$. The underlying 
theoretical models have all $\om=0.3$.} 
\label{tab:om_n1fit}
\end{table} 
\renewcommand{\arraystretch}{1.0}
In table \ref{tab:om_n1fit} we show the results of the fit with $N=1$ 
and no prior on $\om$. Note that we show the quantities $w_i$ and {\em 
not} $\wt_i$. The reason for this is, that if we include $\om$ as 
parameter to fit, the errors on the fitted $w_i$ are not symmetric,
and therefore not Gaussian. Therefore, we can not perform the error
propagation using eqn.~(\ref{eqn:transerr}). We study two models which are
theoretically not evolving, $\Lambda$ and the inverse tracker, and two
models which have an evolving $\wp$, the periodic and SUGRA
potential. First, we note 
that the errorbars on the fitted value of $\om$ are large, and that 
the mean value, in the case of the periodic and inverse tracker 
potential is displaced from the theoretical value by over $25\%$! Although, 
due to the large errorbars the mean value and the theoretical value 
always lie within the $1\sigma$ errorbars. In the case of the periodic 
potential we can recover evolution marginally, since $\wt_1 = 
-1.10^{+0.69}_{-0.83}$. However, at the $99\%$ level we can not gain 
any evidence for evolution. For the SUGRA model we can not reconstruct 
evolution since the obtained value for $\wt_1$ is consistent with no 
evolution ($\wt_1 = 0$) already on the $1\sigma$ level. We conclude,
therefore, that if we fit simultaneously for $\om$ and $w_1$
simultaneously it is poor and has large errorbars.   

\begin{table}[!h]
\centering
\begin{tabular}{ccccc}
 &$ w_0 $& $\delta w_0$ & $\om$ & $\delta\om$ \\ 
\hline
$\Lambda$ & $-0.99$ & $0.06$ & $0.30$ & $0.02$ \\ 
\hline
inv. tracker & $-0.45$ & $0.11$ & $0.37$ & $0.12$ \\
\hline
periodic & $-0.28$ & $0.01$ & $0.00$ (!)& $0.03$ \\
\hline
SUGRA & $-0.91$ & $0.07$ & $0.38$ & $0.03$  
\end{tabular}
\caption{The fit results for $\om$ and $w_0$ using the $N=0$ fit.}
\label{tab:wfit_n0}
\end{table}
In table \ref{tab:wfit_n0} we show the results for the constant $N=0$
fit. First, we note that the errorbars on  $w_0$ and $\om$ are much
smaller than for the linear $N=1$ fit. However, we do not expect that
this will work for models which are evolving. In fact, we recognize
that for the periodic potential we get $\om =0$ where this is only the
``best fit'' value because of the constraint on $\om$. If we release the
constraint for this fit, $\om$ floats to negative and non-physical
values. This is because the periodic potential has an evolving $\wp$ and an $N=0$
fit can not reproduce such an evolving model. The pure $\Lambda$, the
SUGRA and the inverse tracker model seem to give reasonable results,
apart from the SUGRA model results in a too large value for $\om$, where the true value
of $\om$ is outside the $1\sigma$ errorbar. This is again due to the
fact that $\wp$ is evolving for the SUGRA model.
\begin{figure}[!h] 
\setlength{\unitlength}{1cm}
\centerline{\hbox{\psfig{file=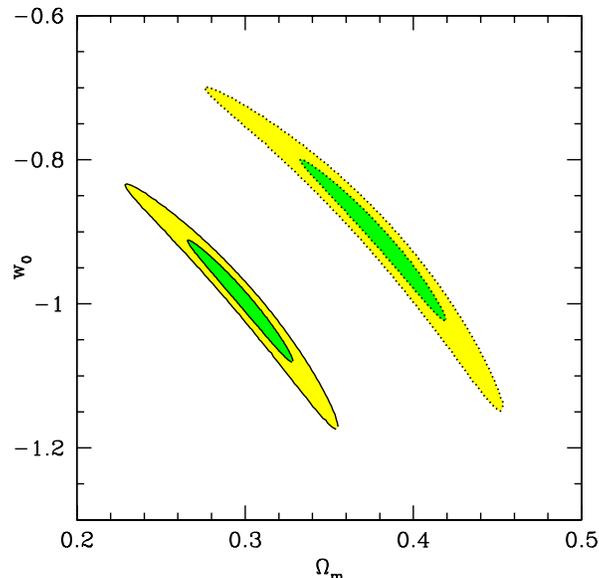,height=8.0cm,width=8cm}}}
\caption{The joint probabilities for $\om$ and $w_0$ for the $N=0$ fit for the $\Lambda$ model (solid line) and the SUGRA model (dotted line). The dark shaded region is the $68.3\%$ confidence level and the brighter shaded region the $99\%$ level.}
\label{fig:w0_om}
\end{figure}
In fig.~\ref{fig:w0_om} we show the joint probability contours in the
$w_0-\om$ plane. The solid line is the pure cosmological constant and
the dotted line the SUGRA model. We note that even for a
marginally evolving model like the SUGRA model that the $N=0$ fit
gives a reasonable result. Note that we did not impose a constraint of
$w_0\ge -1$ for the $N=0$ fit. This constraint basically results in a
cut off of the confidence regions in fig.~\ref{fig:w0_om} below the
$w_0 = -1$ line. To conclude we remark that we either need a fixed prior on
$\om$ to be able to tell whether the equation of state factor $\wp$ is evolving,
or we only consider the constant contribution to $\wp$ which would
allow us to establish both $\om$ and $\wp$ accurately. In section \ref{err} we will now
investigate how different satellite designs and priors on $\om$ may change this behavior.

\section{Impact of experiment design parameters and priors on the
matter content}\label{err}
We will now discuss the impact of the specific SNAP setup on the
estimation on $\wp$. 
\begin{figure}[!h] 
\setlength{\unitlength}{1cm}
\centerline{\hbox{\psfig{file=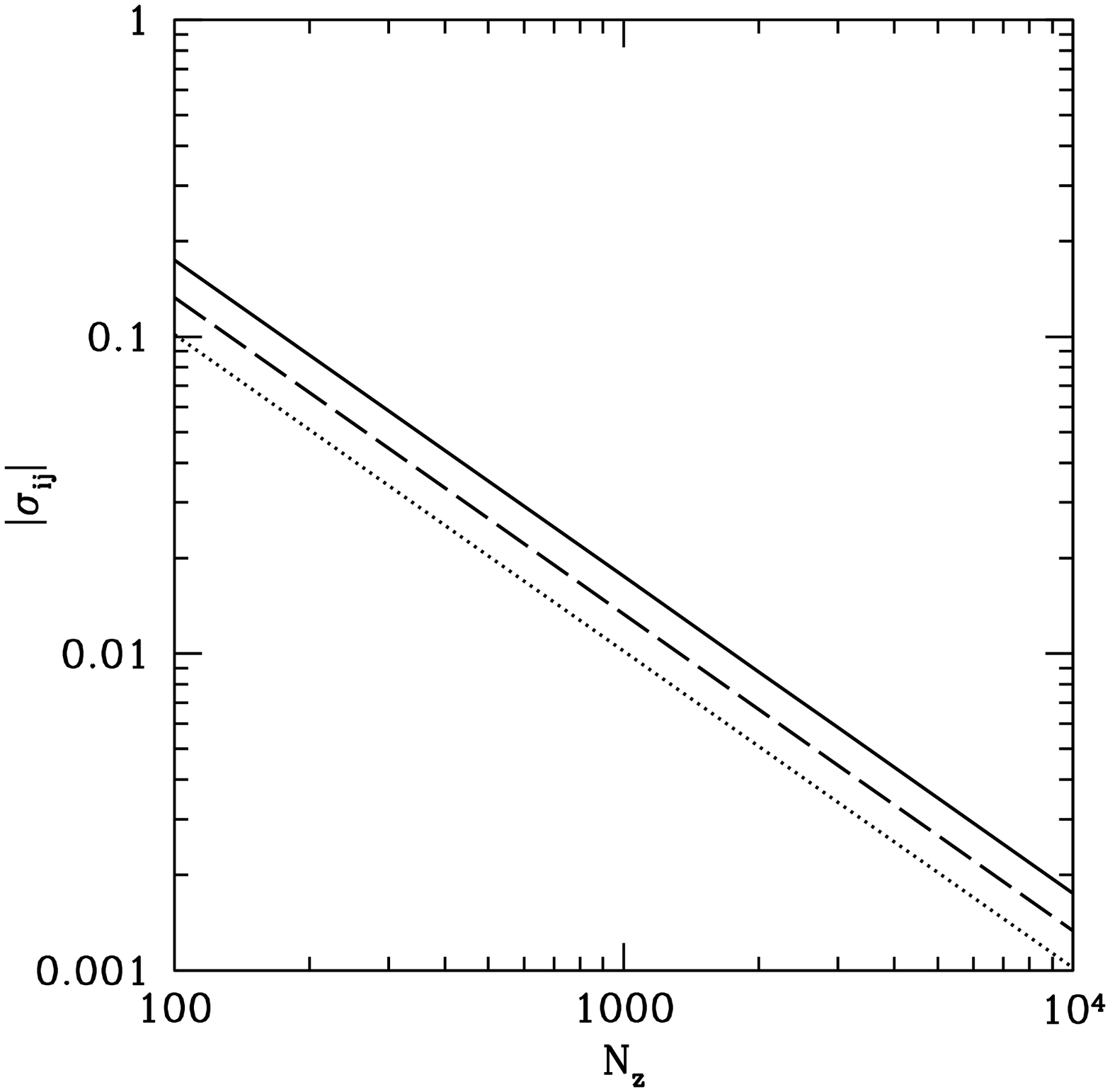,height=8.0cm,width=8cm}\psfig{file=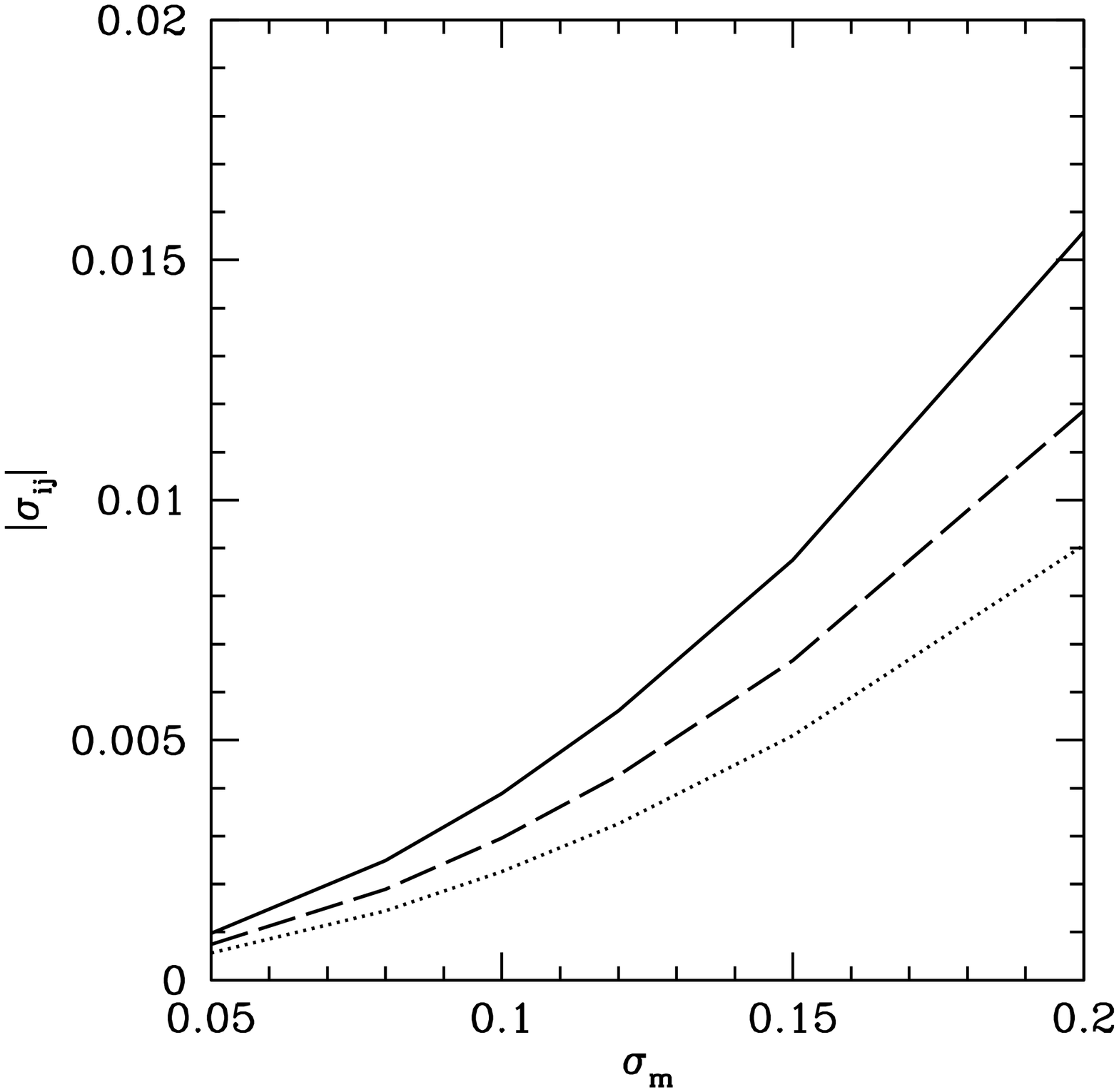,height=8.0cm,width=8cm}}}
\centerline{\hbox{\psfig{file=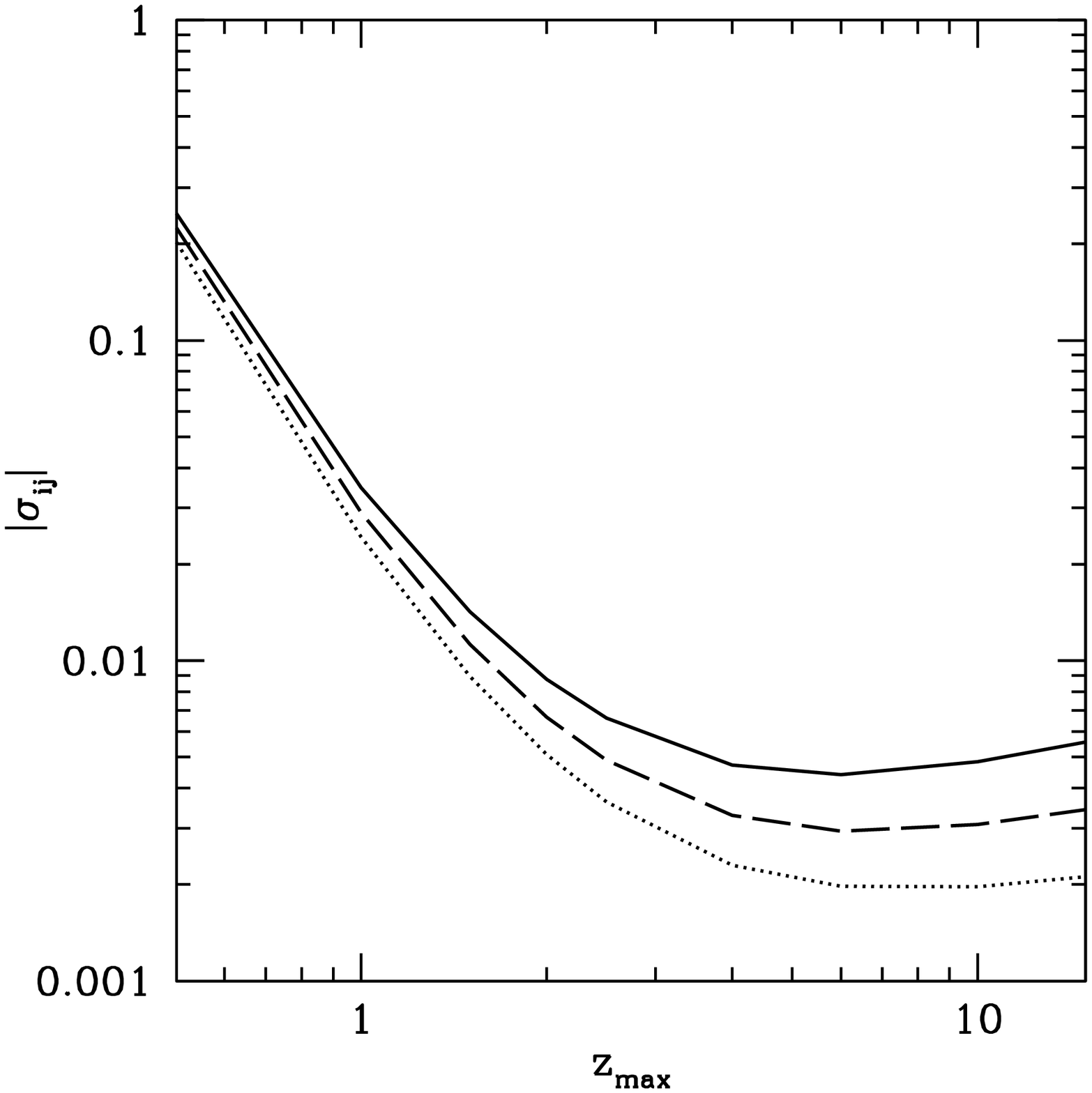,height=8.0cm,width=8cm}\psfig{file=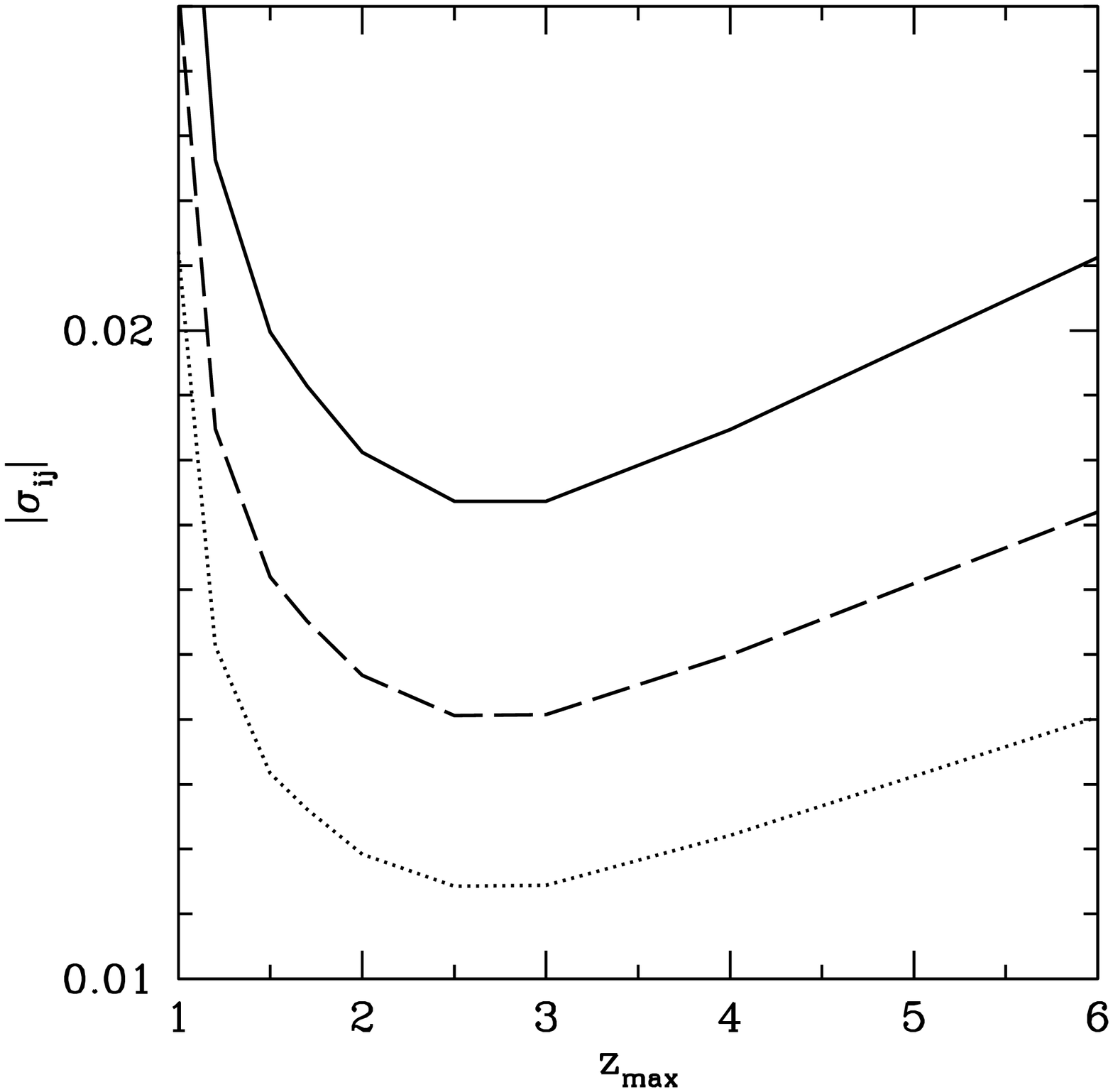,height=8.0cm,width=8cm}}}
\caption{The dependence of the statistical errors $\sigma_{ij}$ on the
SNAP specifications for the linear, $N=1$, fit. The dotted line is
$\sigma_{00}=\delta w_0^2$ the dashed line
is $-\sigma_{01}=\left<\delta w_0 \delta w_1\right>$ and $\sigma_{11}=\delta w_1^2$ the solid line. In the top left
panel we show the dependence on the number of observed redshifts $N_z$
where the maximal redshift is fixed to $z_{\rm max} = 2$ and the
statistical error on the magnitude is $\sm=0.15\, {\rm mag}$. In the top right
panel we show the dependence on $\sm$, where the number of redshifts
is fixed to be $N_z=2000$. In the lower left panel we plot the error
matrix as a function of the maximal observed redshift, where we fix the
number of observed SNe to $2000$ and $\sm=0.15\,
{\rm mag}$. In the lower right panel we show the evolution of the statistical error
with the maximal observed redshift assuming that $90\%$ of the
observations are in the low redshift region $0<z<1.2$.}
\label{fig:zf}
\end{figure}
In fig.~\ref{fig:zf} we show the dependence of the error matrix
$\sigma_{ij}$ on the experimental parameters using the $N=1$ fit for the
periodic potential model with $\om=0.3$ fixed. The
dotted line is $\sigma_{00}$, the dashed line $-\sigma_{01}$ and the
solid line is $\sigma_{11}$. In the
top left panel we show the dependence on the observed number of
redshifts $N_z$ where we keep the magnitude error fixed at $\sm = 0.15\, {\rm mag}$
and the maximal observed redshift is $z_{\rm max} = 2$. We recognize
that the variances scale as expected with $1/N_z$, or the errors with
$1/\sqrt{N_z}$. In the top right panel we show that the dependence on the
absolute magnitude error $\sm$, where the maximal redshift is fixed to
$z_{\rm max}=2$ and the number of observed SNe is $N_z=2000$. We
note that the error matrix is increasing as $\sm$ increases as
expected. The conservative limit on the total dispersion is $\sm =
0.15\, {\rm mag}$ as stated
in table \ref{tab:snap}, which includes using the best
currently known methods of standardizing and calibrating the
luminosity of type Ia SNe with a residual dispersion of $0.12\,
{\rm mag}$ and a measurement uncertainty of
$0.09\, {\rm mag}$ after correcting for extinction and using the color
of the SNe. Optimistically view that with new methods and all the additional
information which is available for the SNe with SNAP might reduce the residual
dispersion from the standardization and calibration to
$0.05\, {\rm mag}$, although we do not expect to the improvement in the
measurement uncertainties
to be more then $0.08\, {\rm mag}$. Therefore, the most
optimistic choice results in $\sm = 0.09\, {\rm mag}$. We see in the top right
panel of fig.~\ref{fig:zf} that if we improve the statistical error
from $\sm = 0.15\, {\rm mag}$ to $\sm =0.09\, {\rm mag}$ the uncertainties on the fit
parameters $w_i$ improve by $70\%$. This means we are able to
reconstruct evolution even for the marginally evolving pure
exponential model (thick long dashed line, fig.~\ref{fig:wall}) at the
$68.3\%$ level with $\wt_1 = -0.14$ and $\delta\wt_1 = 0.09$.

In the lower left panel of fig.~\ref{fig:zf} we plot the
dependence on the maximal observed redshift where we keep the number
of observed SNe redshifts fixed to $2000$ and $\sm =
0.15\,{\rm mag}$. It is evident from the plot that we can gain most accuracy if
the observations are done up to a redshift of $z_{\rm max} \approx 3$.
Beyond this redshift there is no further improvement.  
In the lower right panel we show more realistic result of increasing
the maximal observed redshift on the statistical error. In this plot
we assume that there is a fixed threshold number $\Delta n$ of photons
which have to arrive in the detector in order to observe an
SNe. We neglect the effects of color in this analysis. The
brightness or flux of a SNe is ${\cal F} = {\cal L}/4\pi\dl^2$ and the measured
flux in the detector is ${\cal F} \propto (\Delta n)/(\Delta A \Delta
t)$, where $\Delta A$ is the effective area of the detector. 
From this relation we can work out the time $\Delta t$ for which
we have to measure
the flux from a SNe at a particular redshift in order to observe
it. We assume that for a fraction $p$ of the total observation time $T$
we find SNe in
a low redshift region $0<z<z_l$ and a fraction $(1-p)$ in the high redshift region
$z_l<z<z_{\rm max}$. In order to calibrate the relation we {\em fix}
the low redshift region with $z_l=1.2$ and the number of observed SNe
in this region to ${\cal N}_l = 1850$ as in table \ref{tab:snap}. The
number of observed SNe in the high redshift region is then approximately
\beq
{\cal N}_h \approx {\cal N}_l \frac{1-p}{p}
\frac{\left(1+z_l+z_l^2/3\right)}{\left(1+z_l\right)^2+\left(1+z_l\right)\left(z_{\rm
max} -z_l\right)+\left(z_{\rm max}-z_l\right)^2/3}\, .
\label{sampling}
\eeq
In the lower right panel of fig.~\ref{fig:zf} we assume that for $p$
equivalent to 90\%
of the total observation time we discover SNe in the low redshift
region $0<z<1.2$. If we choose $z_{\rm max} = 1.7$ we observe ${\cal
N}_h = 65$ SNe in the region $1.2 < z < 1.7$ in agreement with
the values given in table
\ref{tab:snap}. For redshifts
$z\le 1.2$ a similar equation to eqn.~(\ref{sampling}) holds, where
in this case the entire observation time is spent in the low redshift
region. If we observe out to a redshift of $z_{\rm max}
= 2.5$ we can observe ${\cal N}_h = 48$ SNe. From fig.~\ref{fig:zf} we
recognize that if we go from $z_{\rm
max}=1.7$ to $z_{\rm max} = 2.5$ we improve the statistical error by
$10\%$. We see from table \ref{tab:wcoef} that such an improvement of
the statistical error on $\wt_1$ is only marginal and we can not
distinguish more models or establish evolution for more models when
compared to the case when
we can just measure out to redshifts of $z_{\rm max} = 1.7$. 

We notice in fig. \ref{fig:w_wfit_mod},
that the ``reconstructed'' $\wp$ has a region where the errorbars are
relatively small, which is for the $N=1$ fit in eqn.(\ref{dfit})
around $z_f\approx 0.2$. We find the position $z_f$ of the feature by
minimizing $\delta\wp$ in eqn.(\ref{wfit_err}) assuming a
symmetric error matrix and a linear fit 
\beq
	1+z_f = -\frac{\sigma_{01}}{\sigma_{11}}\, ,
\eeq
with $\sigma_{ij}$ the covariance matrix from
eqn.(\ref{wfit_sig}). This feature corresponds to the ``sweet spot''
in \cite{Huterer:00b} and is also the redshift around which
\cite{Huterer:00b} perform the $w$-expansion in order to obtain
uncorrelated expansion coefficients. From
this expression we already expect that the position of the feature
will not vary if we change the number of data points $N$ or the
statistical error on the magnitude $\sm$ and this
behavior was established by numerical experiments. 
\begin{figure}[!h] 
\setlength{\unitlength}{1cm}
\centerline{\hbox{\psfig{file=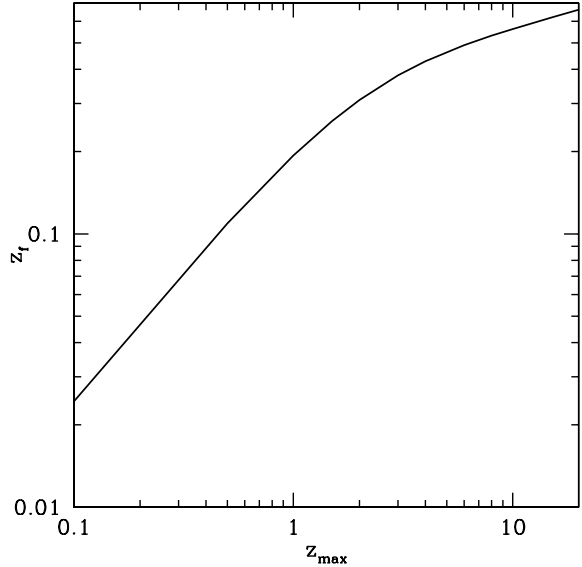,height=8.0cm,width=8cm}}}
\caption{The shift of the minimal error region $z_f$ with the change
of the maximal observed redshift $z_{\rm max}$.}
\label{fig:zfeat}
\end{figure}
In fig.~\ref{fig:zfeat} we plot the dependence of the position of the
feature $z_f$ versus the maximally measured redshift for the periodic
potential. We plotted the behavior up to the very high
redshift $z=20$ and recognize that even for such high redshifts the
feature is still below $z_f = 0.7$. The transition from matter to
vacuum domination (when the energy density of the dark energy field
dominates over the one of matter) for the periodic model is around
$z\approx 0.75$. So even if one could measure SNe at the unrealistic distance
of $z=20$ we still do not have the smallest errorbars in the
interesting $z=0.75$ region. This behavior is similar for {\em all}
the dark energy models we studied. 

We examine now the behavior of the statistical error if we fit for
the parameters $w_0$ and $\om$. 
\begin{figure}[!h] 
\setlength{\unitlength}{1cm}
\centerline{\hbox{\psfig{file=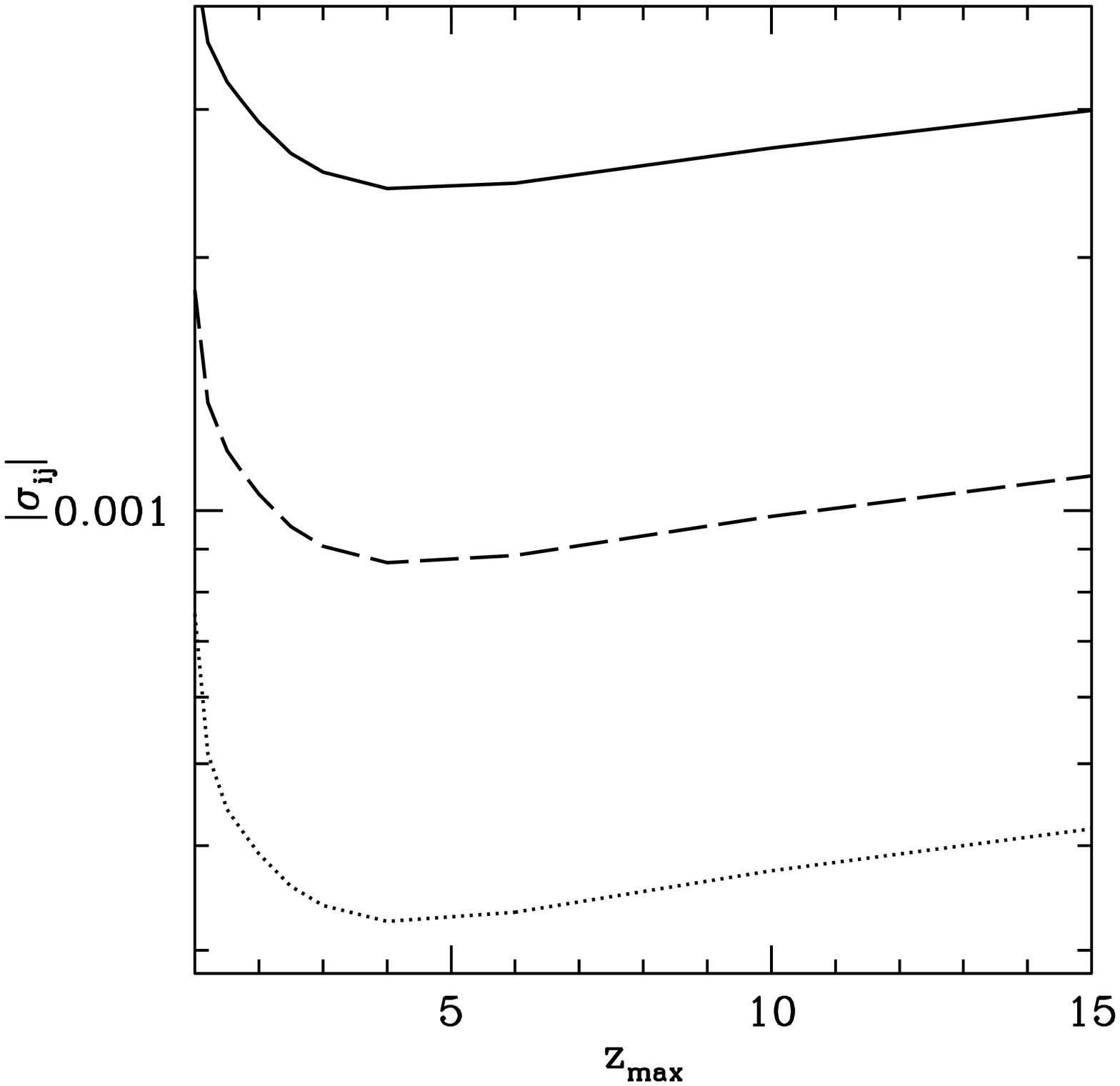,height=8.0cm,width=8cm}\psfig{file=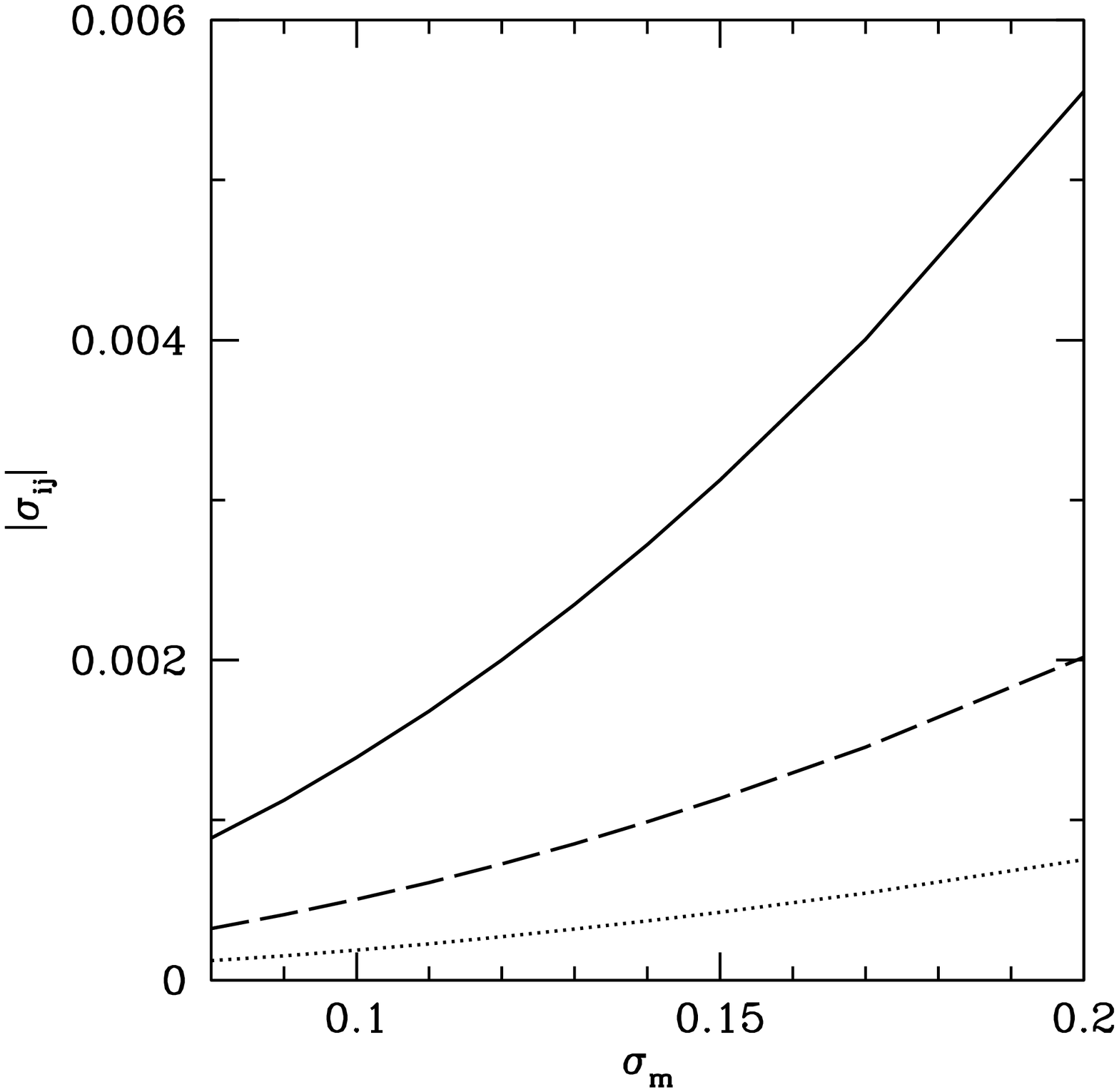,height=8.0cm,width=8cm}}}
\caption{Improvement of the statistical error on $w_0$ and $\om$ for
the fitted $\Lambda$ model. The solid line is $\sigma_{11} =
\delta\Omega_m^2$, the dashed line is $-\sigma_{01}=\left<\delta w_0
\delta\Omega_m\right>$ and the dotted line is $\sigma_{00}=\delta w_0^2$. In the left panel we show the improvement
of the uncertainty on the fit parameters
due to increasing the maximal observed redshift, where we assume that for
$90\%$ of the observation time we find low redshift SNe. In the right
panel we show the dependence of the statistical error on the residual
magnitude dispersion.}
\label{fig:sig_om}
\end{figure}
In fig.~\ref{fig:sig_om} we show the behavior of the
covariance matrix $\sigma_{ij}$ with a varying $z_{\rm max}$ and
$\sm$. In this case $\sigma_{11}$ is the 
square of the error on the best fit $\om$. In the left panel we show
the evolution with the maximal measured redshift where we construct the
sampling rate in the high and low redshift bins with
eqn.~(\ref{sampling}) and ${\cal N}_l = 1850$, $z_l=1.2$.
We recognize that we improve the statistical uncertainty by $14\%$ if
we measure SNe out to redshifts of $z_{\rm max} = 2.5$ instead of
$z_{\rm max}=1.7$. In the right panel we change the residual
dispersion $\sm$ with the rest of the parameters fixed to the
SNAP specification in table \ref{tab:snap}. The improvement on the
statistical error by going from $\sm = 0.15\, {\rm mag}$ to $\sm = 0.09\,
{\rm mag}$ is $64\%$. We will discuss the relevance of this
improvement when we examine different priors on the fit parameters
$w_0$, $w_1$ and $\om$. However we conclude that an improvement of the
measurement uncertainties is far more relevant than the ability to
observe SNe at larger redshifts.

We will now discuss how different priors on the fit parameter $\om$
influence the accuracy.  We have discussed just the most 
extreme cases of priors on the $N=1$ fit with either $\om=0.3$ fixed
or $w_1 = 0$ fixed. 
\begin{figure}[!h]
\centerline{\psfig{file=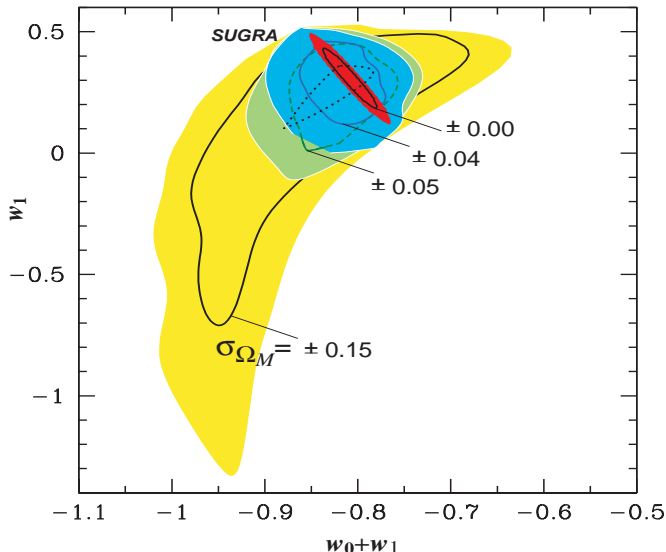,height=2.9in,width=3.5in}}
\caption{Error contours of the SUGRA model with different priors on $\om$
in the $\wt_0$--$w_1$ plane.  The solid line contours are the $39.3\%$
joint probability regions, which project to $1\sigma$ errors on the
axis and shaded regions represent
the $68\%$ or $1\sigma$ {\em joint} probabilities. The range of
increasingly larger contours represents the result using prior
knowledge on $\Omega_m$ with increasingly poorer uncertainty,
$\sigma_{\Omega_m}=\delta\om$. The unlabelled, black dotted curve corresponds to the
projected-1$\sigma$ error contour obtained for more optimistic dataset 
specifications and with a prior of $0.25\le\Omega_m\le0.35$.}
\label{fig:w1w0}
\end{figure}
In fig.~\ref{fig:w1w0} we show the decreasing errorbars in the
$\wt_0-w_1$ plane with different Gaussian priors on $\om$. Note that
we analyzed the full likelihood function and did not assume a Gaussian
shape for the probability distribution. The
current observations provide either just a crude measurement or
upper limits on $\om$. From \cite{Bahcall:98} we obtain $\om =
0.2^{+0.3}_{-0.1}$ which is too crude to result in a significant
improvement on our $w_0$-$w_1$-$\om$ estimation. The X-ray observation
in \cite{Mohr:99} give an upper limit of $\om \le 0.32 \pm 0.01$ and
the Sunyaev Zel'dovich results in \cite{Carlstrom:99} are $\om \le
0.34^{+0.05}_{-0.03}$. Future Sunyaev Zel'dovich surveys
\cite{Carlstrom:99,Holder:99} can possibly determine $\om$ up to an
accuracy of $3\%$ if 
one considers in conjunction with CMB
measurements\cite{Haiman:00}. Future X-ray surveys could determine $\om$ to even higher
accuracy independent of any other measurement\cite{Haiman:00}.

We observe in fig.~\ref{fig:w1w0} that only very tight bounds of $\om\pm\delta\om$ with
$\delta\om \le 0.05$ improve the statistical errorbars which are
stated in table \ref{tab:om_n1fit}. If we
improve the statistical error on the 
magnitude to $\sm=0.09$ and double the number observed SNe we can even
further improve the accuracy (dotted line in fig.~\ref{fig:w1w0}).
\begin{table}[!h]
 \centering
\begin{tabular}{l c | cc}
prior $\delta_{\Omega_m}$ & measurement $\sigma_{\rm mag} $ & $\delta_{w_0}$ 
& $\delta_{w_1}$ \\
\hline
No ${\Omega_m}$ prior; $w_1 = 0$& 0.15 & 0.06 &   \\
0.15 & 0.15 & 0.15 & 0.6 \\
0.05 & 0.15 & 0.06 & 0.2 \\
\,\,\," & {\it 0.09} & {\it 0.05} & {\it 0.12} \\
0.04 & 0.15 & 0.05 & 0.16 \\
0 \,\,\, (fixed ${\Omega_m}$) & 0.15 & 0.03 & 0.12 \\
\end{tabular}
\caption{Statistical measurement 
uncertainties on $w_0$ and $w_1$, given 
supernova magnitude measurement uncertainty, $\sigma_{\rm mag}$, and 
a range of uncertainties, $\delta_{\Omega_m}$, in the 
independent prior knowledge of $\Omega_m$.   (As in fig.~\ref{fig:w1w0} ,
the Supergravity model is used here as the example, but the other
models give comparable results.)}
\label{tab:w1w0}
\end{table}
In table~\ref{tab:w1w0} we summarize our findings for the accuracy of the
SNAP measurement with different priors on $\om$. If we combine these
findings with the results in table \ref{tab:wcoef} we  see that for the
periodic potential the current accuracy on $\om$ is sufficient to
establish evolution at least on the $1\sigma$ level, however for the
SUGRA model we need at least the tight prior with $\delta\om = \pm
0.05$.

A further restrictive prior for the linear fit is $-1\le\wp(z=0)<0$ which
results in the constraint $-1<w_0+w_1<0$. The constraint $-1\le \wp$
excludes non-minimally coupled scalar-tensor theories
\cite{Chiba:99a,Sahni:99,Caldwell:99}, which we have not included in our
discussion. If we analyze the results
for the $N=1$ fit with fixed $\om$ this prior does not improve the
statistical uncertainty. If we do not fix $\om$, the constraint
$-1<w_0+w_1<0$ for the linear fit, or $-1<w_0<0$ for the constant fit
also does not improve the statistical errors. From this we conclude that
either we use a tight measurement of $\om$ from an independent
observation or we can not establish evolution of $\wp$ for most of the
dark energy models.  A further prior we have
used throughout this paper is the assumption that the universe is
flat. For a non-flat universe the luminosity distance is given by
\beq
d_{\rm L}(z) = \frac{c\left(1+z\right)}{\sqrt{\left|\Omega_k
\right|}H_0}{\cal S}\left(\sqrt{\left|\Omega_k
\right|} \int\limits_0^z
\left[\Omega_k(1+z')^2+\Omega_m(1+z')^3+\Omega_\phi(1+z')^{3(w_0+1)}{\rm
e}^{3w_1z'}\right]^{-1/2}\;dz'\right)\, ,
\eeq
with $\Omega_k = 1-\Omega_\phi-\Omega_m$ and ${\cal S}(x)=\sin(x)$ for
$\Omega_k < 0$ and ${\cal S}(x)=\sinh(x)$ for $\Omega_k > 0$.
\begin{figure}[!h] 
\setlength{\unitlength}{1cm}
\centerline{\hbox{\psfig{file=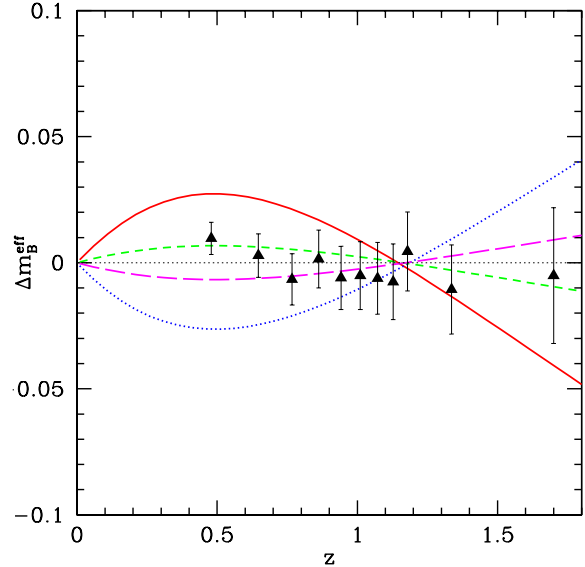,height=8.0cm,width=8cm}}}
\caption{The difference in magnitudes between non-flat models and the
fiducial $\Lambda$ cosmology, where the ratio between $\Omega_m$ and
$\Omega_\phi$ is fixed. The solid line is for $\Omega_k=-0.2$, the
dotted line for $\Omega_k=0.2$, the short dashed line for
$\Omega_k=-0.05$ and the long dashed line for $\Omega_k=0.05$. The data points correspond to the binned SNAP data as in fig.\ref{fig:snap}.}
\label{fig:openmag}
\end{figure}
In fig.\ref{fig:openmag} we show the magnitude difference to the
fiducial $\Lambda$ model for models with non-vanishing curvature
$\Omega_k$, where we fixed the ratio $\Omega_m/\Omega_\phi$. The solid
line is for $\Omega_k=-0.2$ and the dotted line for
$\Omega_k=0.2$. SNAP could clearly distinguish these models on the
$1\sigma$ level. However the current uncertainties on the curvature from
CMB and large scale structure observations are $\Delta\Omega_k= \pm
0.05$ on the $2\sigma$ level \cite{Boomerang:01b,Efstathiou:01}. The
long and short dashed lines in fig.\ref{fig:openmag} correspond to
$\Omega_k = \pm 0.05$ and we see that with the SNAP data we can not
improve the current constraints. The inclusion of a curvature term in the 
analysis presented in this paper in principle increases the
uncertainty of the estimated parameters, however in the light that future CMB
observations like MAP and Planck will provide even tighter constraints
on the curvature (for Planck $\Delta\Omega_k=\pm0.007$
\cite{Bond:97}), we assume, as mentioned before, a fixed prior of
$\Omega_k=0$.  

We turn now to the question how systematic errors influence our
ability to distinguish dark energy models and reconstruct the equation
of state factor $\wp$. The systematic error in the luminosity distance
is related to that in the magnitude $\ss$ by
\beq
	\pm\delta\dl^{\rm sys} = \left(10^{\pm\frac{\ss}{5}}-1\right)\dl\, .
\eeq
Since we marginalize over the magnitude zero point ${\cal M}$
in eqn.(\ref{magh0free}), we expect the systematic error for the
magnitude to be zero for low redshifts. We assume a linear drift of
the systematic error by
\beq
	\ss = \frac{{\hat \ss}}{1.5}z\, ,
\eeq
where ${\hat \ss}$ is the systematic error at a redshift of $z=1.5$. 
\begin{figure}[!h] 
\setlength{\unitlength}{1cm}
\centerline{\hbox{\psfig{file=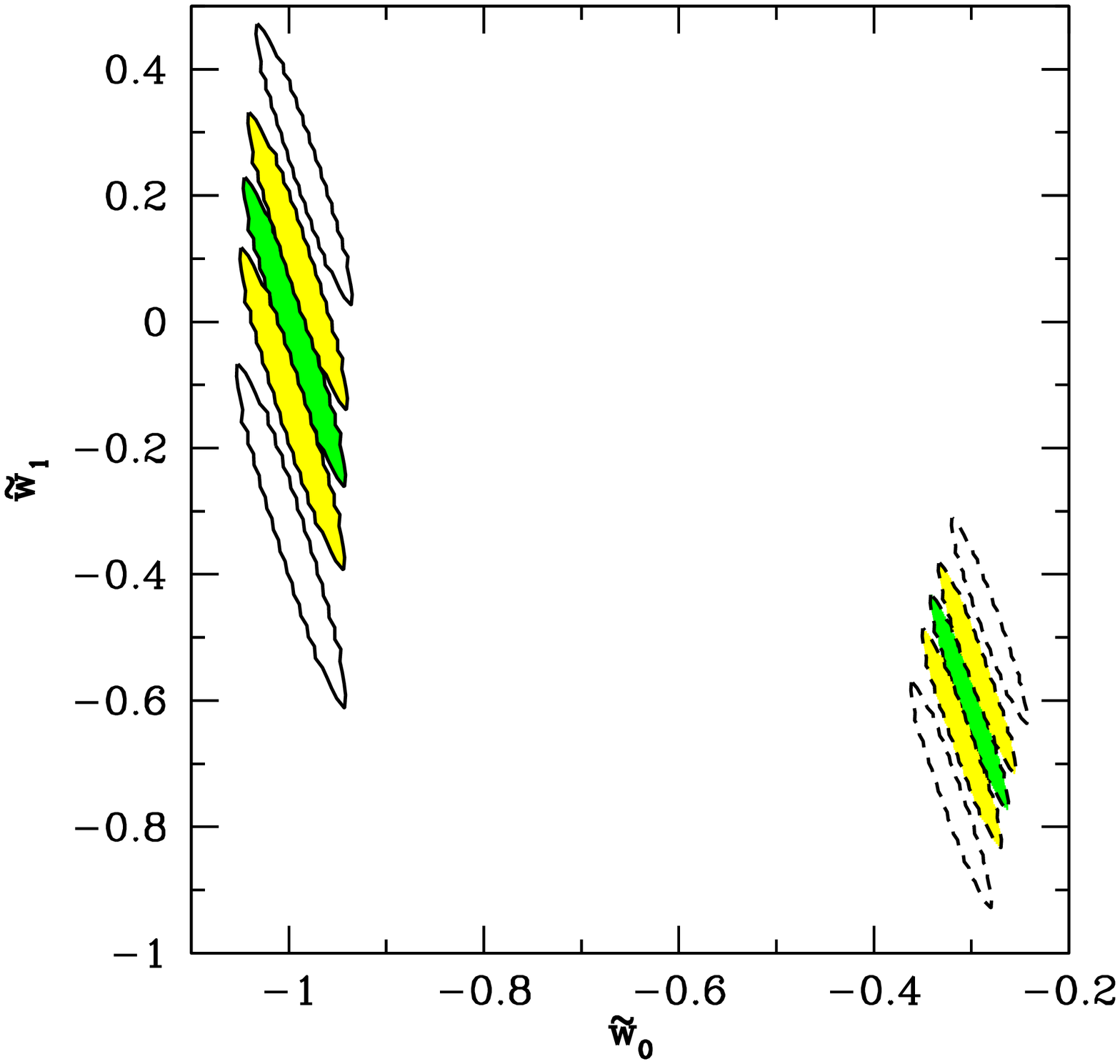,height=8.0cm,width=8cm}\psfig{file=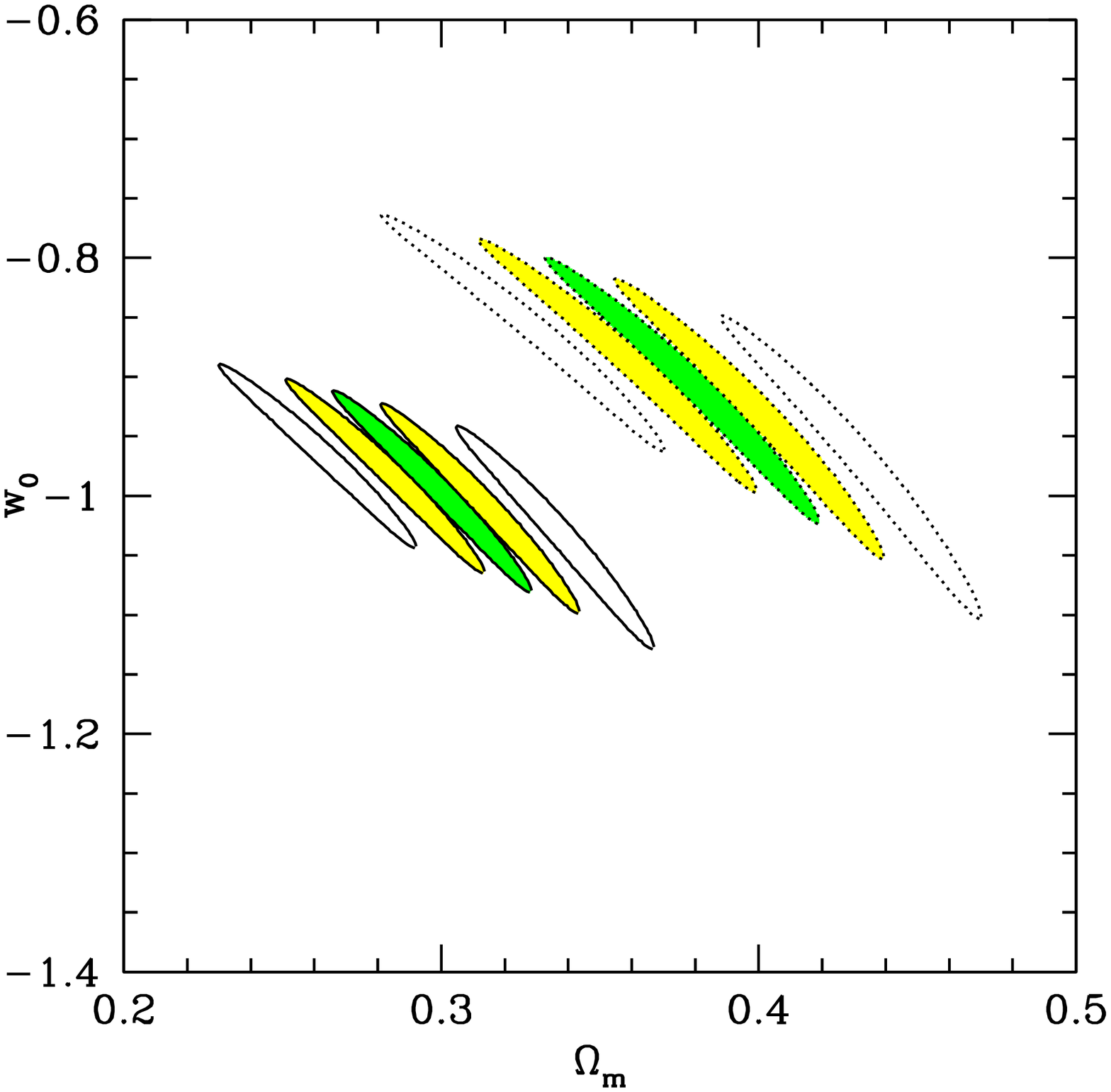,height=8.0cm,width=8cm}}}
\caption{The effects of the linear drifting systematic error on the
fitted parameters. The dark shaded region corresponds to $68.3\%$
joint probability region of the models
without systematic error, the light shaded regions to ${\hat
\ss}=\pm0.02\,{\rm mag}$ and the transparent regions to ${\hat \ss}=\pm
0.05\, {\rm mag}$. The ellipses with a solid margin are the results for the
$\Lambda$ model. In the left panel we show the results for the $N=1$
fit with $\om$ fixed. The dashed margins are from the periodic
potential model. In the right panel the results for the $\om$-$w_0$ fit
are shown, where the dotted lines correspond to the SUGRA model.}
\label{fig:sys}
\end{figure}
In fig.~\ref{fig:sys} we show the influences of the systematic
uncertainty, where we plot the $68.3\%$ joint probability regions on the
estimated parameters with the standard SNAP specifications
from table \ref{tab:snap}. The light shaded regions correspond to a
systematic error of ${\hat \ss}=\pm 0.02\, {\rm mag}$, the transparent regions
to ${\hat \ss}=\pm 0.05\,{\rm mag}$ and the dark shaded region is the result for
{\em no} systematic error. In the left panel we show the result for
the $N=1$ fit with a fixed $\om$. We note that the linear
drifting systematic error leads to a shift in the $\wt_1$ direction,
but the marginalized error on $\wt_0$ changes only slightly. The shift
for the periodic potential model (dashed lines) is much smaller than
for the $\Lambda$ model (solid lines). But if we take into account the
relative sizes of the statistical error depicted by the error
ellipses, we find that the shift due the systematic error is
roughly in agreement with these errors. In both cases the
systematic error with ${\hat \ss}=\pm 0.05\,{\rm mag}$ is only in marginal
agreement with the reconstructed values of $\wt_1$ with no systematic
error. The value of $\wt_1$ for the $\Lambda$ model is $\wt_1 = -0.011$ and the
one for the periodic potential is $\wt_0 = -0.60$ if we do not include
the systematic error. The $68.3\%$ joint
probability regions for ${\hat \ss} = \pm 0.05\,{\rm mag}$ only
marginally overlap with these values
in both cases. However,
the $99\%$ confidence regions are all in agreement even in the shifted
case with the theoretical mean values. We can reconstruct
evolution of the periodic potential model even for ${\hat \ss} =
-0.05\,{\rm mag}$ which results in $\wt_1 = -0.47$ and $\delta\wt_1 = 0.11$
which is still in agreement with $\wt_1\ne 0$ at the $3\sigma$ level. We also
expect that we can distinguish the same models 
without inclusion of a systematic error since the main difference of
the models is along the $\wt_0$ axis as evident from fig.~\ref{fig:w0w1}.
In the right panel we show the
results for the $\om$-$w_0$ fit, where the dotted ellipses correspond
to the results from the SUGRA model. As before the estimate value of
$w_0$ is not as strongly affected by the systematic error as the
fitted value of $\om$. Again the $68.3\%$ confidence
levels for ${\hat \ss} = \pm 0.05\, {\rm mag}$ (transparent regions) are only
marginally in agreement with the fitted  $\om$ for no systematic
error (dark shaded region). We conclude that the influence of the
systematics is of significant importance for the estimated $\om$ or
$\wt_1$ only at the $1\sigma$ level, but is almost negligible for the
estimation of the constant contribution $\wt_0$ to the equation of
state factor $\wp$. In order to answer the question as to whether we can
reconstruct evolution {\em and} $\om$ of the periodic model with the
SNAP experiment we have to take into account systematic errors. First,
we realize that for the standard SNAP configuration from table
\ref{tab:snap} with a systematic error of ${\hat \ss} = 0.02$ and a
statistical uncertainty we get for the best fit parameter, $w_0 =
0.75^{+0.86}_{-0.65}\pm0.04$, $w_1 = -1.07^{+0.69}_{-0.83}\pm 0.04$
and $\om = 0.41^{+0.08}_{-0.21}{}^{+0.02}_{-0.03}$, where the second
error is due to the systematics. We recognize that
we can only reconstruct evolution, which is the inconsistency of $w_1$
with a zero mean, at the $1\sigma$ level and that the
statistical error on $\om$ is nearly of the order of the mean
($50\%$). The limited amount of linear evolution we can reconstruct at the
$3\sigma$ level with this experimental setup is $w_1 \approx \pm 2.4$
and none of the models we studied has such a large linear
evolution. If we are able to increase the precision of the SNAP
satellite to the limit of $\sm = 0.09$ the best fit parameters for the
periodic model are, $w_0 = 0.75^{+0.49}_{-0.42}\pm0.04$, $w_1 =
-1.07^{+0.43}_{-0.48}\pm0.04$ and $\om =
0.41^{+0.05}_{-0.09}{}^{+0.02}_{-0.03}$. We note that for this
setup we can reconstruct evolution at the $2\sigma$ level, but also
that the errorbars on $\om$ are significantly smaller for this setup
as for the conservative SNAP specifications from table
\ref{tab:snap}. In order to establish evolution on the $3\sigma$ level,
we need a linear term of the order $w_1 \approx \pm 1.5$ which none of the
studied dark energy models fulfill. In order to improve the SNAP
proposal even further we assume that we are able to measure twice as
many SNe in each redshift interval as for the proposed SNAP specifications from table
\ref{tab:snap}. For this setup we obtain the following parameters:
$w_0 =0.75^{+0.35}_{-0.30}\pm 0.04$, $w_1 =
-1.07^{+0.31}_{-0.31}\pm0.04$ and $\om =
0.41^{+0.04}_{-0.06}{}^{+0.02}_{-0.03}$. We notice that with the $\om$
measurement we almost reach the systematic error limit and we obtain
relatively tight bounds on the matter contents. Furthermore, we have
reconstructed evolution for the periodic potential at the $3\sigma$
level and the limit on evolution we can measure with this setup is
$w_1 \approx \pm 1$. So we can conclude that if it is possible to
build SNAP with $\sm =0.09$ and measure around $N_z = 3830$ SNe out
to a redshift $z_{\rm max} =1.7$ it is possible to establish evolution
of the equation of state factor $\wp$ on the $3\sigma$ level, if the
slope of the linear evolution is $|w_1|\ge 1$.

\section{Alternatives to the SNAP mission}\label{alt}

We examine now the question of whether there could be an alternative
to SNAP. Therefore, we assume that we can improve the low
redshift results from SCP and measure about ${\cal N}_l = 160$ SNe in
the redshift 
range $z=0.1-0.55$, with a statistical uncertainty of $\sm = 0.20\,
{\rm mag}$ and a systematic error of ${\hat \ss} = 0.05\, {\rm
mag}$, where the reference redshift for the systematic error is
$z=0.5$. For the high redshift region we assume that we can observe
${\cal N}_h = 100$ SNe in a redshift range $z=2-2.5$ with the Next
Generation Space Telescope (NGST) and the same
statistical and systematic error as in the low redshift region.
\begin{figure}[!h] 
\setlength{\unitlength}{1cm}
\centerline{\hbox{\psfig{file=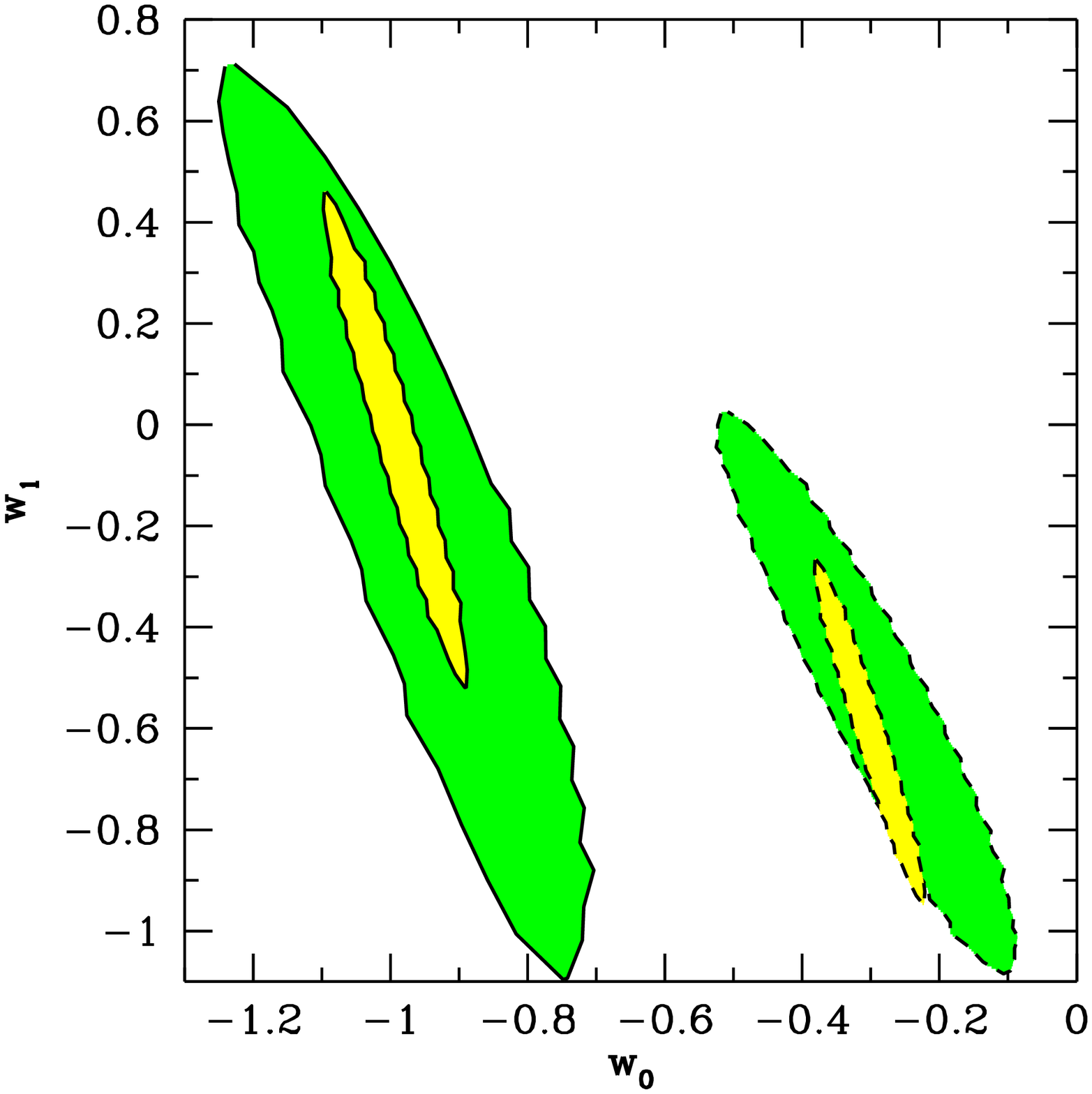,height=8.0cm,width=8cm}\psfig{file=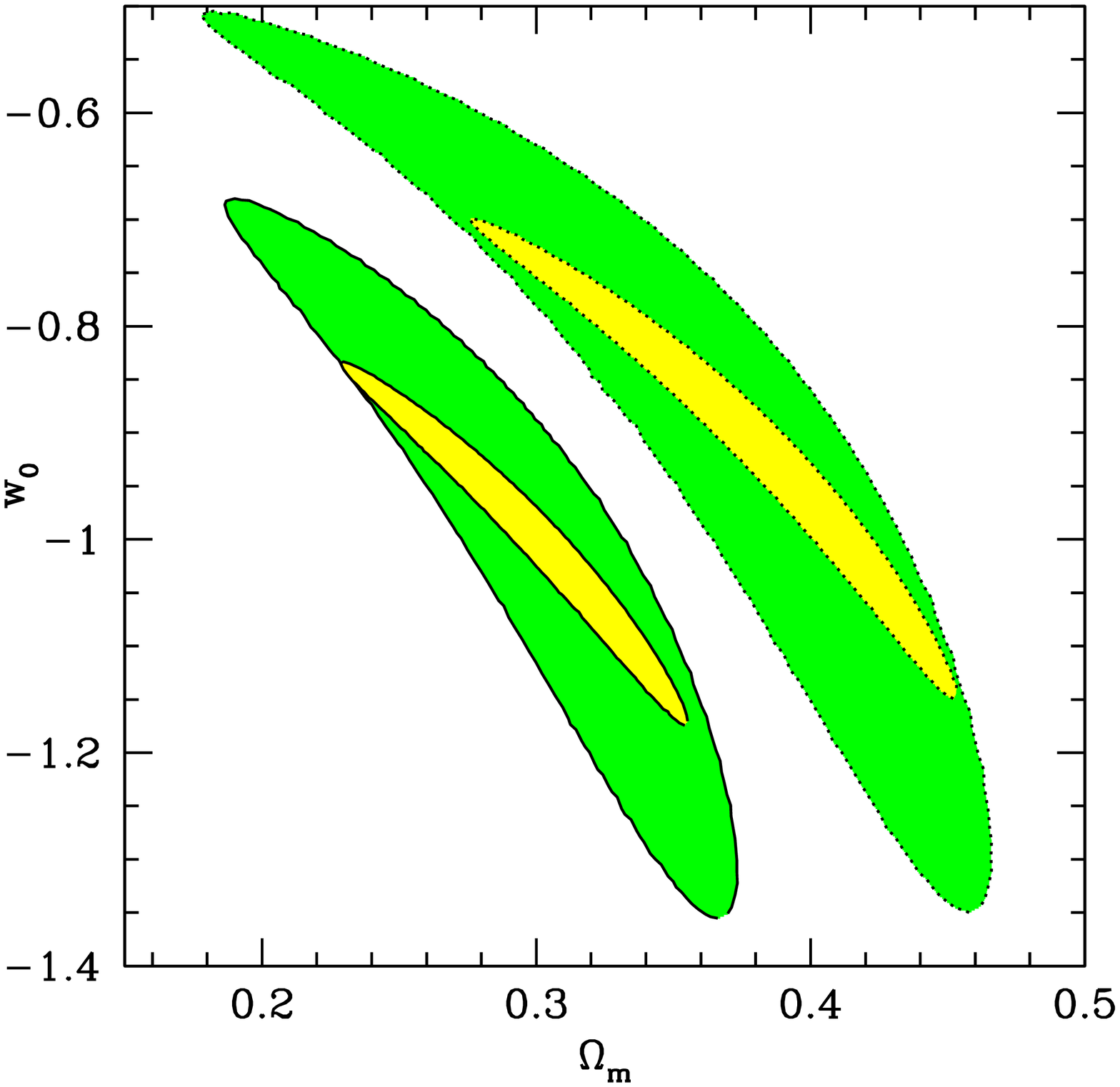,height=8.0cm,width=8cm}}}
\caption{The $99\%$ joint probability contours for $N=1$ (left panel)
and $\om$-$w_0$ fit (right panel), for the SNAP (light shaded) and low
redshift+NGST 
(dark shaded) specifications. In the left panel we plot the $\Lambda$
model (solid lines) and the periodic potential model (dashed
lines). In the right panel there is the SUGRA model (dotted lines) and
the $\Lambda$ model.}
\label{fig:ngst}
\end{figure}
In fig.~\ref{fig:ngst} we show the results of these specifications for
the $\Lambda$ model, the periodic potential (dashed line) and the
SUGRA inspired model (dotted line). In the left panel we plot the
$99\%$ joint probability regions for the linear fit with a fixed prior on
$\om$. With NGST specifications we can not establish evolution even
for the strongly evolving periodic potential model. If we study the
constant fit with no constraints on $\om$ (right panel) we can not
improve current bounds on $\om$ as given in \cite{Bahcall:98}. We have not
included the systematic error of this measurement in the analysis,
which makes the results even more spread out. Altogether it seems
clear that a refined low redshift measurement together with the NGST 
can not provide the same information concerning dark energy models as
SNAP, although NGST can complement SNAP in the high redshift region. This
leaves the question open if there are surveys which exploit 
different physics to gain information about the dark energy content of
the universe. One possibility are future galaxy cluster surveys
\cite{Haiman:00} either with an Sunyaev Zel'dovich or an X-ray survey.
The analysis for these surveys in the context of dark energy has only
been performed for constant $\wp$ models. The $3\sigma$ errors for the
joint probabilities in the $\om$-$w_0$ plane for a X-ray
survey have roughly the same size as for the SNAP specification in
fig.~\ref{fig:w0_om}. It is interesting that these methods seem to
be complementary to the SNAP observations in the sense that the error
contours are nearly perpendicular to each other \cite{Newman:00}. One of the main
drawbacks of this method is the unknown evolution 
of cluster luminosity-temperature relation. How an evolving $\wp$
could influence the estimation of parameters with galaxy cluster
abundance is not clear and should be included in a future analysis. A
different method was proposed in \cite{Newman:00} which exploits
galaxy counts with the planned Deep Extragalactic Evolutionary Probe
(DEEP) survey\cite{DEEP}. This survey gives roughly the same error
contours as the ones we present in fig.~\ref{fig:w0_om}. Again the analysis
of this survey has not included an evolving $\wp$ model. Further
possibilities to constrain dark energy could be the Alcock-Paczynski
test \cite{Alcock:79,Dalal:01}, the evolution of density perturbations
\cite{Starobinsky:98}, gravitational waves \cite{Zhu:01}, lensing surveys or observations of the Lyman-$\alpha$ forest \cite{Tegmark:01}.

\section{Conclusions}\label{conc}
We have investigated the prospects of future supernova experiments to
pin down the nature of the dark energy.  We have emphasized the
importance of choosing a good parameterization scheme for the
different dark energy models, concluding
that parameterizing the equation of state of the dark energy is the
most powerful known method, because the magnitude-redshift
functions $m(z)$ which result give the best fits to most of the dark energy
models actually proposed.  There are two caveats worth noting here.
Firstly, there might be an even better parameterization scheme out
there that has yet to be discovered (that provides even better fits).
Secondly, this scheme is unlikely to provide good fits to all possible 
models (the periodic model is an example).  The ultimate test for a
given model is to simply generate $m(z)$ for that model and compare it 
with the data.

We then use the proposed SNAP satellite as case study for what might
be possible with improved datasets. In three recent publications by
I.~Maor et al. \cite{Maor:00}, P.~Astier \cite{Astier:00} and our own
\cite{Weller:00b} the prospect of the SNAP mission
have been briefly discussed. Maor et al. \cite{Maor:00} argue that
SNAP can not distinguish different dark energy models 
and that it is nearly impossible to reconstruct evolution of the
equation of state of the dark energy component. Ours and P.~Astier's
\cite{Astier:00} results agree with their findings, but as seen
in figs.~\ref{fig:w0_om} and \ref{fig:sys}, we show that if we constrain
the analysis to a constant $\wp$ then it is well within the scope
of SNAP to distinguish dark energy models, even if we do not impose
any priors on $\om$. If it is possible to
exploit the full precision of the SNAP instrument ($\sm = 0.09\,
{\rm mag}$) and to constrain $\om$ to $0.05$, then it is even possible to
reconstruct evolution at the $3\sigma$ level as long as the linear
evolution today is above the $|w_1|>0.6$ threshold. These results are
confirmed by \cite{Astier:00}. As a conclusion we can say that SNAP
certainly has the ability to distinguish dark energy models from a
cosmological constant and possible can put some constraints on the
evolution of the equation of state of the scalar field
component. Whether or not alternative surveys, like a X-ray or SZ
survey \cite{Haiman:00} or the DEEP survey \cite{Newman:00} can
achieve the same accuracy is currently under investigation.

\section*{Acknowledgement}
We wish to thank G.~Aldering, R.~Battye, E.~Copeland, E.~Linder, A.~Lewin,
S.~Perlmutter, V.~Sahni and C.~Skordis for useful discussions. JW and
AA acknowledge support by a 
DOE grant DE-FG03-91ER40674 and UC Davis. JW is supported by
PPARC grant PPA/G/O/1999/00603.

\def\jnl#1#2#3#4#5#6{\hang{#1, {\it #4\/} {\bf #5}, #6 (#2).}}
\def\jnltwo#1#2#3#4#5#6#7#8{\hang{#1, {\it #4\/} {\bf #5}, #6; {\it
ibid} {\bf #7} #8 (#2).}} 
\def\prep#1#2#3#4{\hang{#1, #4.}} 
\def\proc#1#2#3#4#5#6{{#1, in {\it #3 (#4)\/}, edited by #5,\ (#6).}}
\def\book#1#2#3#4{\hang{#1, {\it #3\/} (#4, #2).}}
\def\jnlerr#1#2#3#4#5#6#7#8{\hang{#1 [#2], {\it #4\/} {\bf #5}, #6.
{Erratum:} {\it #4\/} {\bf #7}, #8.}}
\def\prl{Phys.\ Rev.\ Lett.}
\def\pr{Phys.\ Rev.}
\def\pl{Phys.\ Lett.}
\def\np{Nucl.\ Phys.}
\def\prp{Phys.\ Rep.}
\def\rmp{Rev.\ Mod.\ Phys.}
\def\cmp{Comm.\ Math.\ Phys.}
\def\mpl{Mod.\ Phys.\ Lett.}
\def\apj{Ap.\ J.}
\def\apjl{Ap.\ J.\ Lett.}
\def\aap{Astron.\ Ap.}
\def\cqg{Class.\ Quant.\ Grav.} 
\def\grg{Gen.\ Rel.\ Grav.}
\def\mn{MNRAS}
\def\ptp{Prog.\ Theor.\ Phys.}
\def\jetp{Sov.\ Phys.\ JETP}
\def\jetpl{JETP Lett.}
\def\jmp{J.\ Math.\ Phys.}
\def\zpc{Z.\ Phys.\ C}
\def\cupress{Cambridge University Press}
\def\pup{Princeton University Press}
\def\wss{World Scientific, Singapore}
\def\oup{Oxford University Press}
\def\asj{Astron.~J}
\def\imp{Int.\ J.\ Mod.\ Phys.}

\end{document}